\documentclass[twocolumn]{aastex631}

\begin{document}

\newcommand{\vdag}{(v)^\dagger}
\newcommand\latex{La\TeX}

\newcommand{\Msun}{\ensuremath{\textrm{\,M}_{\odot}}}
\newcommand{\Mstar}{M_{\star}}
\newcommand{\Lsun}{\ensuremath{\textrm{\,L}_{\odot}}}
\newcommand{\Mbh}{M_{\rm BH}}
\newcommand{\Mbu}{\ensuremath{\textrm{\,M}_{bulge}~}}
\newcommand{\rat}{\ensuremath{L/L_{edd}}~}
\newcommand{\degs}{\ensuremath{^{\circ}}}
\newcommand{\kms}{\ensuremath{\textrm{\,km s}^{-1}}}
\newcommand{\nth}{\ensuremath{^{\rm th}}}
\newcommand{\chisq}{\ensuremath{\chi^{2}_{red}}}
\newcommand{\chisqlt}{\ensuremath{\chi^{2}_{red} < 1}}
\newcommand{\ergs}{\ensuremath{\textrm{\,erg s}^{-1}}}
\newcommand{\microns}{\ensuremath{\mu{\mbox{m}}}}
\newcommand{\hb}{\ensuremath{\mbox{H}{\beta}}}
\newcommand{\hg}{\ensuremath{\mbox{H}{\gamma}}}
\newcommand{\hd}{\ensuremath{\mbox{H}{\delta}}}
\newcommand{\ha}{\ensuremath{\mbox{H}{\alpha}}}
\newcommand{\oiii}{[\ion{O}{3}] $\lambda$5007}
\newcommand{\oiiid}{\ensuremath{\mbox{[\ion{O}{3}] $\lambda\lambda$4959, 5007}}}
\newcommand{\heii}{\ensuremath{\mbox{\ion{He}{2} $\lambda$4686}}}
\newcommand{\niii}{\ensuremath{\mbox{\ion{N}{3} $\lambda$4100}}}
\newcommand{\niiid}{\ensuremath{\mbox{\ion{N}{3} $\lambda\lambda$4100, 4640}}}
\newcommand{\niid}{\ensuremath{\mbox{[\ion{N}{2}] $\lambda\lambda$6548, 6584}}}
\newcommand{\hei}{\ensuremath{\mbox{\ion{He}{1} $\lambda$5876}}}

\newcommand{\hinkle}[1]{J}

\newcommand{\ebv}{E(B--V)}
\newcommand{\rv}{\ensuremath{\mathrm{R_V}}}

\shorttitle{The Tidal Disruption Event AT\,2019azh}
\shortauthors{Faris et al.}

\title{Light-Curve Structure and \ha\ Line Formation in the Tidal Disruption Event AT\,2019azh}

\author[0009-0007-8485-1281]{Sara Faris}
\affiliation{School of Physics and Astronomy, Tel Aviv University, Tel Aviv 69978, Israel}

\author[0000-0001-7090-4898]{Iair Arcavi}
\affiliation{School of Physics and Astronomy, Tel Aviv University, Tel Aviv 69978, Israel}

\author[0000-0002-7466-4868]{Lydia Makrygianni}
\affiliation{School of Physics and Astronomy, Tel Aviv University, Tel Aviv 69978, Israel}

\author[0000-0002-1125-9187]{Daichi Hiramatsu}
\affiliation{Center for Astrophysics, Harvard \& Smithsonian, 60 Garden Street, Cambridge, MA 02138-1516, USA}
\affiliation{The NSF AI Institute for Artificial Intelligence and Fundamental Interactions, USA}

\author[0000-0003-0794-5982]{Giacomo~Terreran}
\affiliation{Las Cumbres Observatory, 6740 Cortona Drive, Suite 102, Goleta, CA 93117-5575, USA}
\affiliation{Department of Physics, University of California, Santa Barbara, CA 93106-9530, USA}

\author[0000-0003-4914-5625]{Joseph Farah}
\affiliation{Las Cumbres Observatory, 6740 Cortona Drive, Suite 102, Goleta, CA 93117-5575, USA}
\affiliation{Department of Physics, University of California, Santa Barbara, CA 93106-9530, USA}

\author[0000-0003-4253-656X]{D. Andrew Howell}
\affiliation{Las Cumbres Observatory, 6740 Cortona Drive, Suite 102, Goleta, CA 93117-5575, USA}
\affiliation{Department of Physics, University of California, Santa Barbara, CA 93106-9530, USA}

\author[0000-0001-5807-7893]{Curtis McCully}
\affiliation{Las Cumbres Observatory, 6740 Cortona Drive, Suite 102, Goleta, CA 93117-5575, USA}

\author[0000-0001-9570-0584]{Megan Newsome}
\affiliation{Las Cumbres Observatory, 6740 Cortona Drive, Suite 102, Goleta, CA 93117-5575, USA}
\affiliation{Department of Physics, University of California, Santa Barbara, CA 93106-9530, USA}

\author[0000-0003-0209-9246]{Estefania~Padilla~Gonzalez}
\affiliation{Las Cumbres Observatory, 6740 Cortona Drive, Suite 102, Goleta, CA 93117-5575, USA}
\affiliation{Department of Physics, University of California, Santa Barbara, CA 93106-9530, USA}

\author[0000-0002-7472-1279]{Craig Pellegrino}
\affiliation{Department of Astronomy, University of Virginia, Charlottesville, VA 22904, USA}

\author[0000-0002-4924-444X]{K. Azalee Bostroem}
\altaffiliation{LSSTC Catalyst Fellow}
\affiliation{Steward Observatory, University of Arizona, 933 North Cherry Avenue, Tucson, AZ 85721-0065, USA}

\author{Wiam Abojanb}
\affiliation{Atid Peki'in Comprehensive School, Peki'in, 2491400, Israel}

\author[0000-0002-9347-2298]{Marco C. Lam}
\affiliation{Institute for Astronomy, University of Edinburgh, Royal Observatory, Blackford Hill, Edinburgh, EH9 3HJ, UK}

\author[0000-0002-3697-2616]{Lina Tomasella}
\affiliation{INAF – Osservatorio Astronomico di Padova, Vicolo dell’Osservatorio 5, 35122 Padova, Italy}

\author[0000-0001-5955-2502]{Thomas G. Brink}
\affiliation{Department of Astronomy, University of California, Berkeley, CA 94720-3411, USA}

\author[0000-0003-3460-0103]{Alexei V. Filippenko}
\affiliation{Department of Astronomy, University of California, Berkeley, CA 94720-3411, USA}

\author[0000-0002-4235-7337]{K. Decker French}
\affiliation{Department of Astronomy, University of Illinois, 1002 W. Green Street, Urbana, IL 61801, USA}

\author[0000-0002-6576-7400]{Peter Clark}
\affiliation{Institute of Cosmology and Gravitation, University of Portsmouth, Portsmouth PO1 3FX, UK}

\author[0000-0002-4391-6137]{Or Graur}
\affiliation{Institute of Cosmology and Gravitation, University of Portsmouth, Portsmouth PO1 3FX, UK}
\affiliation{Department of Astrophysics, American Museum of Natural History, Central Park West and 79th Street, New York, NY 10024-5192, USA}

\author[0000-0002-8597-0756]{Giorgos Leloudas}
\affiliation{DTU Space, National Space Institute, Technical University of Denmark, Elektrovej 327, 2800 Kgs. Lyngby, Denmark}

\author[0000-0002-1650-1518]{Mariusz Gromadzki}
\affiliation{Astronomical Observatory, University of Warsaw, Al. Ujazdowskie 4,
00-478 Warszawa, Poland}

\author[0000-0003-0227-3451]{Joseph P. Anderson}
\affiliation{European Southern Observatory, Alonso de C\'ordova 3107, Casilla 19, Santiago, Chile}
\affiliation{Millennium Institute of Astrophysics MAS, Nuncio Monsenor Sotero Sanz 100, Off.
104, Providencia, Santiago, Chile}

\author[0000-0002-2555-3192]{Matt Nicholl}
\affiliation{Astrophysics Research Centre, School of Mathematics and Physics, Queens University Belfast, Belfast BT7 1NN, UK}

\author[0000-0003-2375-2064]{Claudia P. Guti\'errez}
\affiliation{Institut d’Estudis Espacials de Catalunya (IEEC), E-08034 Barcelona, Spain}
\affiliation{Institute of Space Sciences (ICE-CSIC), Campus UAB, Carrer de Can Magrans, s/n, E-08193 Barcelona, Spain}

\author[0000-0001-8257-3512]{Erkki Kankare}
\affiliation{Department of Physics and Astronomy, University of Turku, FI-20014 Turku, Finland}

\author[0000-0002-3968-4409]{Cosimo Inserra}
\affiliation{Cardiff Hub for Astrophysics Research and Technology, School of Physics \& Astronomy, Cardiff University, Queens Buildings, The Parade, Cardiff, CF24 3AA, UK}

\author[0000-0002-1296-6887]{Llu\'is Galbany}
\affiliation{Institute of Space Sciences (ICE-CSIC), Campus UAB, Carrer de Can Magrans, s/n, E-08193 Barcelona, Spain}
\affiliation{Institut d’Estudis Espacials de Catalunya (IEEC), E-08034 Barcelona, Spain}

\author[0000-0002-1022-6463]{Thomas Reynolds}
\affiliation{Cosmic Dawn Center (DAWN), Niels Bohr Institute, University of Copenhagen, 2200, Denmark}
\affiliation{Department of Physics and Astronomy, University of Turku, FI-20014 Turku, Finland}

\author[0000-0001-7497-2994]{Seppo Mattila}
\affiliation{Department of Physics and Astronomy, University of Turku, FI-20014 Turku, Finland}
\affiliation{School of Sciences, European University Cyprus, Diogenes Street, Engomi, 1516 Nicosia, Cyprus}

\author[0000-0002-7845-8965]{Teppo Heikkil\"{a}}
\affiliation{Department of Physics and Astronomy, University of Turku, FI-20014 Turku, Finland}

\author[0000-0003-3207-5237]{Yanan Wang}
\affiliation{National Astronomical Observatories, Chinese Academy of Sciences, 20A Datun Road, Beijing 100101, China}
\affiliation{Physics \& Astronomy, University of Southampton, Southampton, Hampshire SO17~1BJ, UK}

\author[0000-0001-6286-1744]{Francesca Onori}
\affiliation{INAF - Osservatorio Astronomico d'Abruzzo, via M. Maggini snc, I-64100 Teramo, Italy}

\author[0000-0002-4043-9400]{Thomas Wevers}
\affiliation{Space Telescope Science Institute, 3700 San Martin Drive, Baltimore, MD 21218, USA}
\affiliation{European Southern Observatory, Alonso de C\'ordova 3107, Casilla 19, Santiago, Chile}

\author[0000-0003-3765-6401]{Eric R. Coughlin}
\affiliation{Department of Physics, Syracuse University, Syracuse, NY 13210, USA}

\author[0000-0002-0326-6715]{Panos Charalampopoulos}
\affiliation{Department of Physics and Astronomy, University of Turku, Vesilinnantie 5, FI-20500, Finland}

\author[0000-0001-5975-290X]{Joel Johansson}
\affiliation{Oskar Klein Centre, Department of Physics, Stockholm University, AlbaNova, SE-10691 Stockholm, Sweden}

\correspondingauthor{Sara Faris}
\email{sarafaris452@gmail.com}

\begin{abstract}

AT\,2019azh is a H+He tidal disruption event (TDE) with one of the most extensive ultraviolet and optical datasets available to date.
We present our photometric and spectroscopic observations of this event starting several weeks before and out to approximately 2 years after the $g$-band's peak brightness and combine them with public photometric data. This extensive dataset robustly reveals a change in the light-curve slope and a possible bump in the rising light curve of a TDE for the first time, which may indicate more than one dominant emission mechanism contributing to the pre-peak light curve. Indeed, we find that the \texttt{MOSFiT}-derived parameters of AT\,2019azh, which assume reprocessed accretion as the sole source of emission, are not entirely self-consistent. We further confirm the relation seen in previous TDEs whereby the redder emission peaks later than the bluer emission. The post-peak bolometric light curve of AT\,2019azh is better described by an exponential decline than by the canonical $t^{-5/3}$ (and in fact any) power-law decline. We find a possible mid-infrared excess around the peak optical luminosity, but cannot determine its origin. In addition, we provide the earliest measurements of the \ha\ emission-line evolution and find no significant time delay between the peak of the $V$-band light curve and that of the \ha\ luminosity. These results can be used to constrain future models of TDE line formation and emission mechanisms in general. More pre-peak 1--2\,days cadence observations of TDEs are required to determine whether the characteristics observed here are common among TDEs. More importantly, detailed emission models are needed to fully exploit such observations for understanding the emission physics of TDEs.
\end{abstract}


\section{Introduction} \label{sec: intro}

Supermassive black holes (SMBHs), with masses of $\gtrsim10^6\,\Msun$, are thought to reside in the center of most (if not all) large galaxies in the local Universe.
While some SMBHs, known as active galactic nuclei (AGNs), accrete material that emits radiation, the majority are quiescent \citep[e.g.,][]{Greene2007, Mullaney2013} and thus difficult to study.

One of the few probes that can be used to study inactive SMBHs is the emission produced in a tidal disruption event (TDE). 
A TDE occurs when a star passes close enough to an SMBH for tidal forces to surpass the star's self-gravity, causing its disruption. In a full disruption, the star is torn apart and approximately half of it becomes gravitationally bound to the SMBH and eventually accretes onto it \citep{Rees1988, EvansKochanek1989, Phinney1989}. 

This transient phenomenon can not only serve to confirm the presence of an SMBH but also offers a promising tool for constraining its mass and perhaps even spin \citep[e.g.,][]{Leloudas2016}. As such, TDEs can potentially provide a more complete picture of the SMBH population. This can, in turn, help address some of the open questions regarding SMBHs, from accretion physics through their sub- and super-Eddington growth mechanisms to their scaling relations with global galaxy properties (such as the famous $M$--$\sigma$ relation; e.g., \citealt{Kormendy2013}). However, a main unresolved challenge lies in mapping TDE emission properties to SMBH characteristics.

The first discovered TDEs were searched for and detected in X-ray observations (e.g., \citealt{Bade1996, Komossa&Greiner1999, Cappelluti2009, Maksym2014}; see \citealt{Saxton2020} for a recent review), as the transient accretion disk was expected to emit at these wavelengths. However, in recent years, wide-field optical transient surveys have been discovering a growing number of TDEs in the optical bands, which are also bright in ultraviolet (UV) wavelengths (e.g., \citealt{Gezari2006, vanVelzen2011, Gezari2012, Arcavi2014}; see \citealt{vanVelzen2020} and \citealt{Gezari2021} for recent reviews). 
This surprising discovery has prompted work on theoretical models of TDEs to explain the optical/UV emission properties of these events. 

Two main mechanisms for producing optical/UV emission in TDEs have been proposed. The first is the reprocessing of X-ray emission from an accretion disk by optically thick material surrounding the disk \citep[e.g.,][]{Guillochon2014, Roth2016, Dai2018}. The second model attributes the optical/UV emission to shocks formed between stellar debris streams as they collide around apocenter before circularizing to form an accretion disk \citep{Piran2015}. Numerical simulations by \cite{Steinberg2024} suggest a possible intermediate scenario whereby circularization can begin already at the pericenter, but the emission responsible for the light-curve peak is driven mainly by stream–disk shocks, which further circularizes the debris.

UV/optical TDEs are characterized by a luminous peak with a typical absolute magnitude of $\sim -20$ in the optical (a few events have been found down to peak magnitudes of $\sim -17$), rise timescales of days to weeks, and a smooth decline in the light curve lasting weeks to years \citep[e.g.,][]{vanVelzen2020,vanVelzen2021}. The blackbody temperature of these events remains high and approximately constant at $T \approx 10^4$\,K \citep[e.g.,][]{Gezari2012, Arcavi2014, vanVelzen2020}. Their bolometric luminosity sometimes follows a decline rate consistent with a $t^{-5/3}$ power law, which aligns with theoretical expectations for the mass return rate \citep{Rees1988, EvansKochanek1989, Phinney1989}. 

Spectroscopically, UV/optical TDEs show a strong blue continuum with broad ($\sim 10^4\,\kms$) \heii\ \citep{Gezari2012, Arcavi2014} and/or broad Balmer emission lines \citep[e.g.,][]{Arcavi2014, Gezari2015, Hung2017}, denoted H- He- or H+He-TDEs, accordingly \citep{vanVelzen2021}. The width of the emission lines was initially attributed to Doppler broadening \citep{Ulmer1999, Bogdanovi2004, Guillochon2013}. However, it was later suggested that at least some of the line broadening is caused by electron scattering \citep{roth&kasen2018}.
Some TDE spectra also exhibit \hei\ and/or heavier elements, such as \oiii\ and \niiid\ (sometimes blended with \heii; \citealt{Blagorodnova2017, Onori2019, Leloudas2019}).
Some of these lines have been attributed to the Bowen fluorescence mechanism \citep{Bowen1934}, whereby extreme UV photons generate a specific cascade of lines. TDEs showing these lines are known as Bowen TDEs.

Some UV/optical TDEs are accompanied by X-ray and/or radio emission \citep[e.g.,][]{Brown2017,Saxton2020,Abolfathi2018,Cendes2022,Liu2022,Bu2023}. The X-rays are attributed to direct accretion emission, while the source of the radio emission is debated. It has been suggested to originate in outflows \citep{Alexander2016}, jets \citep{vanVelzen2016}, and in the interaction between the unbound material and the interstellar medium \citep{Krolik2016}. In addition, delayed radio flares have recently been discovered to occur years after the optical peak in a few TDEs \citep{Horesh2021}. Their nature is also debated.

Here, we present and analyze extensive optical and UV observations, and available mid-infrared (MIR) observations, of the TDE AT\,2019azh. X-ray, UV, and optical observations of this event were studied by \cite{Hinkle2021a}, \cite{vanVelzen2021}, \cite{Liu2022}, and \cite{Hammerstein2023}, and long-duration radio emission by \cite{Goodwin2022} and \cite{Sfaradi2022}. Spectropolarimetry of AT\,2019azh was studied by \cite{Leloudas2022} and found to have the lowest polarization among the sample of TDEs studied. 

We complement published optical and UV data of AT\,2019azh with our own. The combined optical and UV dataset presented here makes AT\,2019azh one of the best-observed TDEs so far at these wavelengths, both photometrically and spectroscopically. We describe our observations in Section \ref{sec: observations} and our analysis in Section \ref{sec: analysis}, discuss our results in Section \ref{sec: discussion}, and summarize in Section \ref{sec:summary}.
We assume a flat $\Lambda$CDM cosmology, with $H_0 = 69.6$\,\kms\,Mpc$^{-1}$, $\Omega_m = 0.286$, and $\Omega_{\Lambda} = 0.714$ \citep{Wright2006, Bennett2014}. 

\section{Observations \&\ Data Reduction} \label{sec: observations}

\subsection{Discovery and Classification}

AT\,2019azh was discovered on 2019 February 22 at 00:28:48 (UTC dates are used throughout this paper) \citep[MJD 58536.02;][]{Stanek2019} by the All-Sky Automated Survey for Supernovae \citep[ASAS-SN;][]{Shappee2014} as ASASSN-19dj with a $g$-band apparent magnitude of $\sim$16.2. The event was also detected by the Gaia photometric science alert team \citep{Hodgkin2021H}\footnote{\url{http://gsaweb.ast.cam.ac.uk/alerts}} as Gaia19bvo, and by the Zwicky Transient Facility \citep[ZTF;][]{Bellm2019} as ZTF17aaazdba and ZTF18achzddr\footnote{The multiple names with pre discovery years are due to random image subtraction artifacts, which are common in galaxy nuclei, erroneously identified as possible transients.}. The location of the event (Gaia J2000 coordinates $\alpha = 08^{\rm hr}13^{\rm m}16.96^{\rm s}$, $\delta = +22^\circ 38' 53.99''$) is consistent with the center of the nearby galaxy KUG 0810+227, which has a redshift of $z = 0.0222240\pm0.0000071$ \citep{Almeida2023}, corresponding to a luminosity distance of 96.6\,Mpc. This galaxy was preselected by \cite{French2018} as a possible TDE host, given its post-starburst properties \citep{Arcavi2014}. 

The first few spectra of AT\,2019azh showed a strong blue continuum without obvious features \citep{TNS2019class, Heikkila2019}. The event was later classified as a TDE by \cite{vanVelzen2019ATel}, based on its brightness, high blackbody temperature of $\sim 30,000$\,K, a position consistent with the center of the galaxy (with an angular offset between the ZTF coordinates of the event and the host nucleus of $0.07\arcsec\pm0.31\arcsec$), multiple spectra showing a strong blue continuum, and lack of spectroscopic features associated with a supernova (SN) or AGN. 

\subsection{Photometry}

We obtained optical follow-up imaging of AT\,2019azh with the Las Cumbres Observatory \citep{Brown2013} global network of 1\,m telescopes starting on MJD 58537.06 in the $BgVri$ bands. Standard image processing was performed using the BANZAI automated pipeline \citep{McCully2018}. We combine our set of images with that of \cite{Hinkle2021a} and perform reference-subtraction to remove host galaxy contamination using the High Order Transform of PSF and Template Subtraction algorithm \citep{Alard_1998, Alard, Becker2015} implemented by the \texttt{lcogtsnpipe} image subtraction pipeline \citep{Valenti2016}\footnote{\url{https://github.com/LCOGT/lcogtsnpipe}}.
We use Las Cumbres Observatory images taken at MJD 59131.40 ($\sim 596$\,days after discovery), after the transient faded, as references.
Photometry was calibrated to the Sloan Digital Sky Survey (SDSS) Data Release 14 \citep{Abolfathi2018} for the $gri$ bands and to the AAVSO Photometric All-Sky Survey (APASS) Data Release 9 \citep{Henden2016} for the $BV$ bands. 

AT\,2019azh was observed by all five ASAS-SN units in the $g$ band, with the first detection recorded at MJD 58529.12. We use the ASAS-SN host subtracted photometry as provided by \cite{Hinkle2021a}.

The Swope \citep{Bowen:73} 1\,m telescope at Las Campanas Observatory observed AT\,2019azh in the $uBgVri$ filters starting at MJD 58549.10. We use the Swope host subtracted photometry as provided by \cite{Hinkle2021a}.

We retrieved host subtracted photometry from the Asteroid Terrestrial-impact Last Alert System \citep[ATLAS;][]{Tonry_2018, Smith_2020} in its $c$ and $o$ bands using the ATLAS public forced photometry server\footnote{\url{https://fallingstar-data.com/forcedphot/}}. AT\,2019azh was first detected by ATLAS on MJD 58529.37. More details regarding ATLAS data processing and photometry extraction can be found in \cite{Tonry_2018} and \cite{Smith_2020}.

We retrieved ZTF host subtracted photometry from the public ZTF forced photometry server\footnote{\url{https://ztfweb.ipac.caltech.edu/cgi-bin/requestForcedPhotometry.cgi}}. The event was detected in the ZTF $g$ and $r$ bands starting from MJD 58512.26. A description of forced photometry processing for ZTF can be found in \cite{Masci_2019}. 

The Neil Gehrels Swift Observatory \citep[hereafter, {\it Swift};][]{Roming2005} observed AT\,2019azh with all its UltraViolet and Optical Telescope (UVOT) filters ($b$, $v$, $u$, $uvw1$, $uvm2$, and $uvw2$), starting on MJD 58544.76 (PIs Arcavi, Hinkle, and Gezari). We take the host subtracted extinction-corrected UVOT photometry from \cite{Hinkle2021b}, which incorporates the new UVOT calibrations\footnote{\url{https://www.swift.ac.uk/analysis/uvot/index.php}} not available in the earlier work by \cite{Hinkle2021a}.

We retrieve the available MIR photometry obtained by the Wide-field Infrared Survey Explorer \citep[{\it WISE};][]{Wright2010} NEOWISE Reactivation Releases \citep{Mainzer2011, Mainzer2014} through the NASA/IPAC Infrared Science Archive. {\it WISE} obtains several images of each object during each observing phase (once every six months). We process these data using a custom Python script. The script filters out any individual observation identified as an upper limit and those with observational issues, such as being obtained close to the sky position of the Moon or suffering from poor frame quality. Weighted averages for each visit are then calculated per filter.
We estimate the host galaxy flux and its uncertainty as the average and variance (respectively) of all pre-TDE observations and then subtract this flux from all observations. 

\begin{deluxetable}{cccccc}[t]
    \label{tab:phot_table}
    \centering
    \caption{host subtracted and Milky Way extinction-corrected photometry and $3\sigma$ nondetection upper limits.}
    \tablehead{
    \colhead{MJD} & \colhead{\begin{tabular}[c]{@{}c@{}}Phase\\ (days)\end{tabular}} & \colhead{Magnitude} &\colhead{Error} & \colhead{Filter} & \colhead{Source} 
    }
    \startdata
58509.23  & -55.93 & $>19.66$ & $\vcenter{\hbox{$\ldots$}}$ & g & ZTF \\
58509.28 & -55.88 & $>19.88$ &  $\vcenter{\hbox{$\ldots$}}$ & r &  ZTF \\
58512.26  & -52.90 &  18.86 &  0.05 &  $r$  &  ZTF   \\
58522.18   & -42.98 &  20.13 &  0.15  &  $g$   &  ZTF \\
58537.07 & -28.09 &  16.19 &  0.02 &  $r$    &  Las Cumbres   \\
58537.07   & -28.09 &  15.69   &  0.09 &  $g$   &  Las Cumbres \\
58571.85 & 6.69 & 17.14 & 0.01 & $W1$ & {\it WISE} \\
58571.85 & 6.69  & 17.47 & 0.01 & $W2$ & {\it WISE} \\
   \enddata
    \tablecomments{This table is published in its entirety in the machine-readable format. A portion is shown here for guidance regarding its form and content.}
\end{deluxetable}

We correct all optical and UV photometry for Milky Way extinction assuming a \cite{Cardelli1989} extinction law with $R_{\rm V}=3.1$ and Galactic extinction of $A_{\rm V}=0.122$\,mag, as retrieved from the NASA Extragalactic Database\footnote{\url{https://ned.ipac.caltech.edu/extinction_calculator}} using the \cite{Schlafly_2011} extinction map. We correct the {\it WISE} MIR photometry for extinction using the \cite{Fitzpatrick1999} extinction law with the corresponding coefficients from \cite{Yuan2013}.
All photometry is presented in the AB system \citep{ABsys}, except for the Las Cumbres $BV$-band data, which are presented in the Vega system.

\begin{figure*}[t!]
    \centering
    \includegraphics[width=\textwidth]{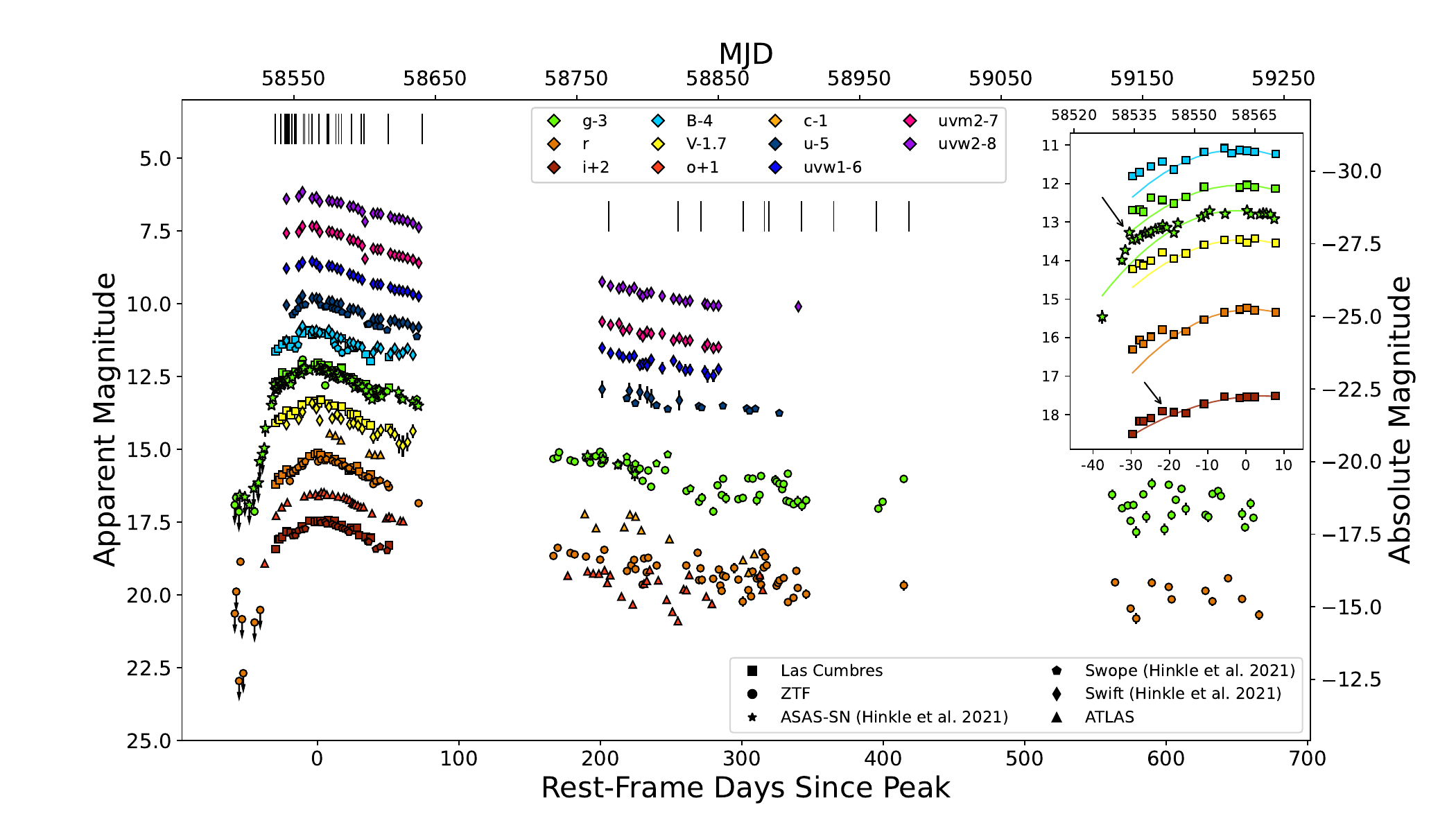}
    \caption{Milky Way extinction-corrected UV and optical light curves of AT\,2019azh from \cite{Hinkle2021a, Hinkle2021b} and this work. Error bars denote 1$\sigma$ uncertainties and are sometimes smaller than the marker size. Markers with arrows indicate 3$\sigma$ nondetection upper limits. Black vertical lines indicate epochs with spectroscopic data. The inset displays the rise of the light curve in the data series, which covers the change in slope (at $\sim -30$\,days from the peak) and possible bump (at $\sim-20$\,days), marked with arrows (nondetections are omitted for clarity). The lines in the inset represent fits to the post-bump light curve, see the text for details.}
    \label{fig:lc}
\end{figure*}

The photometry obtained here from Las Cumbres, ATLAS, and ZTF are presented in Table \ref{tab:phot_table}. This photometry, together with the ASAS-SN and Swope photometry from \cite{Hinkle2021a}, and the {\it Swift} photometry from \cite{Hinkle2021b}, is presented in Figure \ref{fig:lc}. The {\it WISE} photometry is also presented in Table \ref{tab:phot_table} and in Figure \ref{fig:MIR_lc} in Appendix \ref{sec:appendix-MIR}. We present all phases relative to $g$-band peak brightness at MJD $58566.70 \pm 0.52$ (as calculated in Section \ref{sec: analysis}).

\subsection{Spectroscopy}

We obtained spectroscopic observations using the FLOYDS spectrographs \citep{Sand2011} mounted on the Las Cumbres Observatory 2\,m Faulkes Telescope South located at the Siding Spring Observatory in Australia and Faulkes Telescope North located at the Haleakal\=a Observatory in Hawaii, the ESO Faint Object Spectrograph and Camera \citep[EFOSC2;][]{Buzzoni1984} mounted on the 3.58\,m ESO New Technology Telescope (NTT) as part of the extended Public ESO Spectroscopic Survey for Transient Objects (ePESSTO), the Asiago Faint Object Spectrographic Camera (AFOSC) mounted on the Copernico 1.82\,m Telescope in Asiago, Mount Ekar, the Intermediate-dispersion Spectrograph and Imaging System (ISIS) mounted on the 4.2\,m William Herschel Telescope (WHT), the Wide field reimaging CCD camera (WFCCD) mounted on the du Pont 2.5\,m telescope at the Las Campanas Observatory, the Kast Double Spectrograph \citep{kast} mounted on the Shane 3\,m telescope at Lick Observatory, and the Alhambra Faint Object Spectrograph and Camera (ALFOSC) mounted on the 2.56\,m Nordic Optical Telescope (NOT) through the second NOT Un-biased Transient Survey  program\footnote{\url{https://nuts.sn.ie}}.

The FLOYDS spectra were processed and reduced using a custom PYRAF-based pipeline\footnote{\url{https://github.com/LCOGT/floyds_pipeline}}. This pipeline, based on the Image Reduction and Analysis Facility \citep[IRAF;][]{IRAF1, IRAF2} framework, removes cosmic rays and performs wavelength and flux calibration and rectification, flat-field correction, and spectrum extraction.

The Copernico 1.82\,m Telescope spectra were reduced using a custom reduction pipeline based on IRAF tasks. After bias and flat-field correction, spectra were extracted and wavelength calibrated. Nightly sensitivity functions were derived from observations of spectrophotometric standard stars (also used to derive the corrections for the telluric absorption bands). 

The NTT spectra were reduced using the Python-based PESSTO pipeline \citep{Smartt2015}\footnote{\url{https://github.com/svalenti/pessto}}. This pipeline encompasses essential steps, including detector bias calibration, flat-field calibration, cosmic-ray removal, comparison lamp frames, and wavelength and flux calibrations.
The first NTT spectrum, obtained on MJD 58539.16, is publicly available on the Transient Name Server\footnote{\url{http://www.wis-tns.org/}} \citep{TNS2019class}.

The WHT/ISIS spectrum was reduced using custom recipes executed in IRAF. The use of the medium-resolution gratings (R600B and R600R) results in a gap in wavelength coverage between the blue and red arms. Overscan correction, bias subtraction, flat-field correction, and cosmic-ray removal were performed. Wavelength calibration is derived from comparison lamp frames taken at the same position to correct instrument flexure. The optimal extraction algorithm of \citet{Horne1986} is used to extract the one-dimensional spectra. A photometric standard star was observed on the same night to derive the flux calibration.

Observations with the WFCCD on the 2.5\,m du Pont telescope were obtained using a 1.65\arcsec\ (150\,\microns) slit and the blue grism. Average seeing conditions were $\sim$0.5\arcsec. Data were reduced and calibrated using custom Python routines and standard star observations.

The Lick/Kast spectra were taken with the 600/4310 grism, the 300/7500 grating, and the D57 dichroic. All observations were made with the 2.0\arcsec\ slit. This instrument configuration has a combined wavelength range of $\sim$3600--10,700\,\AA, and a spectral resolving power of $R\approx800$. The data were reduced following standard techniques for CCD processing and spectrum extraction \citep{Silverman2012} utilizing IRAF routines and custom Python and IDL codes\footnote{\url{https://github.com/ishivvers/TheKastShiv}}. Low-order polynomial fits to comparison lamp spectra were used to calibrate the wavelength scale, and small adjustments derived from night-sky lines in the target frames were applied. The spectra were flux calibrated and telluric corrected using observations of appropriate spectrophotometric standard stars observed on the same night, at similar airmasses, and with an identical instrument configuration.

The ALFOSC spectrum was reduced using the {\texttt foscgui}\footnote{{\texttt foscgui} is a graphical user interface aimed at extracting SN spectroscopy and photometry obtained with FOSC-like instruments. It was developed by E. Cappellaro. A package description can be found at \url{http://sngroup.oapd.inaf.it/foscgui.html}.} pipeline. The pipeline performs overscan, bias, and flat-field corrections; spectrum extraction; wavelength calibration; flux calibration; and removal of telluric features with IRAF tasks as well as the removal of cosmic-ray artifacts using {\texttt lacosmic} \citep{vanDokkum2001}.

All spectra were obtained with the slit oriented at or near the parallactic angle to minimize slit losses due to atmospheric dispersion \citep{Filippenko1982}. 

We retrieved the spectrum of the host galaxy from SDSS Data Release 18 \citep{Almeida2023}. The spectrum was obtained on 2003 October 30, and covers a wavelength range of 3700--9300\,\AA\ with a spectral resolution of $R \approx 2000$.

\begin{deluxetable}{cccc}[t]
\caption{Log of spectroscopic observations.}
    \label{tab:spec_table}
    \centering
    \tablehead{
     \colhead{\begin{tabular}[c]{@{}c@{}}Phase\\ (days)\end{tabular}} &  \colhead{\begin{tabular}[c]{@{}c@{}}Telescope/\\ Instrument\end{tabular}} &  \colhead{\begin{tabular}[c]{@{}c@{}}Slit Width\\ (\arcsec)\end{tabular}} &  \colhead{\begin{tabular}[c]{@{}c@{}}Exposure \\Time \\ (s)\end{tabular}} 
    }
    \startdata
-32   & NOT/ALFOSC     & 1.3     & 900    \\ 
-28   & NTT/EFOSC2     & 1     & 300    \\ 
-25   & Copernico/AFOSC  & 1.69  & 1800    \\ 
-24   & Copernico/AFOSC  & 1.69  & 1500    \\
-23   & Copernico/AFOSC  & 1.69  & 1200    \\ 
-22   & du Pont/WFCCD  & 1.65  &  2700  \\
-20   & Copernico/AFOSC  & 1.69  & 2400   \\
-20   & du Pont/WFCCD  & 1.65  & 2700 \\
-18   & du Pont/WFCCD  & 1.65  &  2700  \\
-17   & Copernico/AFOSC  & 1.69  & 1800    \\
-12   & Las Cumbres/FLOYDS    & 2     & 1200    \\
-11    & Las Cumbres/FLOYDS    & 2     & 1200   \\
-11    & Copernico/AFOSC    & 1.69     & 2700    \\
-8    & Lick 3\,m/Kast    &  2   &   2400  \\
-6     & NTT/EFOSC2     &  1    & 900     \\
-6     & Las Cumbres/FLOYDS    & 2     & 1200   \\    
-1     & Las Cumbres/FLOYDS    & 2     & 1200   \\ 
+5   & Las Cumbres/FLOYDS    & 2     & 1200    \\
+6    & Lick 3\,m/Kast    &  2   &  2400   \\
+11   & Las Cumbres/FLOYDS    & 2     & 1200     \\
+13   & Las Cumbres/FLOYDS    & 2     & 1200    \\
+15   & Las Cumbres/FLOYDS    & 2     & 1200    \\
+15  & du Pont/WFCCD  & 1.65  &  2700  \\
+22   & Las Cumbres/FLOYDS    & 2     & 1200     \\  
+29   & Las Cumbres/FLOYDS    & 2     & 1200     \\
+31    & Lick 3\,m/Kast    &  2   &  1500  \\
+48    & Lick 3\,m/Kast    &  2   & 1800  \\
+72    & Lick 3\,m/Kast    &  2  &  1800   \\
+204  & Las Cumbres/FLOYDS    & 2     & 3600     \\  
+253  & Las Cumbres/FLOYDS    & 2     & 3600    \\  
+269  & Las Cumbres/FLOYDS    & 2     & 3600      \\ 
+299  & Las Cumbres/FLOYDS    & 2     & 3600    \\ 
+314  & WHT/ISIS      & 1     & 2700    \\ 
+317  & Las Cumbres/FLOYDS    & 2     & 3600   \\ 
+340  & Las Cumbres/FLOYDS    & 2     & 3600   \\ 
+363  & Las Cumbres/FLOYDS    & 2     & 3600 \\
+393  & Las Cumbres/FLOYDS    & 2     & 3600 \\
+416  & Las Cumbres/FLOYDS    & 2     & 3600 \\
    \enddata
    \tablecomments{Phase is given in rest-frame days from the $g$-band peak brightness.}
\end{deluxetable}

We calibrate all spectra of AT\,2019azh (except for the WHT spectrum, owing to its wavelength gap) and that of the host galaxy to photometry and correct the TDE spectra for Milky Way extinction\footnote{The host galaxy spectrum was already corrected for Milky Way extinction, assuming the \cite{Cardelli1989} extinction law and using the all-sky dust maps from Pan-STARRS \citep{Green2018}.} using the \texttt{PySynphot} package \citep{pysynphot2013}\footnote{\url{https://pysynphot.readthedocs.io/en/latest/}}.

\begin{figure*}[t!]
\centering
    \includegraphics[scale=0.95]{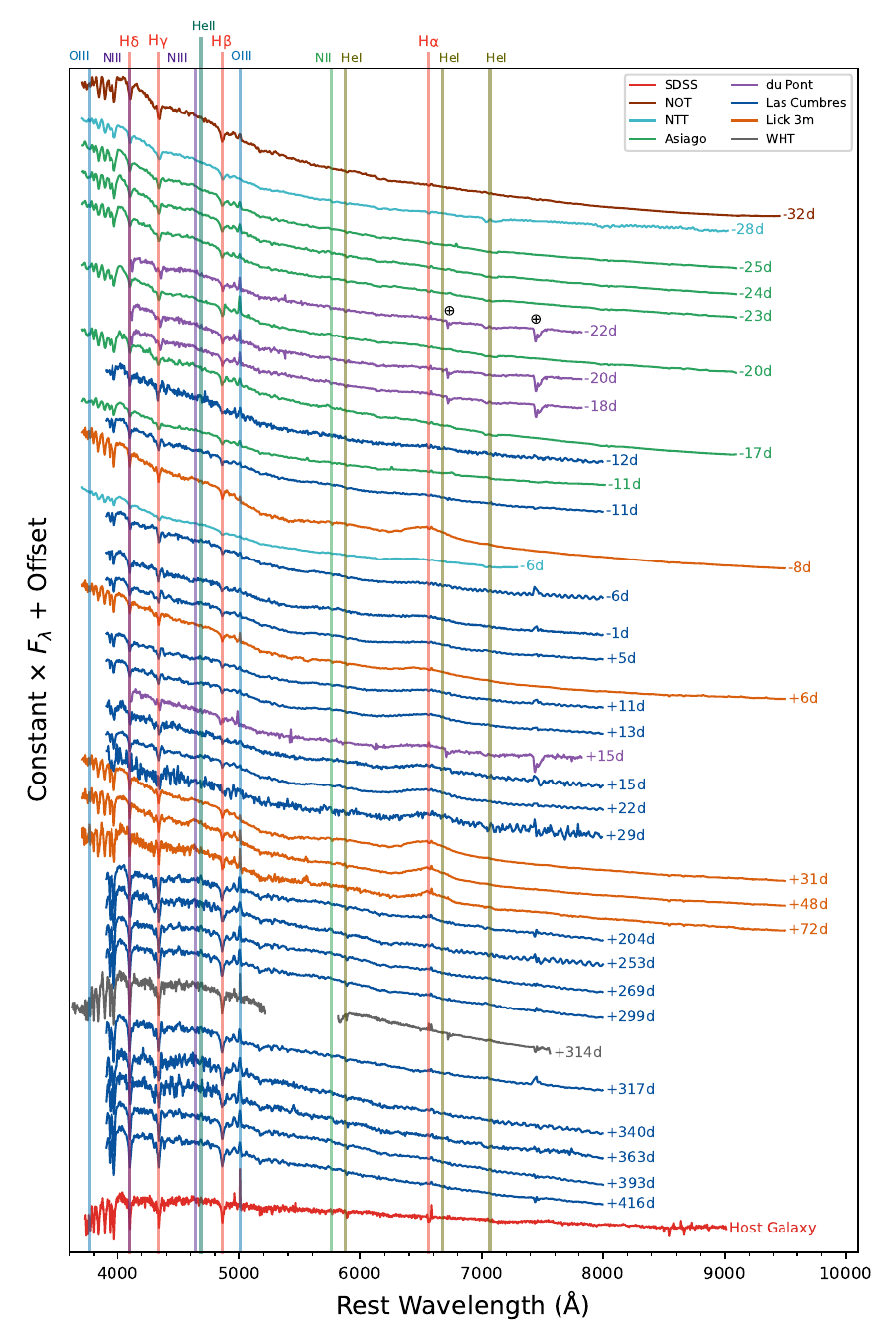}
    \caption{Spectroscopic evolution of AT\,2019azh and the archival host galaxy spectrum from SDSS, after photometric calibration and Galactic extinction correction (except for the WHT spectrum). Notable TDE features, such as broad \ha\ and \heii\ emission lines, are evident in the spectra. We attribute the narrow Balmer absorption and narrow \oiii\ emission lines, seen in all spectra, to the host galaxy. The phase of each spectrum in rest-frame days relative to the $g$-band light-curve peak is indicated and telluric lines are marked.}
    \label{fig:spectra}
\end{figure*}

A log of our spectroscopic observations is provided in Table \ref{tab:spec_table}; all spectra are presented in Figure \ref{fig:spectra} and will be made available through the Weizmann Interactive Supernova Data Repository \citep{Yaron2012}\footnote{\url{https://www.wiserep.org}}.

\section{Analysis} \label{sec: analysis}

\subsection{Photometry}

\subsubsection{Light-Curve Rise}
\label{sec: lightcurve}

The high-cadence pre-peak observations of AT\,2019azh allow us to identify structure in its early optical light curve. First, we identify an abrupt change in the rising slope of the $g$-band light curve at $\sim -30$\,days relative to the peak. We fit the rising $g$-band light curve with a linear function once between MJD $58520$ and $58537$, and once between MJD $58537$ and $58560$, finding a significant change in slope from $0.279\pm0.014$\,mag\,day$^{-1}$ in the first interval to $0.0323\pm0.0042$\,mag\,day$^{-1}$ in the second interval. 

Second, a possible bump at $\sim -20$\,days relative to the peak can be seen in the $BgVri$ bands. While subtle, it is present in all bands that cover that epoch in the Las Cumbres and ASAS-SN data. We fit a second-order polynomial to the photometry after the bump, from -20 to 10\,days relative to the peak, and plot it in the inset of Figure \ref{fig:lc}, extrapolating the fit to the bump epochs. The bump is clearly seen as an excess above this fit. Such structure was not previously robustly identified in a TDE, in part owing to the lack of high-cadence pre-peak observations for most events. However, indications for early light-curve structure were seen in at least two TDEs, which we discuss in Section \ref{sec: discussion}. 

\subsubsection{Light-Curve Peak}

We fit a second-order polynomial to the host subtracted Las Cumbres optical photometry and {\it Swift} UV photometry (except for the {\it Swift} $uvw2$ data, which does not cover enough of the rise to peak brightness) between MJD 58536 and 58596 to determine the peak time and magnitude in each band (the fits are displayed in Figure \ref{fig:peak-fits} in Appendix \ref{sec:peak-fits}). The best-sampled light curve around the peak is that in the $g$ band for which we find a peak time of MJD 58566.70 $\pm$ 0.52 and a peak absolute magnitude of $-19.82 \pm 0.03$. We use this peak time as a reference for all phase information in this paper. 
We also check the cross-correlation offset between the $g$ light curve and the light curves in the bands mentioned above, in the same time range, using the PyCCF package\footnote{\url{http://ascl.net/1805.032}} \citep{Peterson1998}.

\begin{figure}[t]
    \centering
    \includegraphics[width=0.5\textwidth]{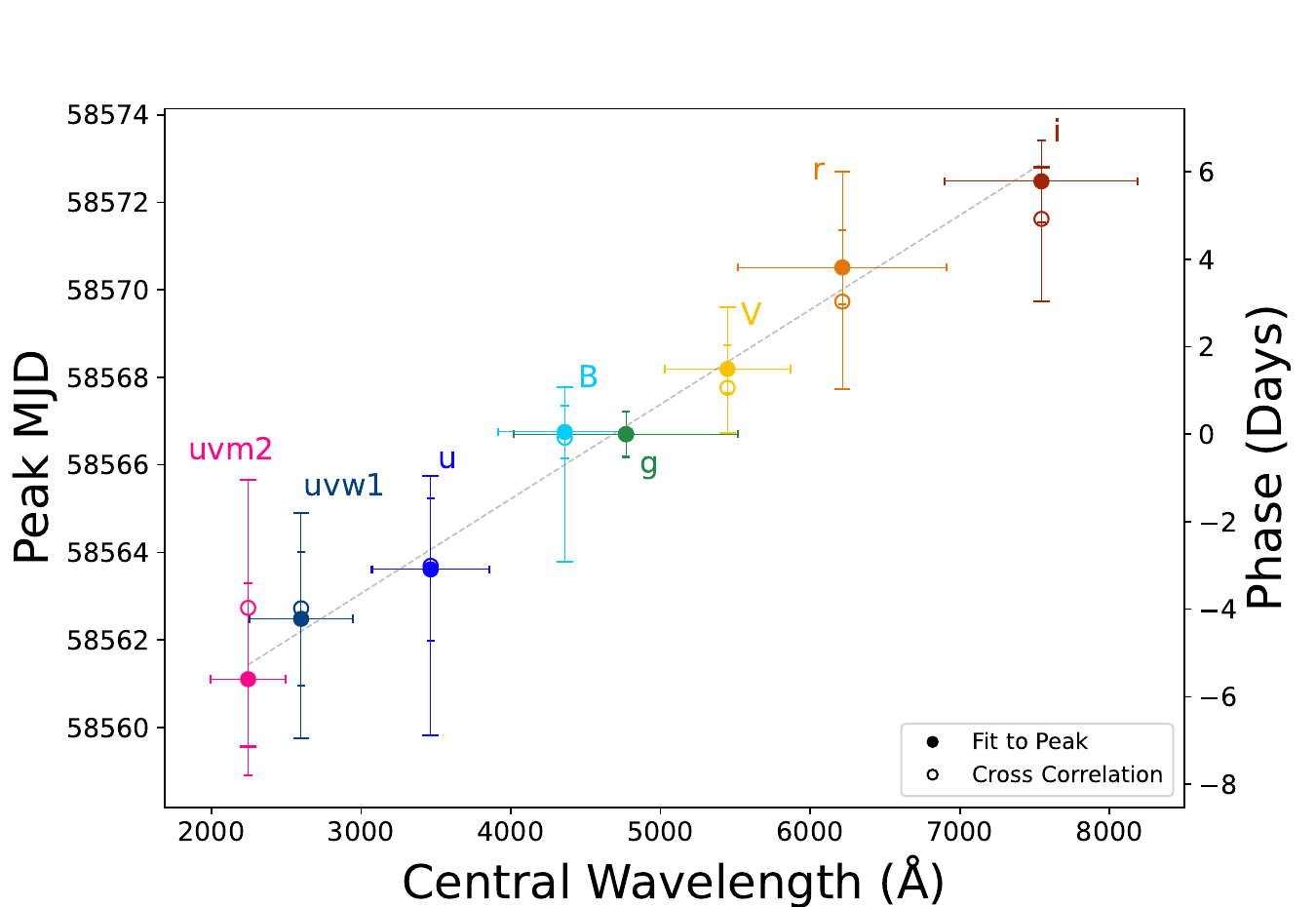}
    \caption{Light-curve peak MJD in various bands. Bluer bands peak earlier than redder bands. Filter widths are indicated with horizontal lines. The dashed gray line indicates a linear fit to the data.}
    \label{fig:peak-band}
\end{figure}

Table \ref{tab:peak_mjd_mag} details the peak time and apparent magnitude from the fit to peak in each band. Figure \ref{fig:peak-band} illustrates the peak times of each band in relation to their central wavelengths. The $uvm2$, $uvw1$, and $u$-band central wavelengths and filter widths are taken from \cite{Poole2008}, while the central wavelengths and filter widths for the rest of the bands are from the Las Cumbres Observatory website\footnote{\url{https://lco.global/observatory/instruments/filters/}}. We find consistent results between the peak time fit method and the cross-correlation method. Both show a monotonic peak time vs. wavelength relation (also found in other TDEs; see Section \ref{sec: discussion}), with the peak-fit method results having a Pearson correlation coefficient of 0.993, and a best-fit linear slope of $(2.16\pm0.10)\times10^{-3}$\,day\,\AA$^{-1}$.

\begin{deluxetable*}{ccccccc}
    \label{tab:peak_mjd_mag}
    \centering
    \caption{Peak MJD and magnitude, determined by fitting a second-order polynomial to the Las Cumbres and {\it Swift} UV photometry around peak brightness.}
    \tablehead{
    \colhead{Band} & \colhead{\begin{tabular}[c]{@{}c@{}} Central \\ Wavelength \\ (\AA) \end{tabular} } & \colhead{\begin{tabular}[c]{@{}c@{}} Filter Width \\ (\AA) \end{tabular} } &\colhead{\begin{tabular}[c]{@{}c@{}}Peak MJD\end{tabular}}  & \colhead{\begin{tabular}[c]{@{}c@{}}Phase \\ (days) \end{tabular} } & \colhead{\begin{tabular}[c]{@{}c@{}}Peak Magnitude\end{tabular}} & \colhead{\begin{tabular}[c]{@{}c@{}}Cross-Correlation \\ Delay (days) \end{tabular}} } 
    \startdata
$uvm2$   & 2246 & 498  & 58561.10  $\pm$  2.20 &  -4.70  $\pm$  2.30  &  -20.49  $\pm$  0.02  &  $-3.97^{+3.17}_{-2.93}$ \\ 
$uvw1$   & 2600 &693  & 58562.48 $\pm$ 1.52 & -4.22$ \pm $1.66  &   -20.30$ \pm $0.02  & $-3.98^{+2.97}_{-2.05}$ \\
$u$    &  3465  & 785 & 58563.61 $\pm$ 1.63 & -3.16 $\pm$ 1.77 &   -20.09 $\pm$ 0.02  & $-3.01^{+3.88}_{-2.05}$ \\
$B$    & 4361  &890 & 58566.75  $\pm$ 0.61 & 0.05 $\pm$ 0.92  &  -19.91 $\pm$ 0.02  & $-0.08^{+2.84}_{-1.15}$  \\
$g$    & 4770  &1500  & 58566.70 $\pm$ 0.52 & 0 & -19.82 $\pm$ 0.03 & 0    \\ 
$V$    & 5448  &840 & 58568.19 $\pm$ 0.54 & 1.49 $\pm$ 0.87  & -19.83 $\pm$ 0.02 & $1.06^{+1.04}_{-1.84}$   \\
$r$    & 6215  &1390 & 58570.51 $\pm$ 0.85 & 3.81 $\pm$ 1.09  &   -19.68 $\pm$ 0.03  &$3.03^{+2.00}_{-2.97}$  \\
$i$    & 7545 &1290 & 58572.48 $\pm$ 0.93 & 5.78 $\pm$ 1.15 &-19.47 $\pm$ 0.02   & $4.92^{+1.88}_{-1.18}$  \\ 
   \enddata
\tablecomments{ Phases are given relative to the $g$-band peak, and cross-correlation delays are given relative to the $g$  light curve.}
\end{deluxetable*}

We find a significant MIR flare at 5.15\,days after the $g$-band peak, with a $W1 - W2$ color of $-0.32$\,mag, which is well below the AGN threshold of $W1 - W2 = 0.8$ determined by \cite{Stern2012}. 

We calculate the expected MIR flux of a blackbody with the best-fit temperature and radius from day 8.09 after the $g$-band peak (the closest blackbody fit to the time of the {\it WISE} detections; see Section \ref{sec:blackbody}) using the \texttt{synphot} package \citep{synphot} with the {\it WISE} $W1$ and $W2$ filter bandpasses from \cite{Wright2010}. We find that such a blackbody would produce a $W1$ and $W2$ AB magnitude of $-16.32\pm0.07$ (the difference between the $W1$ and $W2$ magnitudes is negligible at the assumed temperature of $2.46\pm0.15\times10^4$\,K derived in Section \ref{sec:blackbody}). The MIR detection extinction-corrected $W1$ and $W2$ AB magnitudes are $-17.76\pm0.13$ and $-17.45\pm0.11$, respectively, which are $\sim 1.1$--1.5\,mag brighter than the blackbody emission inferred from the optical and UV data. This excess may be due to a prompt dust echo, as observed, for example, by \cite{Newsome2023}, but we cannot verify this without further data. We leave further analysis of the MIR emission from AT\,2019azh to future work.

\subsubsection{Blackbody Fits} \label{sec:blackbody}

We fit the UV/optical photometry of AT\,2019azh with a blackbody spectrum through the \texttt{SuperBol} fitting package\footnote{\url{https://superbol.readthedocs.io/en/latest/}} \citep{Nicholl2018}, which uses the least-squares fitting method\footnote{We convert the UVOT magnitudes to the Vega system, as required by SuperBol, using the conversions in \url{https://swift.gsfc.nasa.gov/analysis/uvot_digest/zeropts.html}.}. 
Here, we exclude ATLAS observations since the $c$- and $o$-band filters overlap with other filters, making them not fully independent observations. We restrict the fitting to epochs with available UV observations, as this helps reduce systematic errors when fitting blackbodies hotter than $\sim 30,000$\,K with optical data alone \citep{Arcavi_2022}, while linearly interpolating the optical light curves where necessary.
We then calculate the bolometric luminosity using the Stefan-Boltzmann law, \(L_{\rm bol} = 4\pi{R^2} \sigma_{\rm SB}{T^4}\), with $ \sigma_{\rm SB}$ the Stefan-Boltzmann constant, and $R$ and $T$ the blackbody radius and temperature from the fit, respectively\footnote{Here, we are not including any X-ray emission outside of the blackbody inferred from the optical and UV flux. Such emission is negligible around optical peak but comparable to what we measure at late times \citep{Hinkle2021a}.}.

\begin{figure}[t]
    \centering
     \hspace*{-0.3cm}
    \includegraphics[width=0.5\textwidth]{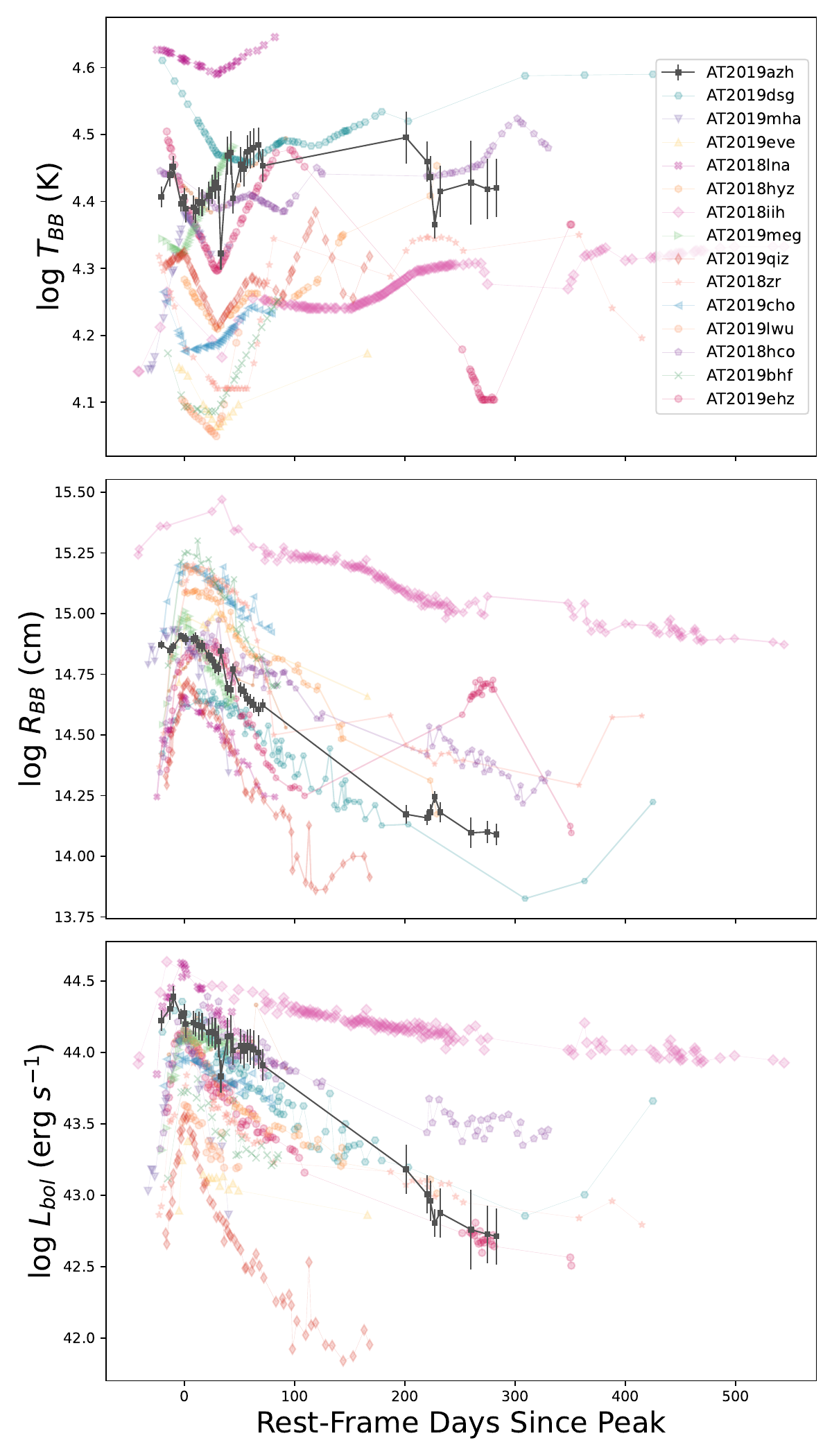}
    \caption{Blackbody temperature (top), radius (middle), and inferred bolometric luminosity (bottom) evolution obtained using \texttt{SuperBol} for AT\,2019azh (black), compared to a sample of TDEs from \cite{vanVelzen2021}. The temperature, radius, and bolometric luminosity of AT\,2019azh are consistent with those of other UV/optical TDEs. The \cite{vanVelzen2021} measurements assume a parametric time evolution and hence are smoother.}
    \label{fig:superbol}
\end{figure}

The evolution of the blackbody temperature, radius, and resulting bolometric luminosity are given in Table \ref{tab:blackbody} and presented in Figure \ref{fig:superbol} in comparison to 15 other TDEs from \cite{vanVelzen2021}\footnote{We compare to this sample since it is one of the largest samples of homogeneously analyzed TDE photometry to date.}. 
As with other TDEs, AT\,2019azh exhibits constant high ($\sim 25,000$\,K) temperatures with values at the high end, but consistent with the sample of \cite{vanVelzen2021}.
Its blackbody radius evolution is also consistent with that of other TDEs and falls in the middle of the comparison sample. 
The bolometric luminosity of AT\,2019azh is on the high end of the comparison sample, but still consistent with it.
Our results are also roughly consistent with those of \cite{Hinkle2021b}, but we obtain slightly lower temperatures and bolometric luminosities, especially at late times, compared to them.

\begin{deluxetable}{cccc}
\label{tab:blackbody}
\caption{Blackbody temperature and radius, and resulting bolometric luminosity.}
\tablehead{
  \colhead{Phase} &  \colhead{\begin{tabular}[c]{@{}c@{}} $T_{\rm BB}$ \\  ($10^4$\,K) \end{tabular}}& \colhead{\begin{tabular}[c]{@{}c@{}} $R_{\rm BB}$ \\  ($10^{14}$\,cm) \end{tabular}} & \colhead{\begin{tabular}[c]{@{}c@{}} $L_{\rm bol}$ \\  ($10^{44}\ergs$) \end{tabular}}
  }
\startdata
-21.94 & 2.55  $\pm$  0.13 & 7.45  $\pm$  0.43 & 1.66  $\pm$  0.40 \\
-13.25 & 2.75  $\pm$  0.15 & 7.05  $\pm$  0.45 & 2.03  $\pm$  0.52 \\
-10.59 & 2.83  $\pm$  0.16 & 7.33  $\pm$  0.45 & 2.46  $\pm$  0.62 \\
-3.75 & 2.49  $\pm$  0.10 & 8.08  $\pm$  0.38 & 1.80  $\pm$  0.33 \\
-0.76 & 2.55  $\pm$  0.12 & 7.91  $\pm$  0.43 & 1.88  $\pm$  0.40 \\
1.37 & 2.54  $\pm$  0.17 & 7.70  $\pm$  0.63 & 1.77  $\pm$  0.56 \\
1.58 & 2.36  $\pm$  0.16 & 7.98  $\pm$  0.68 & 1.42  $\pm$  0.45 \\
8.09 & 2.46  $\pm$  0.15 & 7.87  $\pm$  0.56 & 1.61  $\pm$  0.45 \\
10.40 & 2.43  $\pm$  0.15 & 7.93  $\pm$  0.59 & 1.55  $\pm$  0.44 \\
13.92 & 2.51  $\pm$  0.18 & 7.38  $\pm$  0.63 & 1.55  $\pm$  0.51 \\
16.57 & 2.50  $\pm$  0.17 & 7.36  $\pm$  0.61 & 1.51  $\pm$  0.49 \\
22.83 & 2.56  $\pm$  0.18 & 6.71  $\pm$  0.56 & 1.38  $\pm$  0.45 \\
25.35 & 2.62  $\pm$  0.19 & 6.46  $\pm$  0.55 & 1.40  $\pm$  0.47 \\
28.66 & 2.68  $\pm$  0.22 & 6.06  $\pm$  0.57 & 1.34  $\pm$  0.50 \\
31.59 & 2.63  $\pm$  0.21 & 5.93  $\pm$  0.55 & 1.19  $\pm$  0.44 \\
33.64 & 2.10  $\pm$  0.17 & 7.00  $\pm$  0.70 & 0.68  $\pm$  0.26 \\
39.42 & 2.94  $\pm$  0.28 & 4.93  $\pm$  0.51 & 1.29  $\pm$  0.56 \\
42.47 & 2.97  $\pm$  0.32 & 4.85  $\pm$  0.56 & 1.30  $\pm$  0.64 \\
44.72 & 2.54  $\pm$  0.19 & 5.91  $\pm$  0.47 & 1.03  $\pm$  0.35 \\
51.63 & 2.85  $\pm$  0.25 & 4.86  $\pm$  0.47 & 1.10  $\pm$  0.45 \\
54.62 & 2.81  $\pm$  0.25 & 4.79  $\pm$  0.47 & 1.02  $\pm$  0.41 \\
57.74 & 2.98  $\pm$  0.25 & 4.45  $\pm$  0.40 & 1.10  $\pm$  0.42 \\
60.33 & 3.00  $\pm$  0.28 & 4.34  $\pm$  0.44 & 1.08  $\pm$  0.45 \\
63.72 & 3.02  $\pm$  0.29 & 4.21  $\pm$  0.44 & 1.05  $\pm$  0.46 \\
67.50 & 3.05  $\pm$  0.26 & 4.03  $\pm$  0.38 & 1.00  $\pm$  0.39 \\
71.57 & 2.84  $\pm$  0.23 & 4.19  $\pm$  0.38 & 0.81  $\pm$  0.30 \\
201.24 & 3.13  $\pm$  0.40 & 1.49  $\pm$  0.19 & 0.15  $\pm$  0.09 \\
220.51 & 2.88  $\pm$  0.29 & 1.44  $\pm$  0.15 & 0.10  $\pm$  0.05 \\
223.89 & 2.73  $\pm$  0.28 & 1.52  $\pm$  0.16 & 0.09  $\pm$  0.04 \\
227.94 & 2.32  $\pm$  0.16 & 1.76  $\pm$  0.14 & 0.06  $\pm$  0.02 \\
232.99 & 2.60  $\pm$  0.33 & 1.52  $\pm$  0.21 & 0.08  $\pm$  0.04 \\
260.19 & 2.68  $\pm$  0.56 & 1.25  $\pm$  0.26 & 0.06  $\pm$  0.05 \\
275.68 & 2.62  $\pm$  0.39 & 1.26  $\pm$  0.19 & 0.05  $\pm$  0.04 \\
283.51 & 2.63  $\pm$  0.38 & 1.23  $\pm$  0.19 & 0.05  $\pm$  0.03 \\
\enddata
\tablecomments{Phases are given relative to $g$-band peak brightness.}
\end{deluxetable}

\begin{figure}[t]
    \centering
     \hspace*{-0.3cm}
    \includegraphics[width=0.5\textwidth]{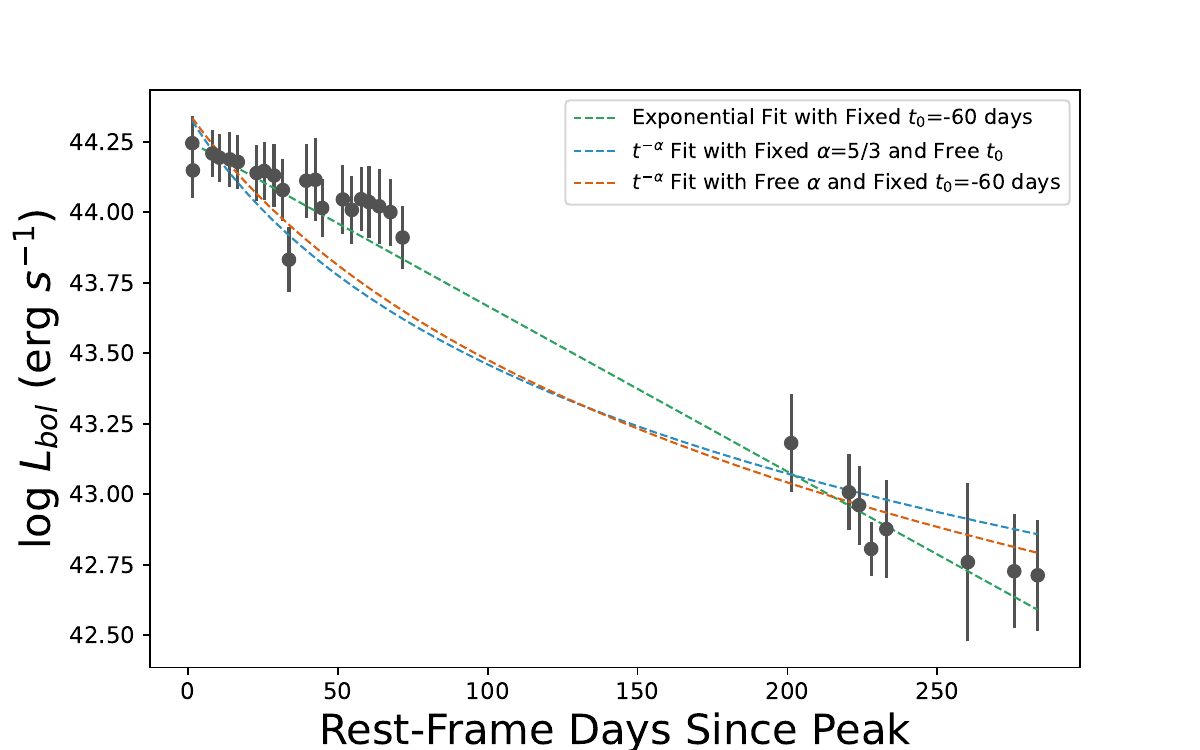}
    \caption{Power-law and exponential fits to the bolometric luminosity decline. The data are better represented by an exponential decline rather than a power law.}
    \label{fig:decay_fit}
\end{figure}

We fit the post-peak bolometric light curve with a power law of the form $L\propto\left(\frac{t-t_0}{\tau}\right)^{-\alpha}$ and an exponential decline of the form $L \propto e^{-\frac{t-t_0}{\tau}}$. We perform the power-law fit in three different ways: once with the power-law index fixed to the canonical $\alpha=5/3$ value and free $t_0$, once with free $\alpha$ and fixed $t_0$ (set to the best-fit value of $-60$\,days from the peak, found by MOSFiT below), and once with free $\alpha$ and free $t_0$. The latter fit requires an unphysical $t_0$ of order $10^5$\,days before the peak to match the data, and the other two power-law fits (yielding $\alpha = 2.06 \pm 0.11$ and $t_{0} = -41.49 \pm 5.87$\,days) are unable to match the data at all. The exponential decline, on the other hand, does match the data well. The different fits are shown in Figure \ref{fig:decay_fit}. We conclude that the bolometric light-curve decline of AT\,2019azh is better described by an exponential than a power law, similar to what was seen for ASASSN-15oi \citep{Holoien2016b} and iPTF16fnl \citep{Blagorodnova2017}. Specifically, it does not fit the canonical $t^{-5/3}$ decline quoted for some TDEs.

\subsubsection{TDE Model Fits}

As mentioned in Section \ref{sec: intro}, there are currently two main models for the source of UV/optical emission in TDEs: reprocessing of X-rays from a rapidly formed accretion disk, and shock emission from debris stream collisions during the circularization process. We fit our photometry to the X-ray reprocessing model with the Modular Open Source Fitter for Transients \citep[\texttt{MOSFiT};][]{MOSFiT2018}, and to the stream collision model with the \texttt{TDEMass} package \citep{Ryu2020}. 

The \texttt{MOSFiT} TDE model \citep{Mockler2019} is based on hydrodynamical simulations for converting the mass-fallback rate from the disrupted star to a bolometric flux. This conversion is related to the accretion rate through the viscous timescale $T_{\rm viscous}$, and it assumes a constant efficiency parameter $\epsilon$. The reprocessing layer is assumed to be a simple blackbody photosphere with radius $R_{\rm phot}$.

The free parameters of the model are the BH mass ($\Mbh$); the mass of the disrupted star ($\Mstar$); the viscous timescale ($T_{\rm viscous}$); the efficiency ($\epsilon$); the blackbody photospheric radius $R_{\rm phot} \propto R_{\rm ph,0} \times L^{l}$ (where $R_{\rm ph,0}$ and $l$ are free parameters and $L$ is the bolometric luminosity); the scaled impact parameter ($b$), which is a proxy for the physical impact parameter $\beta \equiv R_{\rm t}/R_{\rm p}$ (with $R_{\rm t}$ the tidal radius and $R_{\rm p}$ the orbit pericenter); the time of first fallback ($t_{\rm exp}$); the host galaxy column density ($n_{\rm H}$); and a white-noise parameter ($\sigma$). 
We use the default priors from \texttt{MOSFiT}, as given by \cite{Mockler2019}.

\begin{figure*}[t!]
    \centering
    \includegraphics[width=\textwidth]{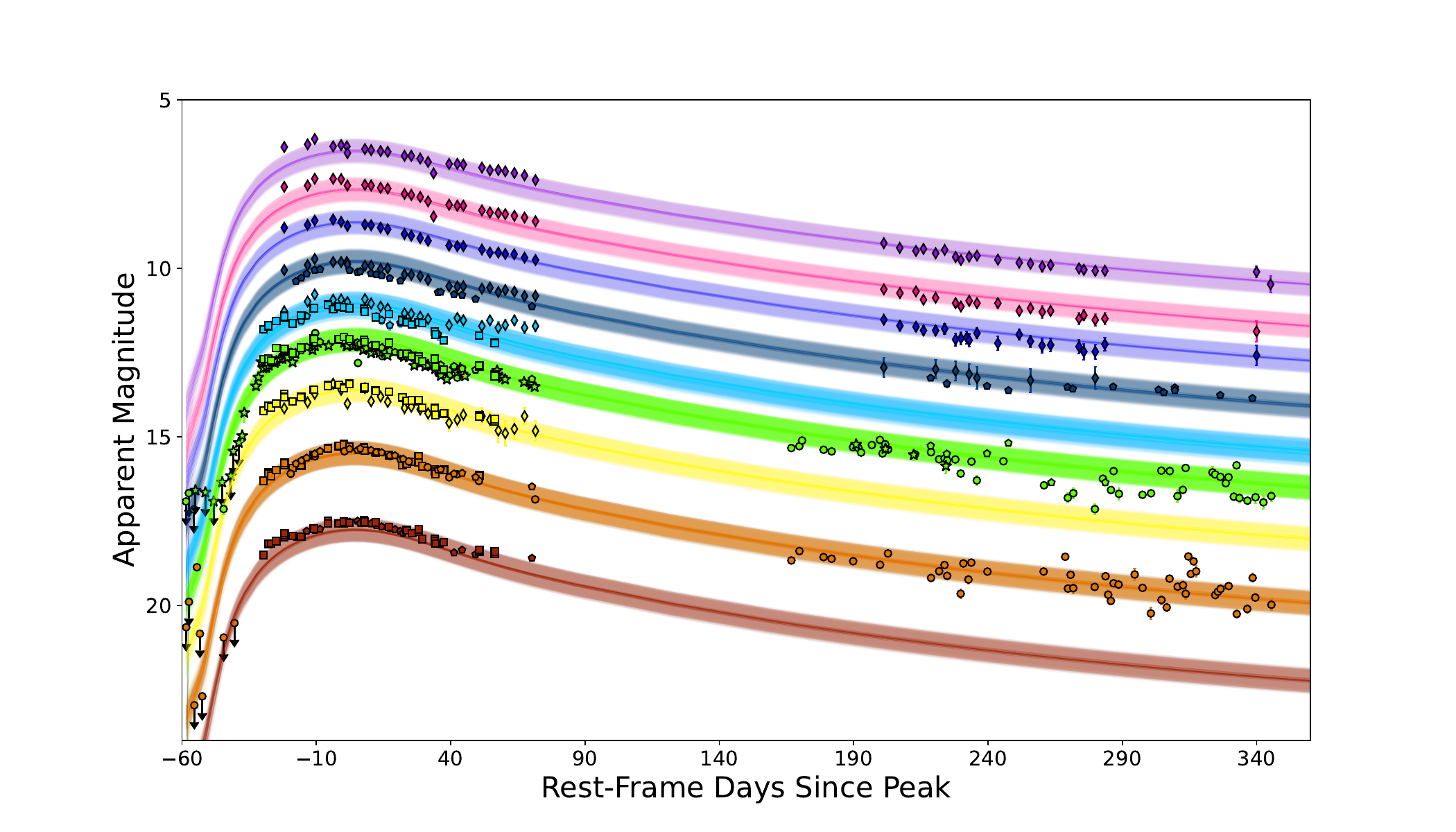}
    \caption{\texttt{MOSFiT} nested sampling fits to AT\,2019azh photometry. Solid lines represent the medians of samples of light curves, while the shaded regions indicate the variance within each sample of models. Overall, the ensembles of models provide a reasonably good fit to the photometric data, but the best-fit parameters are not fully self-consistent (see Section \ref{sec:mosfit_self_consistency}). Arrows indicate 3$\sigma$ nondetection upper limits. Colors, markers, and offsets are the same as in Figure \ref{fig:lc}.}
    \label{fig:mosfit-lc-nested} 
\end{figure*}

We utilize the nested sampling method\footnote{This method is typically employed for models with 10 or more parameters, as is the case here.}, implemented by \texttt{DYNESTY} \citep{Speagle2020}, for the fit. As with the blackbody fits, here we also exclude the ATLAS bands. We further exclude observations more than 1\,yr after discovery because the assumption of a blackbody photosphere made by \texttt{MOSFiT} might not be valid at such late times if the reprocessing material starts to become optically thin. We use the default \texttt{MOSFiT} termination criterion of a PSRF (potential scale reduction factor) of 1.1.

No formal goodness of fit metric is produced by this fitting procedure. The model matches the observations reasonably well {\rm in some regions and deviates from the data in others}, as can be seen in Figure \ref{fig:mosfit-lc-nested}. 
Table \ref{tab:param-nested} presents the best-fit parameters obtained from the fit; the posterior distributions, which are well converged, are displayed in Figure \ref{fig:mosfit-corner} in Appendix \ref{sec:appendix-mosfit}. 
The efficiency parameter approaches its maximum allowed value, which affects the stellar mass parameter owing to their degeneracy \citep{Mockler2021}. 
The impact parameter is $b = 0.99^{+0.01}_{-0.03}$, suggesting that the star is almost fully disrupted.

As \cite{Mockler2019} pointed out, this model includes several simplifications of the complex physics involved.
For instance, assuming solar-composition polytropes instead of more realistic stellar density profiles that take into account the stellar metallicity, age, and evolutionary stage, could introduce systematic uncertainties in determining the stellar mass. 
\cite{Mockler2019} quantified these and other systematic uncertainties arising from some of the model simplifications, and we include these uncertainties in the total error estimates in Table \ref{tab:param-nested}.

\begin{deluxetable}{cccc}[t]
    \label{tab:param-nested}
    \centering
    \caption{Best-fit parameters obtained from the \texttt{MOSFiT} fit with 1$\sigma$ confidence intervals.}
     \tablehead{
\colhead{Parameter}  & \colhead{Best-Fit Value} & \colhead{Total Error} & \colhead{Units}
}
\startdata 
log({$\Mbh$}) & {$7.21^{+0.02}_{-0.02}$} & $\pm0.20$ & {\Msun} \\
{$\Mstar$}  & {0.1000$^{+0.0002}_{-0.0002}$} & $\pm0.66$ & {\Msun} \\
{log($T_{\rm viscous}$)} & $0.44^{+0.14}_{-0.42}$  & $\pm0.43$ & d \\ 
log($\epsilon$)  & $-0.47^{+0.05}_{-0.08}$  & $\pm0.68$ \\
log($R_{\rm ph,0}$)  & $0.38^{+0.07}_{-0.05}$  & $\pm0.4$  \\
$l$  & $1.72^{+0.05}_{-0.06}$ & $\pm0.2$ \\ 
$b$  &  $0.99^{+0.01}_{-0.03}$  &  $\pm$ 0.35 \\ 
{$t_{\rm exp}$}  & $-6.95^{+1.24}_{-1.00}$ & $\pm15$ & day \\ 
log({$n_{\rm H}$})  & $20.66^{+0.03}_{-0.04}$  &  & cm$^{-2}$ \\ 
log($\sigma$) & $-0.45^{+0.01}_{-0.01}$ \\ 
\enddata
\tablecomments{The ``Total Error'' column includes systematic errors estimated by \cite{Mockler2019} due to some of the simplifying assumptions in the model.}
\end{deluxetable}

In \texttt{TDEMass} \citep{Ryu2020}\footnote{\url{https://github.com/taehoryu/TDEmass}}, the mass of the disrupted star and the disrupting SMBH are estimated by numerically solving two nonlinear equations (Eqs. 11 and 12 of \citealt{Ryu2020}) and interpolating within precalculated tables of the peak bolometric luminosity ($L_{\rm obs}$) and the temperature at this peak ($T_{\rm obs}$). 
The equations include two parameters that determine the size and energy dissipation area of the emitting region: $c_{1}$, related to the apocenter distance for the orbit of the most tightly bound debris, and $\Delta\Omega$, the solid angle of the area where shocks dissipate a significant amount of energy. 
The values of these parameters are not well constrained, and the default model values of $c_{1}=1$ and $\Delta\Omega = 2\pi$ are assumed.

From our \texttt{SuperBol} fit, we find a peak luminosity of $L_{\rm obs} = 2.46\pm0.62\times10^{44}$\,\ergs\ and a temperature at this peak of $T_{\rm obs} = 28,300\pm1550$\,K. With these values, we obtain from \texttt{TDEMass} a BH mass of $\Mbh = 2.5^{+0.29}_{-0.24} \times 10^{6}\,\Msun$  and a stellar mass of $\Mstar = 4.8^{+4}_{-2.5}\,\Msun$. 
Figure \ref{fig:tdemass} in Appendix \ref{sec:appendix-tdemass} displays the degeneracy between these two parameters. We compare these results to those found through \texttt{MOSFiT} in Section \ref{sec: discussion}, though we do not expect them to agree since each model assumes a different emission mechanism responsible for the observed light curve. 

\subsection{Spectroscopy}

\subsubsection{Coronal Emission Lines}

We use a custom analysis code (Clark et al. 2024, in preparation) to check for the presence of narrow [\ion{Fe}{7}], [\ion{Fe}{10}], [\ion{Fe}{11}], and [\ion{Fe}{14}] coronal emission lines in our spectra. Such lines are seen in extreme coronal line emitters \citep[e.g.,][]{Komossa2008, Wang2012, Yang2013}, a subset of which is associated with TDEs \citep[e.g.,][]{Onori2022, Clark2023, Short2023, Callow2024} occurring in gas-rich environments. We find no significant evidence for such features in any of our spectra.

\subsubsection{Other Emission Lines}
\label{sec:spec-subtract}

To identify and study the broad emission lines, we follow the spectral analysis process outlined by \cite{Charalampopoulos2022} for removing host galaxy and continuum contributions to the emission line profiles (after performing the photometric calibration and Galactic extinction correction as detailed in Section \ref{sec: observations}).
We exclude from this analysis the du Pont spectra to avoid telluric contamination, and all spectra taken after the seasonal gap (day 205 after peak and onward) given that the broad emission lines are very weak at such late times. 

\begin{deluxetable*}{cccc}
\label{tab:halpha}
\caption{\ha\ luminosity, FWHM, and central wavelength offset of AT\,2019azh.}
\centering
\tablehead{
  \colhead{Phase} & \colhead{\begin{tabular}[c]{@{}c@{}}\ha\ Luminosity \\ ($10^{40}\ergs$) \end{tabular}} &  \colhead{\begin{tabular}[c]{@{}c@{}}\ha\ FWHM  \\ (\kms) \end{tabular}} & \colhead{\begin{tabular}[c]{@{}c@{}}\ha\ Central Wavelength Offset  \\ (\kms) \end{tabular}}}
\startdata
-32 & 1.88  $\pm$  0.17 & 16589.22  $\pm$   1171.02  & -475.83  $\pm$   487.42  \\
-28 & 1.88  $\pm$  0.21 & 13954.00  $\pm$  1231.55 & -466.12  $\pm$  522.95 \\
-25 & 2.15  $\pm$  0.18 & 26471.07  $\pm$  1809.76 & -4986.51  $\pm$  768.48 \\
-24 & 3.37  $\pm$  0.21 & 21255.54  $\pm$  1000.60 & -5735.38  $\pm$  424.88 \\
-23 & 3.31  $\pm$  0.19 & 22257.40  $\pm$  963.45 & -6126.03  $\pm$  409.11 \\
-20 & 3.81  $\pm$  0.14 & 21156.45  $\pm$  584.78 & -5662.58  $\pm$  248.32 \\
-17 & 6.40  $\pm$  0.28 & 21238.53  $\pm$  717.22 & -5751.50  $\pm$  304.55 \\
-12 & 5.62  $\pm$  0.23 & 21790.47  $\pm$  693.67 & -3340.03  $\pm$  294.55 \\
-11 & 7.14  $\pm$  0.46 & 21490.06  $\pm$  1090.57 & -3528.82  $\pm$  463.09 \\
-11 & 10.59  $\pm$  0.23 & 19292.27  $\pm$  319.33 & -3488.98  $\pm$  135.60 \\
-8 & 12.63  $\pm$  0.16 & 14136.78  $\pm$  136.217 & -1165.36  $\pm$  57.84 \\
-6 & 14.45  $\pm$  0.36 & 17393.32  $\pm$  347.60 & -3145.95  $\pm$  147.60 \\
-6 & 13.69  $\pm$  0.31 & 19052.18  $\pm$  330.17 & -2678.93  $\pm$  140.20 \\
-1 & 15.43  $\pm$  0.31 & 17610.98  $\pm$  269.39 & -2797.66  $\pm$  114.39 \\
+5 & 14.56  $\pm$  0.20 & 18962.47  $\pm$  200.04 & -2396.22  $\pm$  84.94 \\
+6 & 11.84  $\pm$  0.14 & 17968.18  $\pm$  162.57 & -2685.32  $\pm$  69.03 \\
+11 & 10.18  $\pm$  0.15 & 17536.29  $\pm$  199.07 & -1490.40  $\pm$  84.53 \\
+13 & 13.39  $\pm$  0.16 & 17677.46  $\pm$  157.96 & -1792.32  $\pm$  67.07 \\
+15 & 7.72  $\pm$  0.19 & 14905.48  $\pm$  286.30 & -1711.47  $\pm$  121.57 \\
+22 & 9.80  $\pm$  0.10 & 17390.90  $\pm$  141.47 & -1640.46  $\pm$  60.07 \\
+29 & 10.26  $\pm$  0.45 & 13992.84  $\pm$  465.88 & 491.88  $\pm$  197.83 \\
+31 & 7.19  $\pm$  0.08 & 14193.45  $\pm$  124.82 & -1065.69  $\pm$  53.00 \\
+48 & 6.38  $\pm$  0.06 & 12932.58  $\pm$  95.83 & -166.10  $\pm$  40.69 \\
+72 & 5.05  $\pm$  0.11 & 10420.17  $\pm$  167.85 & 664.59  $\pm$  71.28 \\
\enddata
\tablecomments{Phases are given relative to the $g$-band peak brightness.}
\end{deluxetable*}

\begin{figure*}[t!]
    \centering
    \includegraphics[width=\textwidth]{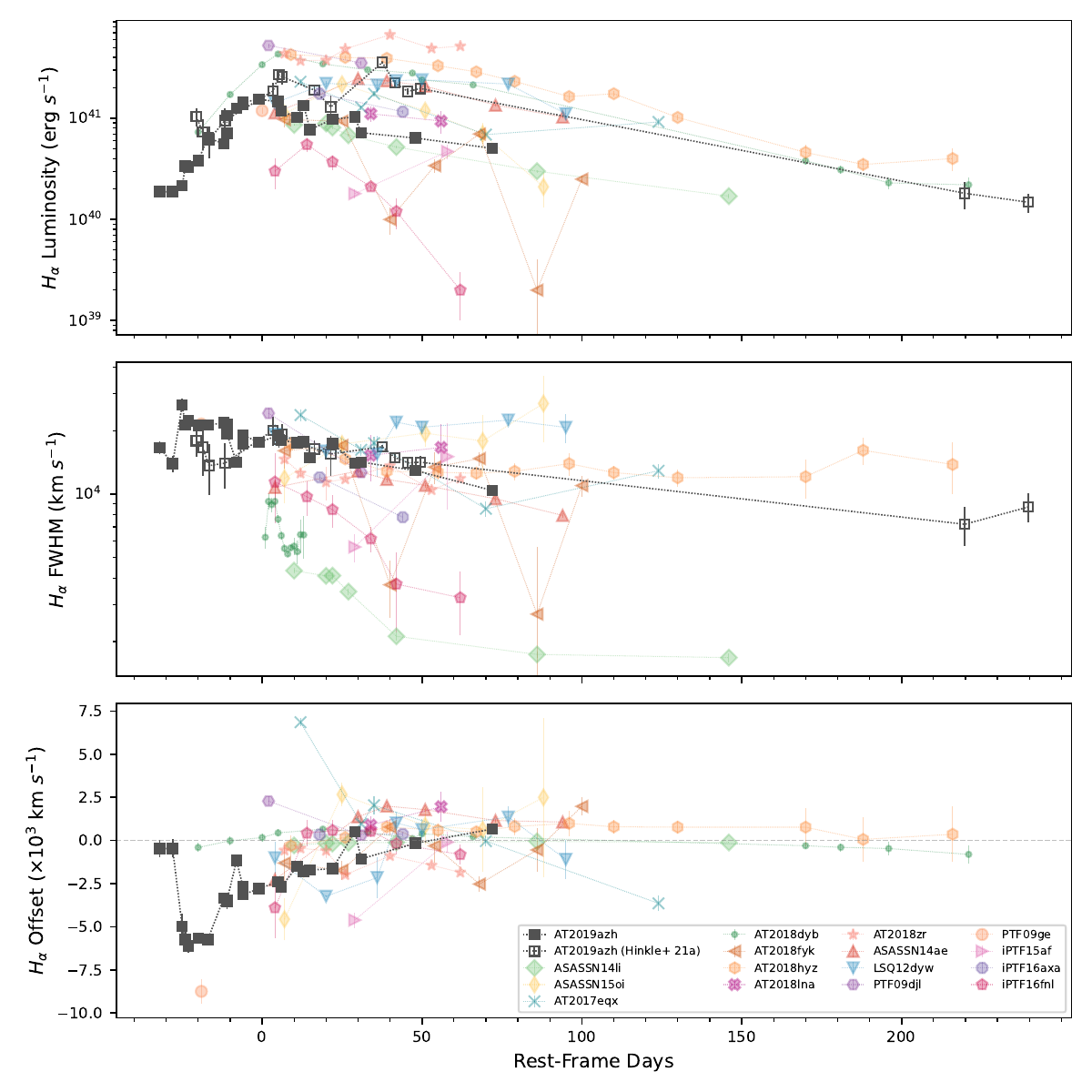}
    \caption{Evolution of \ha\ line luminosity (top), FWHM (center), and central wavelength offset (bottom) of AT\,2019azh from this work compared to those of \cite{Hinkle2021a} and a sample of 15 TDEs from \cite{Charalampopoulos2022}. AT\,2019azh is plotted relative to its $g$-band peak, while the comparison data are plotted relative to their peak or discovery date \citep[see][for details]{Charalampopoulos2022}.}
    \label{fig:ha-lum-fwhm}
\end{figure*}

First, we subtract the host galaxy spectrum from each TDE spectrum after resampling the host spectrum to the wavelengths of the TDE spectrum using the \texttt{SciPy} \texttt{interp1d} function\footnote{\url{https://docs.scipy.org/doc/scipy/reference/generated/scipy.interpolate.interp1d.html}}. Since different spectra are taken under different seeing conditions, and the TDE spectra are taken with varying slit widths and angles, while the SDSS host spectrum was obtained through a fiber, there will be different host galaxy contributions to each TDE spectrum. Thus, it is impossible to completely remove host galaxy emission from the TDE spectra. Here, we attempt to minimize host galaxy contamination, but some residuals likely remain (see below). 

Next, we identify line-free regions in the host subtracted spectra to fit and remove the spectral continuum using a third-order polynomial. We use line-free regions outlined by \cite{Charalampopoulos2022} as a basis, while tailoring them to match the AT\,2019azh spectra. The selected line-free rest-frame wavelength ranges are 3900--4000, 4220--4280, 5100--5550, 6000--6100, and 6800--7000\,\AA. 

An example of this spectral processing procedure, as performed on the spectrum from 13\,days after peak brightness, is provided in Figure \ref{fig:spec-phase13} in Appendix \ref{sec:appendix-spectra}. All spectra after host and continuum removal, for which this process was conducted, are presented in Figure \ref{fig:spectra-final} in Appendix \ref{sec:appendix-spectra}.

Broad emission lines of \ha, \heii, and \hei\ are evident, as in other UV/optical TDEs \citep[e.g.,][]{Gezari2012, Arcavi2014, Holoien2016}. 
The broad \heii\ emission line appears already in the first spectrum, remains relatively strong and broad until the seasonal gap at 72\,days after the light-curve peak, and weakens in the spectra obtained after the gap (see Figure \ref{fig:spectra-final} in Appendix \ref{sec:appendix-spectra}).
The broad \ha\ emission line strengthens at early times and later significantly weakens and narrows in the post-seasonal gap spectra. 
This behavior was observed in some other TDEs (e.g., \citealt{Gezari2012, Holoien2014, Holoien2016}), and is discussed further in Section \ref{sec:ha}.

In addition to the broad emission lines, narrow Balmer $\hb$ and $\hg$ emission lines are seen in the host- and continuum-subtracted spectra. These lines likely originate from oversubtraction of the host galaxy spectrum (as they also appear in the SDSS host spectrum as narrow absorption lines; see Figure \ref{fig:spectra}). 
We also find a strong, narrow \oiii\ absorption line in the host subtracted spectra (see Figure \ref{fig:spec-phase13} in Appendix \ref{sec:appendix-spectra}), which is probably also an oversubtracted host galaxy emission line.

\subsubsection{\ha\ Line Evolution}
\label{sec:ha}
 
Following \cite{Charalampopoulos2022}, we quantify the evolution of the \ha\ emission line, as it is a relatively isolated line. For each host- and continuum-subtracted spectrum, we fit the \ha\ emission line with a Gaussian using the nonlinear least-squares method of the \texttt{LMFIT}\footnote{\url{https://lmfit.github.io/lmfit-py/}} package. 
We use the same initial guesses for the center (6563\,\AA) and width (150\,\AA, corresponding to a Doppler velocity of $\sim 10,000$\,\kms), for all spectra. All Gaussian fits, after normalizing the peak of the feature, are shown in Figure \ref{fig:ha-fits} in Appendix \ref{sec:appendix-spectra}.

The evolution of the \ha\ line luminosity is presented in the top panel of Figure \ref{fig:ha-lum-fwhm}, along with data from 15 other TDEs obtained from \cite{Charalampopoulos2022}, which were measured using the same methodology as described here and which constitute the largest sample of homogeneously analyzed TDE spectra to date. Around peak brightness, the \ha\ luminosity is similar to that of the comparison sample. The post-peak slight decay of the \ha\ luminosity is also consistent with the rest of the sample. However, the extensive pre-peak spectral observations of AT\,2019azh reveal the initial formation of this emission line in a TDE for the first time. These observations can be used to constrain future models of spectral line formation in TDEs. We also compare our results with those of \cite{Hinkle2021a}, which agree at early times but not at late times. This might be due to the different analysis method used by \cite{Hinkle2021a}, where, for example, the continuum is removed differently than here. 

We also measure the evolution of the full width at half-maximum (FWHM) intensity of the Gaussian fits to the \ha\ emission line of AT\,2019azh, and compare them to those of the \cite{Charalampopoulos2022} sample and to the results of \cite{Hinkle2021a} in the middle panel of Figure \ref{fig:ha-lum-fwhm}. 
The FWHM of the \ha\ emission line of AT\,2019azh is at the upper range of the \cite{Charalampopoulos2022} sample, and it shows a clear gradual decline. Here, our results are consistent with those of \cite{Hinkle2021a}.

Finally, in the bottom panel of Figure \ref{fig:ha-lum-fwhm}, we compare the evolution of the \ha\ best-fit Gaussian central wavelength offset from the rest wavelength with that of the same sample from \cite{Charalampopoulos2022}. 
While the line centers of the first two spectra are consistent with zero offset, a blueshift rapidly develops and slowly returns back to zero offset within a few months. The magnitude of the offset is consistent with those of other events in the \cite{Charalampopoulos2022} sample, but AT\,2019azh is the only event showing this kind of evolution. 
In addition, from Figure 7 it appears that there is an anticorrelation between the \ha\ FWHM and offset. Indeed, we find such an anticorrelation (with a Pearson coefficient of $-0.876$; see Figure \ref{fig:offset-fwhm} in Appendix \ref{sec:appendix-spectra}). However, events in the \cite{Charalampopoulos2022} sample do not show similar behavior, therefore this anticorrelation does not seem like a universal property of TDEs.

The \ha\ luminosity, FWHM, and central-wavelength offset values for AT\,2019azh are presented in Table \ref{tab:halpha}.

\begin{figure}[t]
\hspace*{-0.5cm}
    \includegraphics[width=0.55\textwidth]{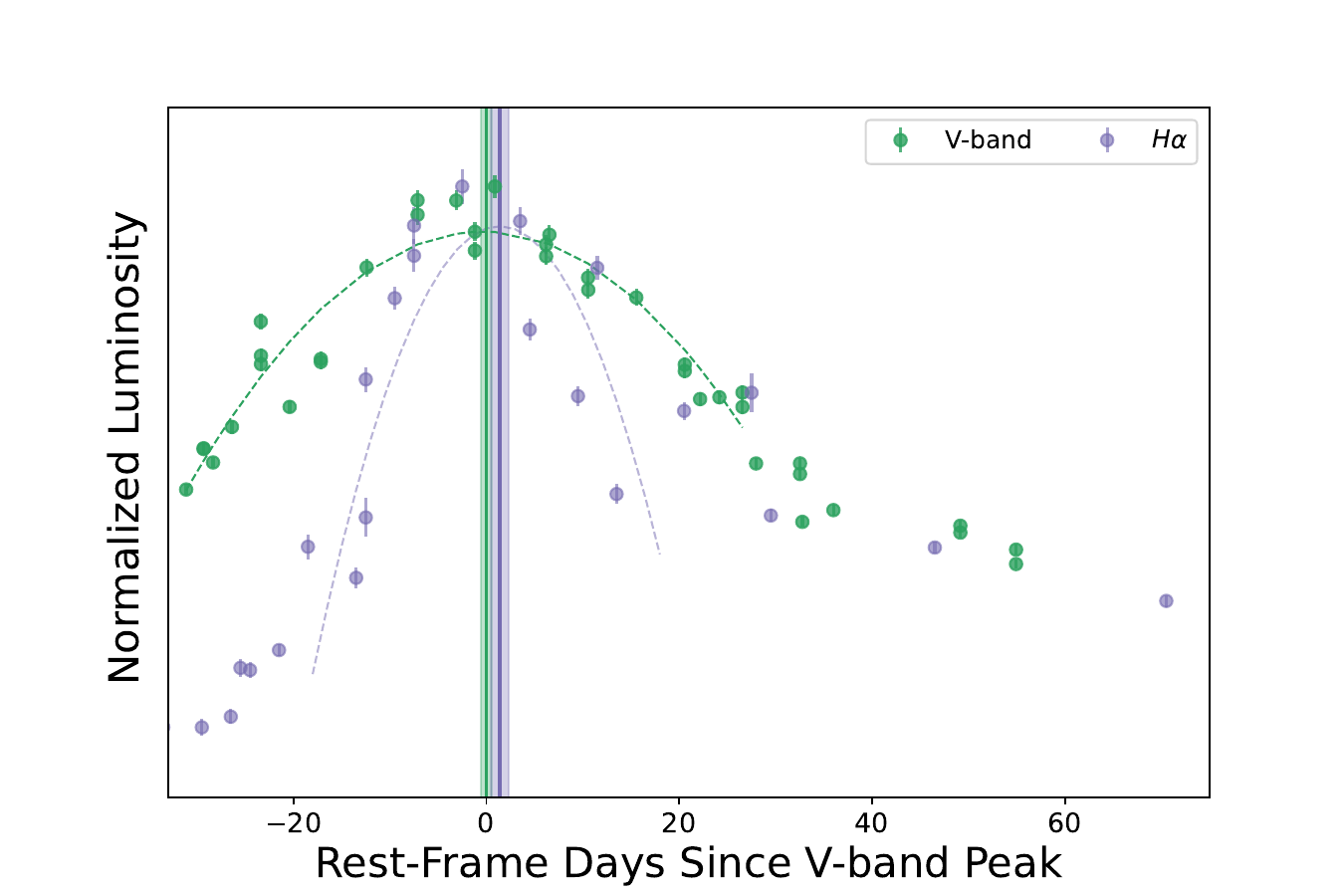}
    \caption{\ha\ luminosity evolution (purple), compared to the $V$-band light curve of AT\,2019azh (green). The colored dashed lines are the parabolic fits to the data around peak brightness, from which the peak times were determined. The vertical solid lines and shaded regions mark the peak times for the $V$-band and \ha\ peaks, and their $1\sigma$ uncertainties, respectively. The \ha\ peak time is consistent with that of the $V$-band peak.}
    \label{fig:comphalg}
\end{figure}

\cite{Charalampopoulos2022} showed that TDEs exhibit a time lag between their light curve (i.e., continuum emission) and \ha\ luminosity peaks. Figure \ref{fig:comphalg} compares the evolution of the \ha\ luminosity and the $V$-band light curve for AT\,2019azh.
To determine the peak time of the \ha\ line luminosity for AT\,2019azh, we fit a second-order polynomial to the \ha\ luminosity from $-18$ to +18\,days since the $g$-band peak and find that the peak occurred on MJD $58569.59 \pm 1.07$, or $\Delta t = 1.40 \pm {0.93}$\,days after the $V$-band light-curve peak.

\section{Discussion} 
\label{sec: discussion}

\subsection{Spectroscopic Classification of AT\,2019azh}

\begin{figure}[t]
    \hspace*{-0.5cm}
   \centering  \includegraphics[width=0.55\textwidth]{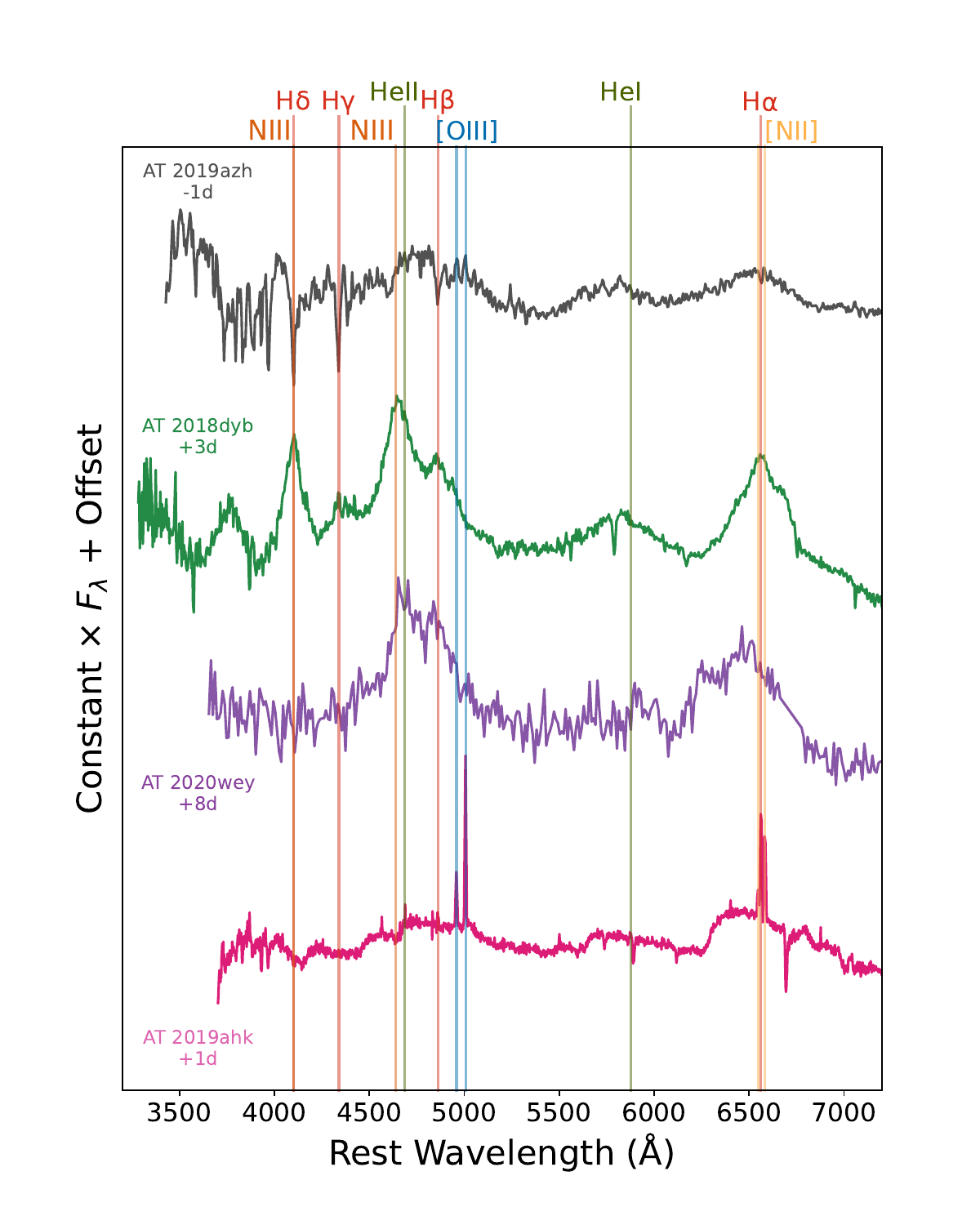}
\caption{Spectral comparison around peak luminosity of AT\,2019azh (host subtracted spectrum) with the well-observed Bowen TDE AT\,2018dyb \citep[][not host subtracted]{Leloudas2019}, the H+He-TDE AT\,2020wey \citep[][host subtracted]{Charalampopoulos2023}, which also showed possible early light-curve structure, and AT\,2019ahk \citep[][not host subtracted]{Holoien_2019b}, which did not show such structure in its very densely sampled light-curve rise. All spectra are continuum-subtracted. The similarity of the AT\,2019azh spectrum to that of AT\,2020wey classifies it as a H+He-TDE. The spectrum of AT\,2019ahk displays distinctive narrow emission lines, implying it might be an AGN-related flare rather than a classical TDE. Days relative to the light-curve peak are shown next to each spectrum.}
    \label{fig:comp-spec}
\end{figure}

In Figure \ref{fig:comp-spec} we compare the continuum-subtracted spectra around the peak of AT\,2019azh to those of the Bowen TDE AT\,2018dyb \citep{Leloudas2019}, the H+He-TDE AT\,2020wey \citep{Charalampopoulos2023}, which also showed a possible early light-curve bump, and AT\,2019ahk \citep{Holoien_2019b}, which does not show such structure despite having a very densely sampled early-time light curve (see below)\footnote{We subtract the continuum of each spectrum following the procedure detailed in Section \ref{sec:spec-subtract}.}. AT\,2019azh does not show \niiid\ emission like those seen in AT\,2018dyb, meaning it is not a Bowen TDE. Its broad H and \heii\ emission features are similar to those of AT\,2020wey, making it a H+He-TDE. AT\,2019ahk shows strong AGN-like narrow spectral emission lines, not seen in most TDEs, implying it might be a different type of flare.

\subsection{Peak Luminosity Time Delays}

In Section \ref{sec: lightcurve} we measured a time delay in the peak luminosity between the different bands, with the redder bands peaking later than the bluer ones (Figure \ref{fig:peak-band}). Such behavior has been seen in other TDEs, such as AT\,2018zr \citep[PS18kh;][]{Holoien_2019}, AT\,2019ahk \citep[ASASSN-19bt;][]{Holoien_2019b}, and AT\,2018dyb \citep[ASASSN-18pg;][]{Holoien_2020}, where it has been attributed to the blackbody temperature evolution around the peak. \cite{Wang2023} also find that the optical emission lags behind the UV emission in the peculiar nuclear transient AT\,2019avd. They interpret this lag as evidence for the optical emission being reprocessed UV emission.
This phenomenon is also observed in AGNs \citep[e.g.,][]{Shappee2014}, where it is attributed to an accretion disk emission model \citep[e.g.,][]{Cackett2007}, according to which the inner, hotter accretion disk is illuminated by X-rays first, with the illumination progressing outward, causing variations in the light curve to manifest initially in the bluer bands associated with the inner disk, followed by the redder bands.
An opposite time delay was measured for the TDE ASASSN-14li by \cite{Pasham2017}. There, the UV lagging behind the optical is interpreted as evidence for the stream collision scenario.

\subsection{Early Light-Curve Structure}

\begin{figure}[t]
    \centering
    \includegraphics[width=0.5\textwidth]{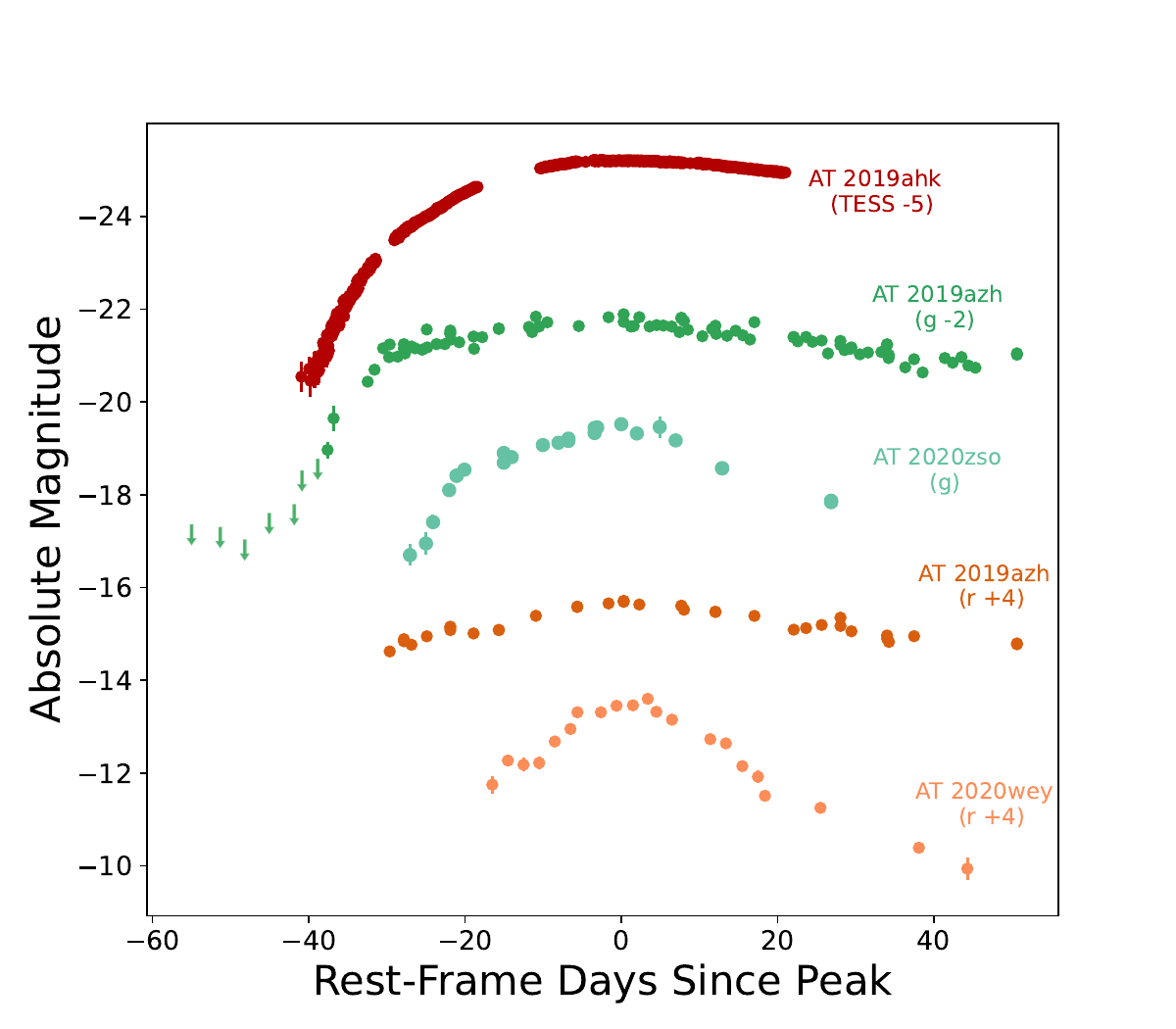}
     \caption{Comparison between the light curves of AT\,2019azh (ASAS-SN and Las Cumbres data only, shown for clarity) to those of AT\,2019ahk, the TDE with the most densely sampled light curve to date \citep{Holoien_2019b}, AT\,2020wey \citep{Charalampopoulos2023}, and AT\,2020zso. AT\,2019ahk lacks any pronounced early-time light-curve structure like that seen in AT\,2019azh, despite having high-cadence TESS observations. In contrast, the light curve of the TDE AT\,2020wey shows a possible early bump in its $r$-band light curve, and AT\,2020zso exhibits a change in its rising slope. Arrows indicate 3$\sigma$ nondetection upper limits.}
    \label{fig:19bt-2020wey-2020zso}
\end{figure}

Our high-cadence photometric observations also reveal, for the first time, both a change in light-curve slope and a possible bump in the rising light curve of a TDE. The most densely sampled rising light curve of a TDE is that of AT\,2019ahk \citep[ASASSN-19bt;][]{Holoien_2019b}, which was observed with a 30 minute cadence using TESS. Its light-curve rise is smooth (Figure \ref{fig:19bt-2020wey-2020zso}), in stark contrast to that of AT\,2019azh. However, AT\,2019ahk might not be a spectroscopically classical TDE. As mentioned, it stands out in Figure \ref{fig:comp-spec} owing to its strong and narrow \oiiid\ and \niid\ emission lines, not commonly seen in TDE spectra. Furthermore, the host galaxy of AT\,2019ahk is in the Seyfert region of the Baldwin–Phillips–Terlevich \citep{Baldwin1981} diagram \citep[see Figure 2 in][]{Holoien_2019b}, indicating the presence of an AGN as an ionizing source. AT\,2019ahk might thus be related to an AGN flare rather than a typical optical/UV TDE.

ASASSN-14ko \citep{Holoien2014b} is another nuclear transient with a bump reported in its rising phase \citep{Huang2023}. However, it displays spectra more similar to those of AGNs \citep{Payne2021} and periodicity in its outbursts \citep{Payne2021, Payne2022, Payne2023} which are not seen in the class of TDEs to which AT\,2019azh belongs. Hence, we do not consider it further here.

AT\,2020wey, on the other hand, is a classical spectroscopically classified TDE (Figure \ref{fig:comp-spec}), which does show a possible bump in its early $g$- and $r$-band light curves \citep[Figure \ref{fig:19bt-2020wey-2020zso};][]{Charalampopoulos2023}. Unfortunately, this part of the light curve of AT\,2020wey was not observed at a sufficiently high cadence to robustly characterize this feature.
Finally, AT\,2020zso is a TDE, which shows an abrupt change in its light-curve rise slope \citep[Figure \ref{fig:19bt-2020wey-2020zso}; photometry taken from][]{Wevers2022}. 

No TDE emission model predicts specific light-curve features such as these. This could point toward the presence of more than one emission source as responsible for the UV/optical TDE emission.
The photometry presented here could be used to test future TDE emission models.

\subsection{Estimates of the SMBH Mass}

\cite{Wevers2020} derived the SMBH mass of the host of AT\,2019azh using the $M$--$\sigma$ relation from \cite{Gu2009} with the velocity dispersion measured from the WHT spectrum presented here. They find an SMBH mass of $\Mbh = 2.29\pm2.27\times10^{6}\,\Msun$. This mass is consistent with that found by \texttt{TDEMass} but is a factor of $\sim 7$ smaller than that found by \texttt{MOSFiT}. We do not consider this definitive evidence favoring one model or the other since the host galaxy-derived SMBH mass strongly depends on the choice of scaling relation and the spectral resolution used to infer the velocity dispersion, as is evident in the comparison to other works.

We present a summary of SMBH mass estimates obtained here and in other works using the two light-curve models and host galaxy scaling relations, along with the corresponding Eddington ratios for the peak bolometric luminosity, in Table \ref{tab:bh-mass}. Our results are consistent with those of \cite{Hammerstein2023} for the TDEMass estimates, and marginally consistent (at the $\lesssim2\sigma$ level) with the MOSFiT estimates from that work. Our results are not consistent with those of \cite{Hinkle2021a} or \cite{Nicholl2022}. The discrepancy with \cite{Hinkle2021a} could be due to their use of the pre-corrected UVOT calibrations introduced later by \cite{Hinkle2021b}.

\begin{deluxetable*}{lcccccccc}[t]
    \label{tab:bh-mass}
    \centering
    \caption{SMBH mass estimates and corresponding Eddington ratios for the peak bolometric luminosity.}
     \tablehead{  
     & \multicolumn{2}{c}{\texttt{TDEMass}} &  \multicolumn{2}{c}{\texttt{MOSFiT}}  & \multicolumn{2}{c}{Host Galaxy}\\
     \cmidrule(lr){2-3} \cmidrule(lr){4-5} \cmidrule(lr){6-7} 
    & \colhead{\begin{tabular}[c]{@{}c@{}}$\Mbh$ \\ ($10^6 \Msun$) \end{tabular}} & \colhead{$L_{\rm bol}$/$L_{\rm Edd}$} & \colhead{\begin{tabular}[c]{@{}c@{}}$\Mbh$ \\ ($10^6 \Msun$) \end{tabular}} & \colhead{$L_{\rm bol}$/$L_{\rm Edd}$} & \colhead{\begin{tabular}[c]{@{}c@{}}$\Mbh$ \\ ($10^6 \Msun$) \end{tabular}} & \colhead{$L_{\rm bol}$/$L_{\rm Edd}$}
    }
    \startdata
    This work & $2.50^{+0.29}_{-0.24}$ &  $0.78^{+0.22}_{-0.21}$  & $16.22^{+0.75}_{-0.75}$ &  $0.12^{+0.03}_{-0.03}$ & $2.29\pm2.27$\tablenotemark{a} & $0.85\pm0.87$ \\
    \cite{Hammerstein2023} & $ 2.19^{+0.14}_{-0.00}$& $0.85^{+0.18}_{-0.03}$ & $26.91^{+6.81}_{-5.30}$ & $0.07^{+0.02}_{-0.02}$ & $-$ &$-$\\
    \cite{Hinkle2021a}  & $0.73^{+0.24}_{-0.10}$  &$6.75^{+2.23}_{-0.95}$ & $7.8^{+3.9}_{-4.1}$ & $0.63^{+0.32}_{-0.33}$  &$\sim 12.59$\tablenotemark{b} &$\sim 0.34$ \\
    \cite{Liu2022} &$-$&$-$&$-$&$-$& $23.0^{+13.0}_{-12.0}$\tablenotemark{c} & $\sim 0.06$\\ 
    \cite{Nicholl2022} & $-$&$-$ & $5.01^{+0.70}_{-0.81}$ & $-$ &$-$ &$-$ \\
    \enddata
\tablenotetext{a}{Value from \cite{Wevers2020} using the WHT spectrum presented here and the \cite{Gultekin2009} scaling relation.}
\tablenotetext{b}{Using the SDSS DR14 spectrum and the \cite{Gultekin2009} scaling relation.}
\tablenotetext{c}{Using the \cite{Reines2015} scaling relation.}
\tablecomments{Eddington ratios are calculated for each source using their respective SMBH masses and peak bolometric luminosities.}
\end{deluxetable*}

\subsection{Estimates of the Disrupted Star Mass}

The mass derived for the disrupted star also differs substantially between the models, with \texttt{TDEMass} preferring a star roughly 1 order of magnitude more massive than \texttt{MOSFiT} ($4.8^{+4}_{-2.5}\,\Msun$ vs. $0.10^{+0.02}_{-0.03}\,\Msun$, respectively). In addition to the different model assumptions, this difference could be driven by the \texttt{MOSFiT} prior of a Kroupa initial mass function. \cite{Hinkle2021a} find a similar stellar mass in their \texttt{MOSFiT} fit as in ours, but a much higher one in their \texttt{TDEMass} fit than ours. As mentioned previously, this comparison may not be entirely accurate because of the {\it Swift} calibration updates \citep{Hinkle2021b}, not available to \cite{Hinkle2021a}, which could influence their bolometric luminosity calculations.
\cite{Hammerstein2023} also estimated the stellar mass using these two methods. Our \texttt{TDEMass}-based stellar mass is consistent with their findings, while our \texttt{MOSFiT}-based stellar mass is not. This might be due to differences in the priors used for the efficiency parameter, which is degenerate with the stellar mass.
The efficiency parameter inferred from \texttt{MOSFiT} in our analysis is close to the maximum limit of the prior (see Figure \ref{fig:mosfit-corner} in Appendix \ref{sec:appendix-mosfit}). This relatively high efficiency might be additional evidence for contributions to the emission from the stream collision process.

\subsection{Self-Consistency of \texttt{MOSFiT} Parameters}
\label{sec:mosfit_self_consistency}

For the best-fit stellar mass of $0.1\,\Msun$ and BH mass of $10^{7.21}\, \Msun$ given by \texttt{MOSFiT} (see Table \ref{tab:param-nested}), assuming $R_{\star} \approx R_{\odot}(\Mstar/\Msun) \approx 0.1 R_{\odot}$ \citep{demircan91}, the canonical tidal radius is $R_{\rm t} \approx R_{\star}\left(\Mbh/\Mstar\right)^{1/3} \approx 1.6\,G\Mbh/c^2$. In order to be tidally disrupted and not directly captured by the BH, the tidal radius must be outside of the direct-capture radius (which is larger than the horizon radius for all BH spins $a<1$, and is $4GM/c^2$ for $a=0$; e.g., \citealt{will12}). The direct-capture radius is a function of the BH spin, the square of the specific angular momentum of the star, and the projection of the angular momentum onto the spin axis of the BH, and is minimized at a value of\footnote{We assume $a>0$ in this expression, i.e., the stellar angular momentum is aligned with the BH spin \citep[e.g.,][]{will12, dorazio19, coughlin22}.} $R_{\rm dc} = GM/c^2(1+\sqrt{1-a})^2$. Requiring that $R_{\rm t} \geq R_{\rm dc}$ here sets $a \gtrsim  0.93$ for \emph{any} star to be tidally disrupted and not directly captured.
In fact, even with a spin value of $a \approx 1$, it is statistically improbable for the star to be injected into the loss cone, tidally disrupted, and not directly captured; if we assume that stars entering the loss cone are isotropically distributed at large radii and we are in the pinhole regime, such that two-body interactions result in a large relative change in the square of the specific angular momentum of the star on a per orbit basis (e.g., \citealt{frank76, lightman77, merritt13, stone16}), then the formalism described by \citet{coughlin22} predicts for the parameters here that the fraction of TDEs (i.e., the fraction of stars injected into the loss cone that are tidally disrupted and not directly captured) is $\sim 0.6$\%. This makes the result highly unlikely.

If the event were a partial TDE, these spin constraints and low probabilities would be somewhat ameliorated. Since a partial TDE occurs at $\sim 2 r_{\rm t}$ for a low-mass star (\citealt{Guillochon2013, mainetti17, Miles2020}; see also \citealt{Gafton2015} in the relativistic case at a comparable BH mass, in particular their Figure 3), the corresponding limit on the BH spin 
is $a \gtrsim 0.39$. The probability of being tidally disrupted and not directly captured for a BH spin of $a = 0.999$ is then $\sim 5.1\%$. Additionally, \cite{Coughlin2019} and \cite{Miles2020} found that the fallback rate asymptotically declines as $\propto t^{-9/4}$ (rather than the canonical $t^{-5/3}$) in the case of a partial TDE, which is also more consistent with the best-fit power law of $t^{-2.05}$ for AT\,2019azh (assuming that the fallback rate closely tracks the accretion luminosity \citep[e.g.,][]{Mockler2019, Nicholl2022}. However, \texttt{MOSFiT} finds a best-fit scaled impact parameter of $b=0.99_{-0.03}^{+0.01}$, indicating a full disruption. This means that the best-fit $\Mbh$, $\Mstar$, and $b$ from \texttt{MOSFiT} are not self-consistent \textbf{with a fully accretion-powered picture}.

\textbf{Using the extreme upper and lower values for $\Mstar$ and $\Mbh$, respectively, allowed by the total errors listed in Table \ref{tab:param-nested} amealiorates the problem, requiring $a \gtrsim  0.14$ for a full disruption. In addition, if some of the early emission is contributed by outer shocks, this would decrease the rise-time of the accretion-powered part of the light curve, reducing $\Mbh$ even further, and making a full disruption more likely. We conclude that \texttt{MOSFiT} can only marginally fit the data self-consistently assuming a fully accretion-powered light curve, and that at least some contribution from an additional power source at early times is necessary.}

If we perform the same calculations using the values from \texttt{TDEMass} (i.e., $\Mbh = 2.5^{+0.29}_{-0.24} \times 10^{6}\,\Msun$, $\Mstar = 4.8^{+4}_{-2.5}\,\Msun$), where in this case $R_{\star} \approx 3 R_{\odot}$ \citep{demircan91}, we find $R_{\rm t} \approx 44.3\,G\Mbh/c^2$. Here, $R_{\rm t} \geq R_{\rm dc}$ for any black hole (BH) spin, \textbf{making this model entirely self-consistent}.

\subsection{Time Lag Between \ha\ Emission and the Continuum}
 
Our time lag of $1.40 \pm{0.93}$\,days between the $V$-band light curve and \ha\ luminosity is inconsistent with that of \cite{Hinkle2021a} who measure a $\sim 23$\,days time lag. This stems mainly from a difference between our determination of the light-curve peak and theirs. Their light-curve peak was measured at MJD $58548^{+6.30}_{-2.60}$ (roughly 19\,days before ours). This peak was determined by \cite{Hinkle2021a} as the median value obtained from fitting a second-order polynomial to 10,000 realizations of bolometric light curves, generated from bolometrically corrected ASAS-SN $g$-band data, with the bolometric corrections inferred from blackbody fits. The peak light-curve time determined by \cite{Hammerstein2023} of MJD $58566^{+1.16}_{-1.75}$ is closer to ours.

\section{Summary and Conclusions}
\label{sec:summary}

AT\,2019azh is a H+He-TDE and is one of the best-observed UV/optical TDEs to date, having extensive spectroscopic coverage and multiwavelength photometric coverage starting several weeks before peak brightness (Figure \ref{fig:lc}). These observations reveal the following for the first time:
\begin{enumerate}
    \item A robust change in slope and possible bump in the early light curve of a TDE.
    \item The early evolution of the \ha\ emission line in a TDE. 
\end{enumerate}
Unfortunately, no models exist today that can be compared to these observational characteristics; however, they could be used to constrain future models of TDE emission sources and line formation. Relatively high cadence (1--2 days) observations of TDEs are required to test if the light-curve structure observed in AT\,2019azh is a common feature of rising TDE light curves.

We detect a possible MIR excess beyond what is expected from the optical/UV blackbody at those wavelengths. This excess, detected 5.15\,days after the $g$-band peak, might be due to a prompt dust echo. However, we are not able to determine its origin without additional observations.

The post-peak bolometric decline of AT\,2019azh is not well described by a $t^{-5/3}$ power law, or by any power law, but is better fit by an exponential.  
We find no significant delay between the peak of the $V$-band light curve and the \ha\ luminosity in AT\,2019azh. 

High-cadence pre-peak observations of more TDEs will be able to determine how common the features seen here are among the TDE population. In addition, more detailed modeling of TDE emission is needed to match the quality of current TDE observations and to help constrain the emission mechanism(s) in TDEs. This is an essential step before we can use TDEs to robustly measure SMBH properties.

~\\
\centerline{Acknowledgments}

We thank B. Mockler for helpful advice on using MOSFiT. 
S.F., I.A., and L.M. acknowledge support from the European Research Council (ERC) under the European Union’s Horizon 2020 research and innovation program (grant agreement No. 852097). S.F. and I.A. acknowledge further support from the Israel Science Foundation (grant No. 2108/18).
The Las Cumbres Observatory group is supported by National Science Foundation (NSF) grants AST-1911225 and AST-1911151.
P. Clark and O.G. were supported by the Science \& Technology Facilities Council (grants ST/S000550/1 and ST/W001225/1). 
G.L. was supported by a research grant (19054) from VILLUM FONDEN.
M.N. is supported by the ERC under the European Union’s Horizon 2020 research and innovation program (grant agreement No. 948381) and by UK Space Agency grant No. ST/Y000692/1.
C.P.G. acknowledges financial support from the Secretary of Universities and Research (Government of Catalonia) and by the Horizon 2020 Research and Innovation Program of the European Union under the Marie Sk\l{}odowska-Curie program.
The SNICE research group acknowledges financial support from the Spanish Ministerio de Ciencia e Innovaci\'on (MCIN), the Agencia Estatal de Investigaci\'on (AEI) 10.13039/501100011033, and the European Social Fund (ESF) ``Investing in your future'' under the 2019 Ram\'on y Cajal program RYC2019-027683-I and the PID2020-115253GA-I00 HOSTFLOWS project, from Centro Superior de Investigaciones Cient\'ificas (CSIC) under the PIE project 20215AT016, and the program Unidad de Excelencia Mar\'ia de Maeztu CEX2020-001058-M, and from the Departament de Recerca i Universitats de la Generalitat de Catalunya through the 2021-SGR-01270 grant.
S.M. and T.R. acknowledge support from the Research Council of Finland project 350458.
A.V.F.'s group at UC Berkeley has been supported by the Christopher R. Redlich Fund, William Draper, Timothy and Melissa Draper, Briggs and Kathleen Wood, {Sanford Robertson} (T.G.B. is a Draper-Wood-Robertson Specialist in Astronomy), and numerous other donors.
F.O. acknowledges support from MIUR, PRIN 2020 (grant 2020KB33TP) ``Multimessenger astronomy in the Einstein Telescope Era (METE).''
E.R.C. acknowledges support from the National Science Foundation through grant AST-2006684.
P.C. acknowledges support via an Academy of Finland grant (340613; PI: R. Kotak).
This work was funded in part by ANID, Millennium Science Initiative, ICN12\_009.

This work is based in part on observations collected at the Las Cumbres Observatory, the Copernico 1.82\,m Telescope (Asiago Mount Ekar, Italy) operated by the Italian National Astrophysical Institute -- INAF, Osservatorio Astronomico di Padova, the European Organization for Astronomical Research in the Southern Hemisphere, Chile, as part of ePESSTO under ESO program ID 199.D-0143(T) (PIs: S. Smartt, C. Inserra), the Nordic Optical Telescope, owned in collaboration by the University of Turku and Aarhus University, and operated jointly by Aarhus University, the University of Turku, and the University of Oslo, representing Denmark, Finland, and Norway, the University of Iceland and Stockholm University at the Observatorio del Roque de los Muchachos, La Palma, Spain, of the Instituto de Astrofisica de Canarias, and on observations made under programme W/2019B/P7 with the William Herschel Telescope operated on the island of La Palma by the Isaac Newton Group of Telescopes in the Spanish Observatorio del Roque de los Muchachos of the Instituto de Astrofísica de Canarias.
The NOT observations were obtained through the NUTS Collaboration supported in part by the Instrument Center for Danish Astrophysics (IDA).
A major upgrade of the Kast spectrograph on the Shane 3\,m telescope at Lick Observatory was made possible through generous gifts from William and Marina Kast as well as the Heising-Simons Foundation. Research at Lick Observatory is partially supported by a generous gift from Google. 
We thank for their assistance the staff at the various observatories where data were obtained.

This research has made use of the NASA/IPAC Infrared Science Archive, which is funded by the National Aeronautics and Space Administration (NASA) and operated by the California Institute of Technology. This publication also makes use of data products from NEOWISE, which is a project of the Jet Propulsion Laboratory/California Institute of Technology, funded by the Planetary Science Division of NASA.
This publication makes use of data products from the WISE, which is a joint project of the University of California, Los Angeles, and the Jet Propulsion Laboratory/California Institute of Technology, funded by NASA.\\
~\\
Supporting research data are available on reasonable request from the corresponding author.

\bibliography{main}
\bibliographystyle{aasjournal}

\appendix 
\restartappendixnumbering

\section{MIR Light Curve}
\label{sec:appendix-MIR}

\begin{figure}[t]
    \centering
    \includegraphics[width=0.5\textwidth]{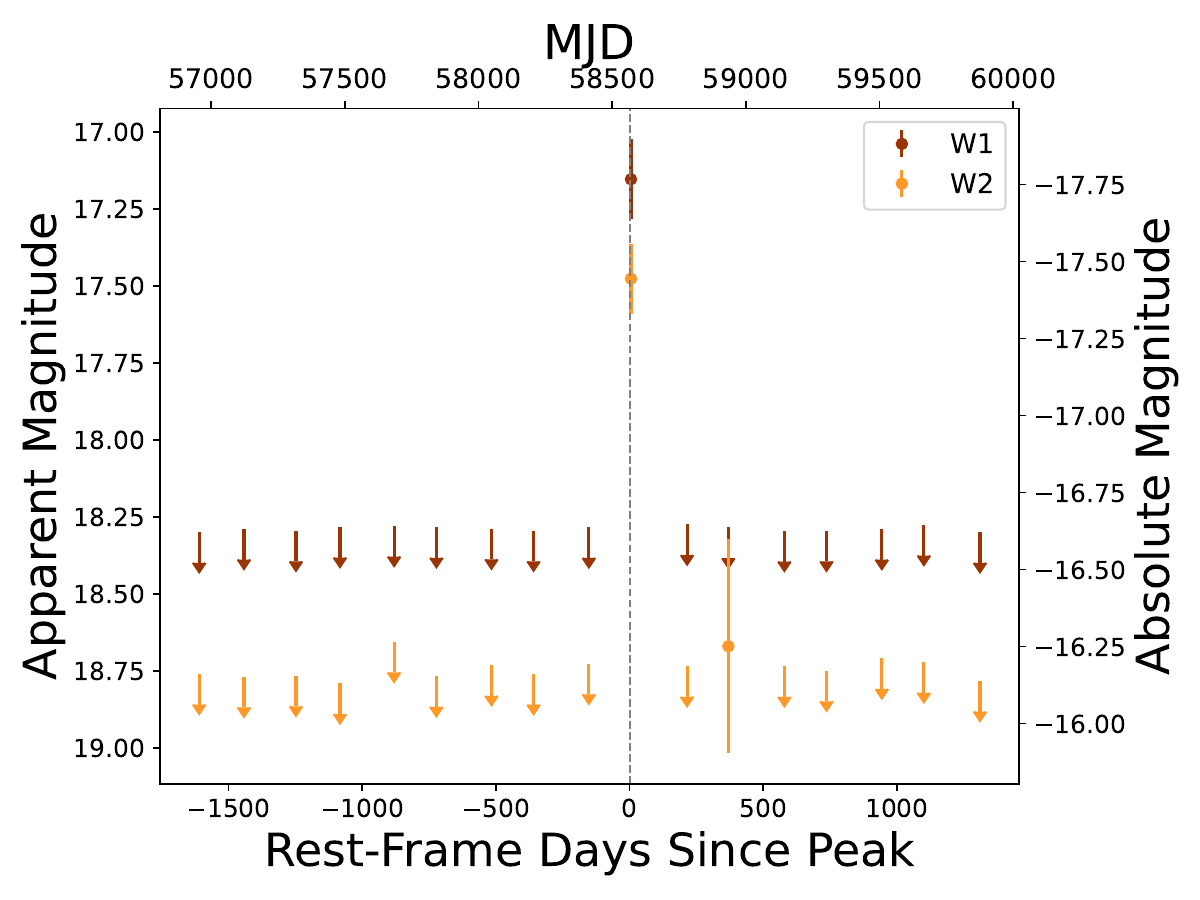}
    \caption{Milky Way extinction-corrected MIR light curves of AT\,2019azh from {\it WISE} in days relative to the $g$-band peak. Arrows indicate 3$\sigma$ nondetection upper limits and the dashed vertical line indicates the $g$-band peak.}
    \label{fig:MIR_lc}
\end{figure}

Figure \ref{fig:MIR_lc} displays the MIR light curves of AT\,2019azh. A significant MIR flare appears at 5.15\,days after the $g$-band peak.

\restartappendixnumbering
\section{UV/optical Peak Fits}
\label{sec:peak-fits}

\begin{figure}[t]
    \centering
    \includegraphics[width=0.5\textwidth]{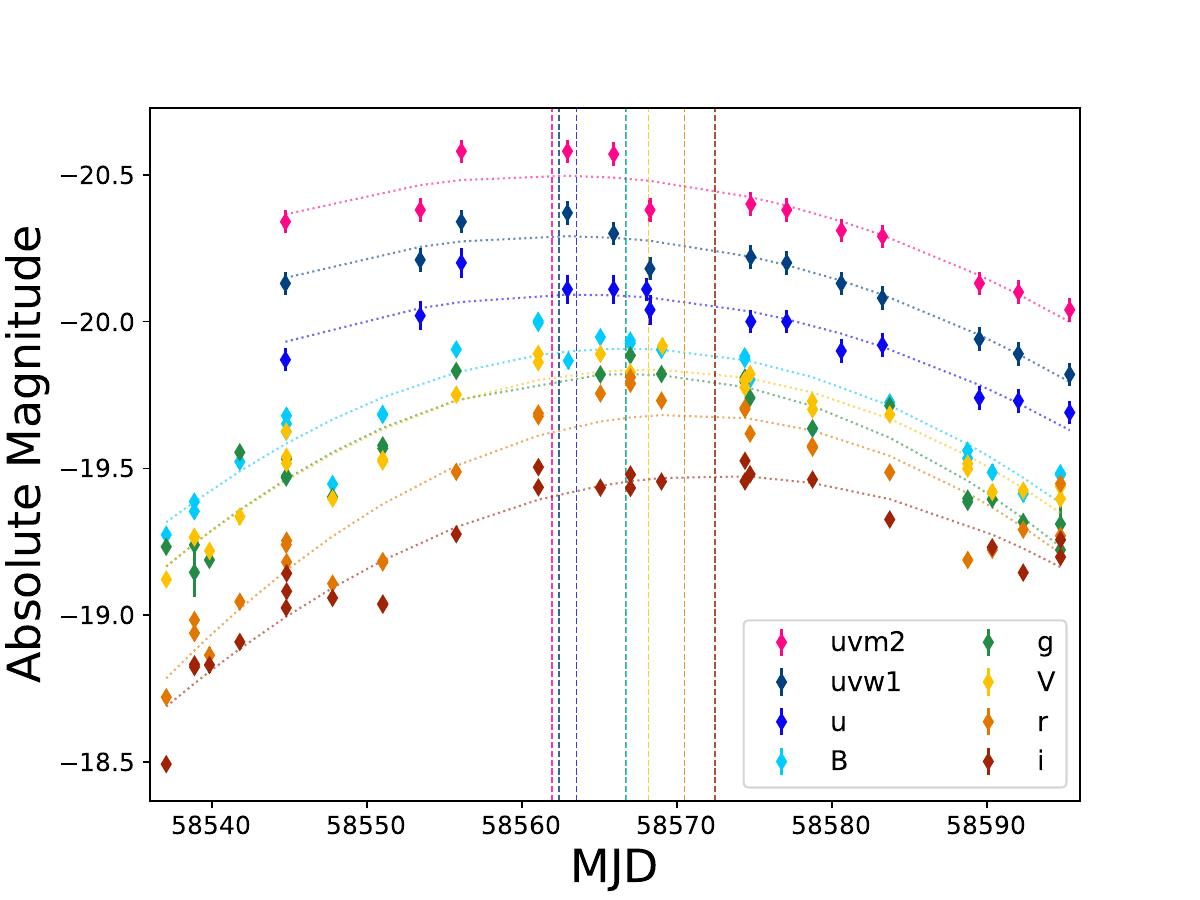}
    \caption{Parabolic fits of the host subtracted Las Cumbres optical photometry and {\it Swift} UV photometry of AT\,2019azh around peak brightness. The dashed vertical lines indicate the time of peak for each band from the best-fit parabola.}
    \label{fig:peak-fits}
\end{figure}

Figure \ref{fig:peak-fits} displays the second-order polynomial fits around the peak of the light curve for different bands.

\restartappendixnumbering
\section{MOSFiT Best-Fit Parameters}
\label{sec:appendix-mosfit}

\begin{figure*}[b]
    \centering
    \includegraphics[width=\textwidth]{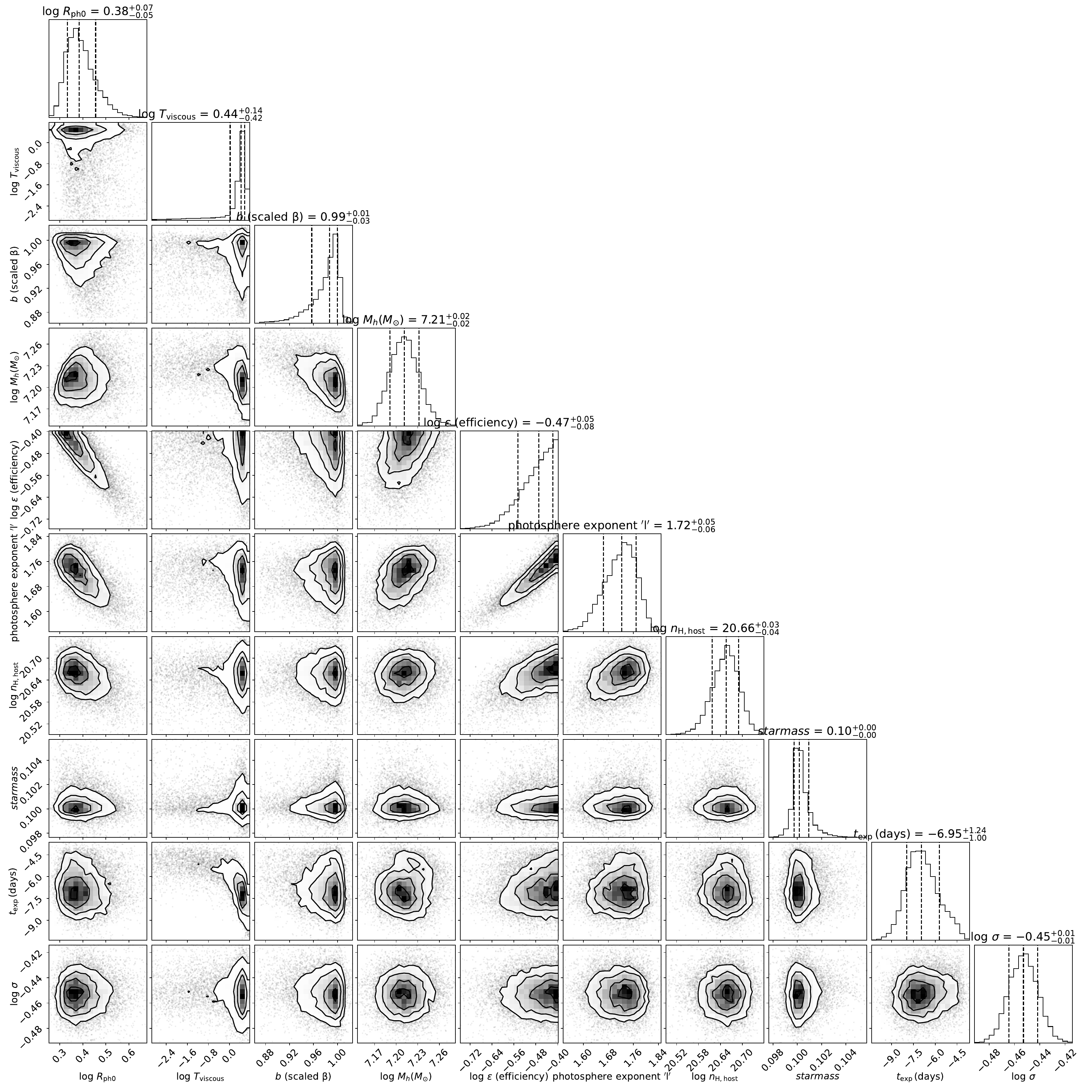}
    \caption{Corner plot showing the posterior parameter distributions for the \texttt{MOSFiT} model fit. 1$\sigma$ confidence intervals are marked.}
    \label{fig:mosfit-corner}
\end{figure*}

Figure \ref{fig:mosfit-corner} shows the two-dimensional \texttt{MOSFiT} posterior parameters distributions. The model fit can be seen to be well converged.

\restartappendixnumbering
\section{TDEMass Parameters} 
\label{sec:appendix-tdemass}

\begin{figure}[b]
    \centering
    \hspace*{-0.4cm}
    \includegraphics[width=0.5\textwidth]{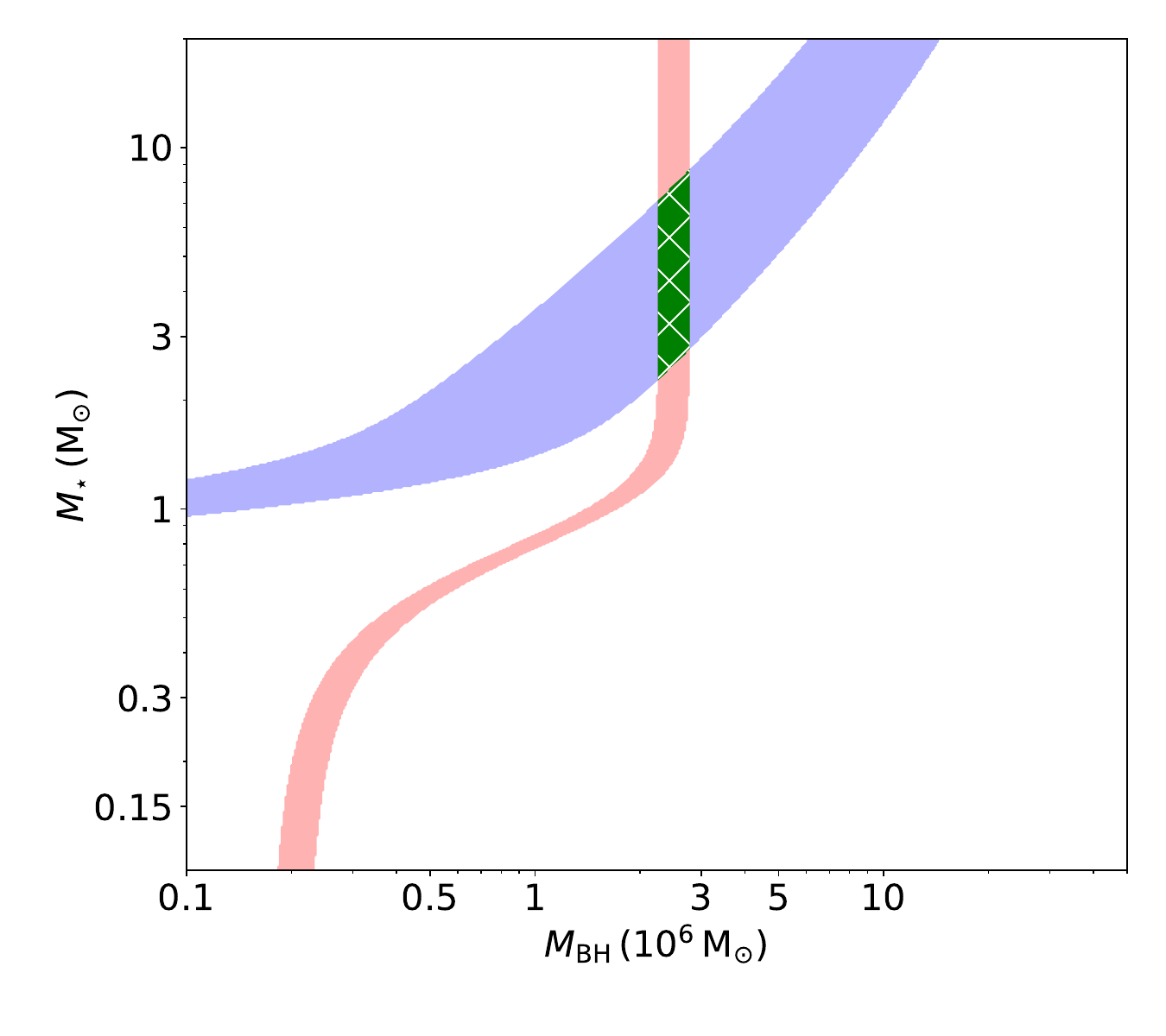}
    \caption{Inferred BH and star masses from \texttt{TDEMass}, with the blue shaded region indicating the range of solutions for the peak bolometric luminosity, and the red shaded region indicating the range of solutions for the blackbody temperature at this peak. The green region indicates the overlapping solutions.}
    \label{fig:tdemass} 
\end{figure}

Figure \ref{fig:tdemass} shows the inferred SMBH and star masses from \texttt{TDEMass}, based on the peak bolometric luminosity and the blackbody temperature at the peak from \texttt{SuperBol}.

\restartappendixnumbering
\section{Spectral Data Processing} 
\label{sec:appendix-spectra}

\begin{figure*}[t!]    
    \centering
    \includegraphics[width=0.8\textwidth]{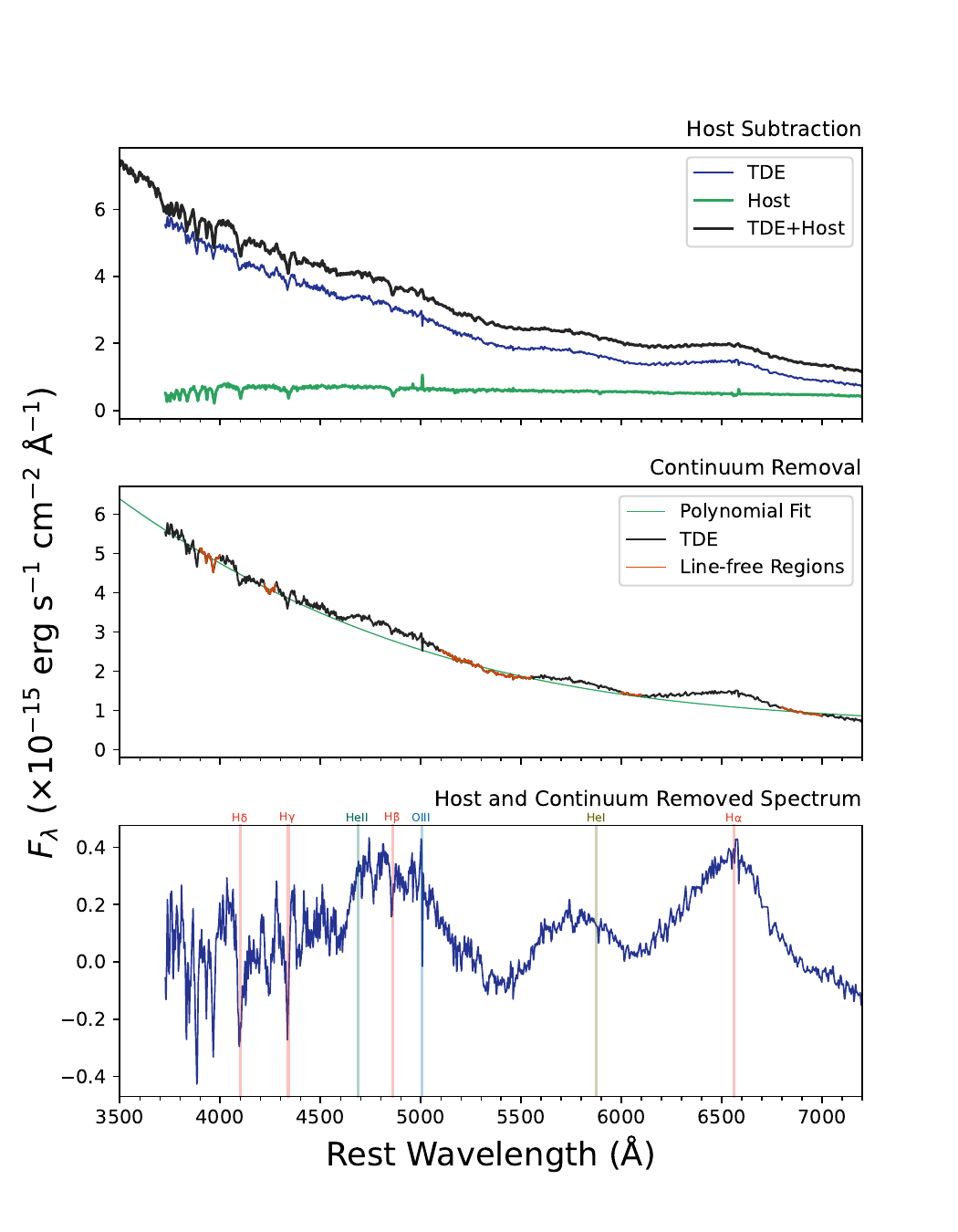}
    \caption{{\it Top:} archival resampled SDSS host galaxy spectrum (green), the photometrically calibrated spectrum of AT\,2019azh, including host contamination, 13\,days after light-curve peak (black), and the host-subtracted TDE spectrum (blue). {\it Middle:} host subtracted spectrum (black) with selected line-free regions (red) used for polynomial fitting (green). {\it Bottom:} host and continuum-subtracted spectrum, showing broad \ha, \heii, and \hei\ emission lines. The narrow \oiii\ and Balmer lines are likely oversubtracted host galaxy lines.}
    \label{fig:spec-phase13}
\end{figure*}

Figure \ref{fig:spec-phase13} displays the step-by-step data-processing procedure applied on a spectrum obtained 13\,days after the light-curve peak, as outlined in Section \ref{sec: analysis}. The same procedure is applied to all spectra of AT\,2019azh, apart from the du Pont and the WHT spectra for the reasons detailed in Section \ref{sec: analysis}.
 
\begin{figure*}[t]
    \centering
    \includegraphics[scale=0.9]{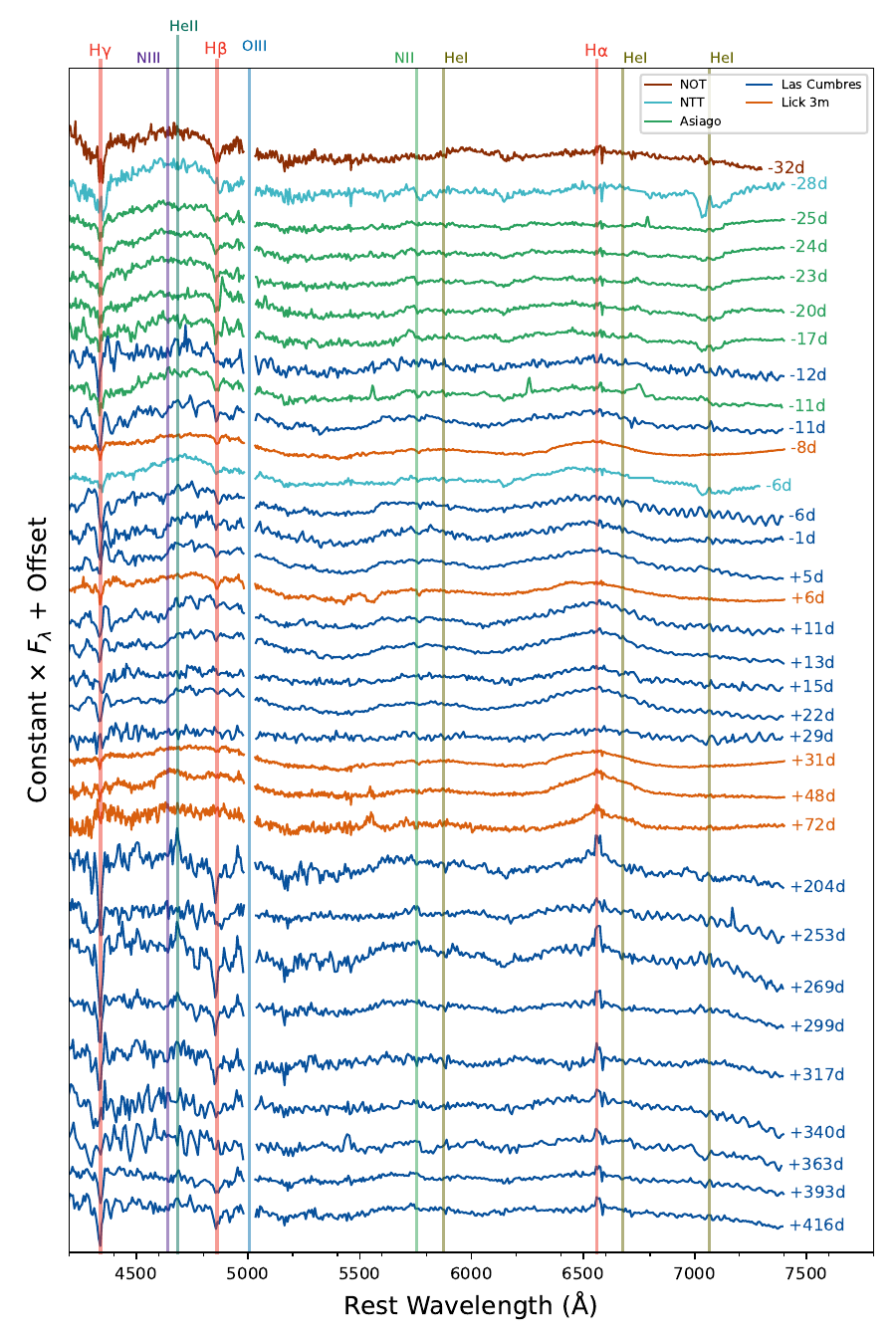}
    \caption{Spectroscopic evolution of AT\,2019azh after host galaxy and continuum removal. \oiii\ lines are masked for display purposes. The phase of each spectrum relative to the $g$-band light-curve peak is indicated beside it.}
    \label{fig:spectra-final}
\end{figure*}

Figure \ref{fig:spectra-final} shows the spectroscopic evolution of AT\,2019azh, after photometric calibration, host subtraction, and continuum removal, as described in Section \ref{sec: analysis}. 

Figure \ref{fig:ha-fits} shows the Gaussian fits of the \ha\ line performed on the host galaxy and continuum-subtracted spectra. Masking the feature around 6100--6200\,\AA\ does not significantly alter the fits, indicating that they are not strongly affected by this feature. In the $-32$\,d spectrum, removing the data blueward of 6200\,\AA\ also does not significantly alter the fit, indicating that it is not strongly affected by the emission seen there.

\begin{figure}
    
\gridline{\fig{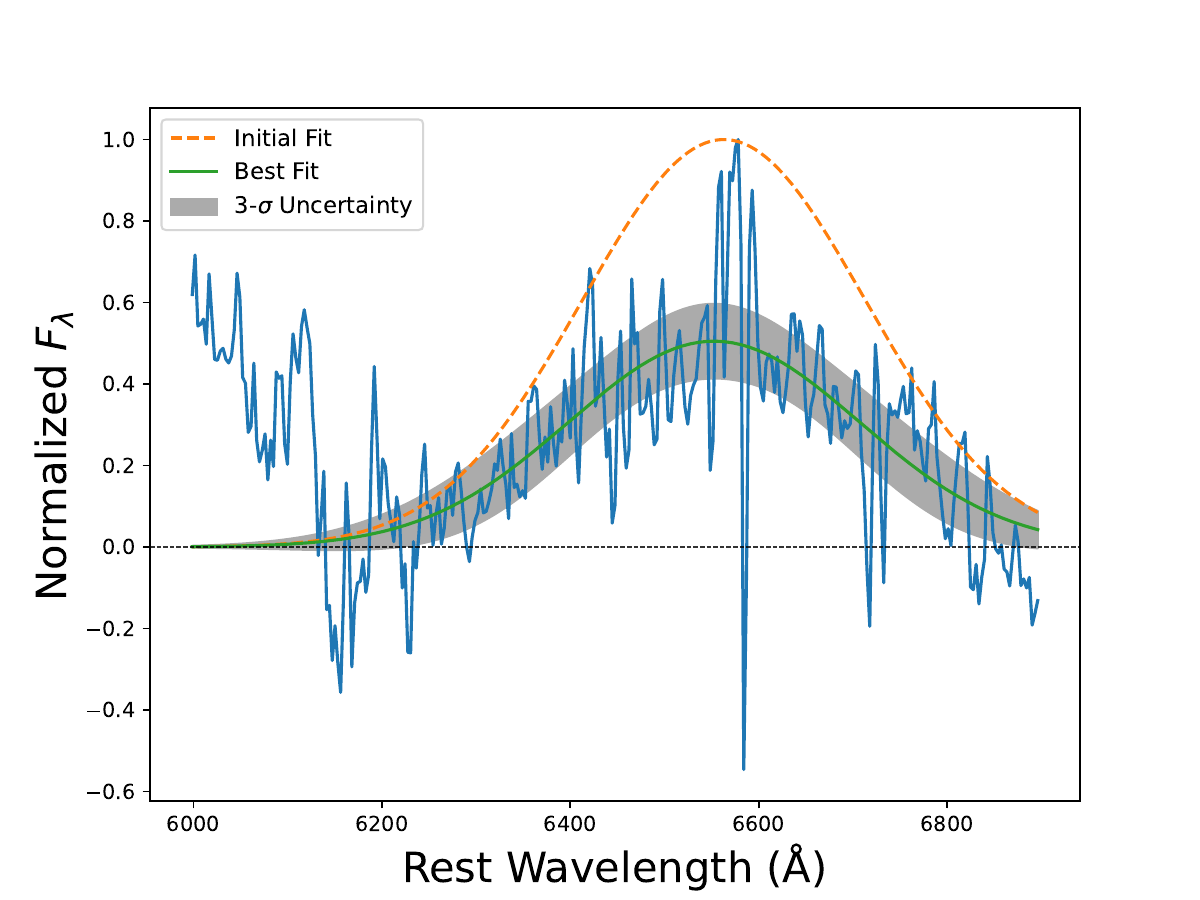}             {0.35\textwidth}{Phase: -32d}
          \fig{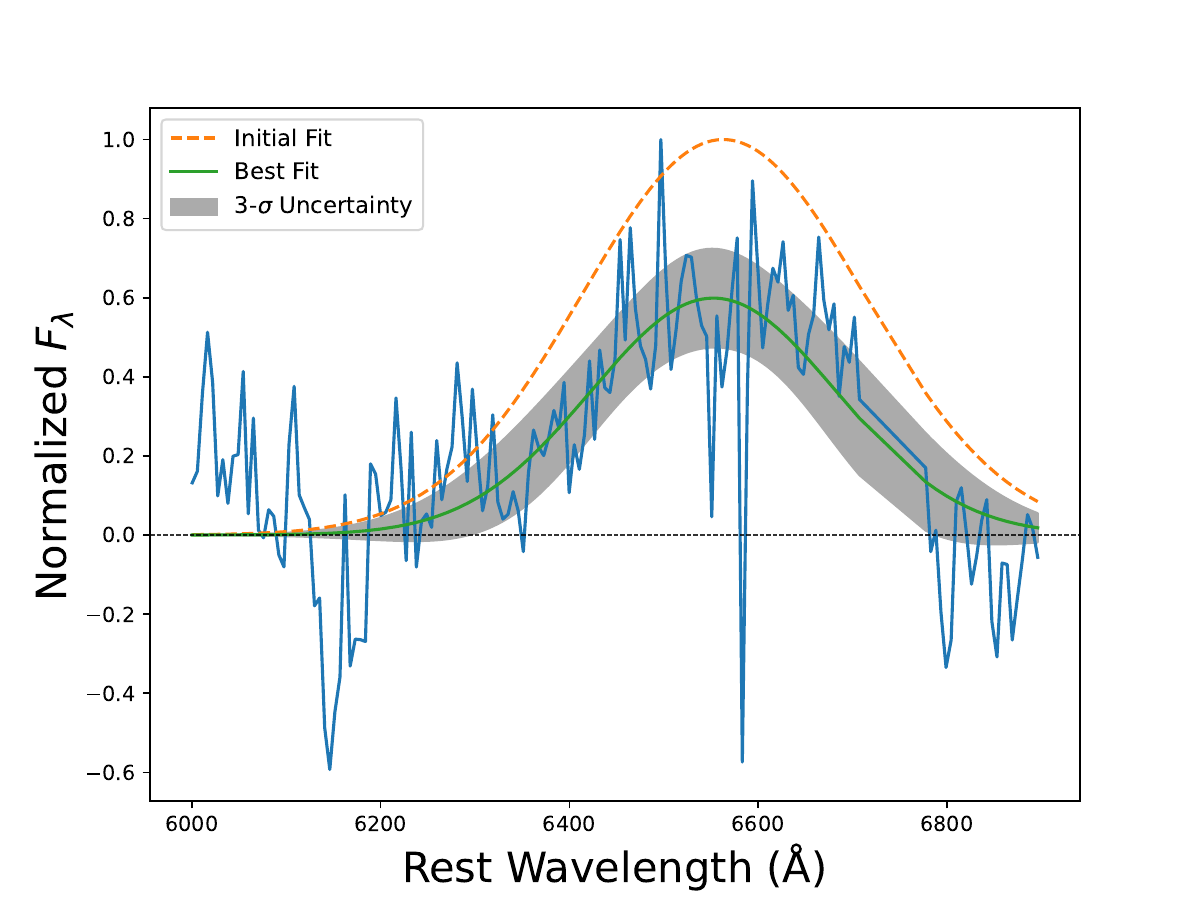}{0.35\textwidth}{Phase: -28d}
          \fig{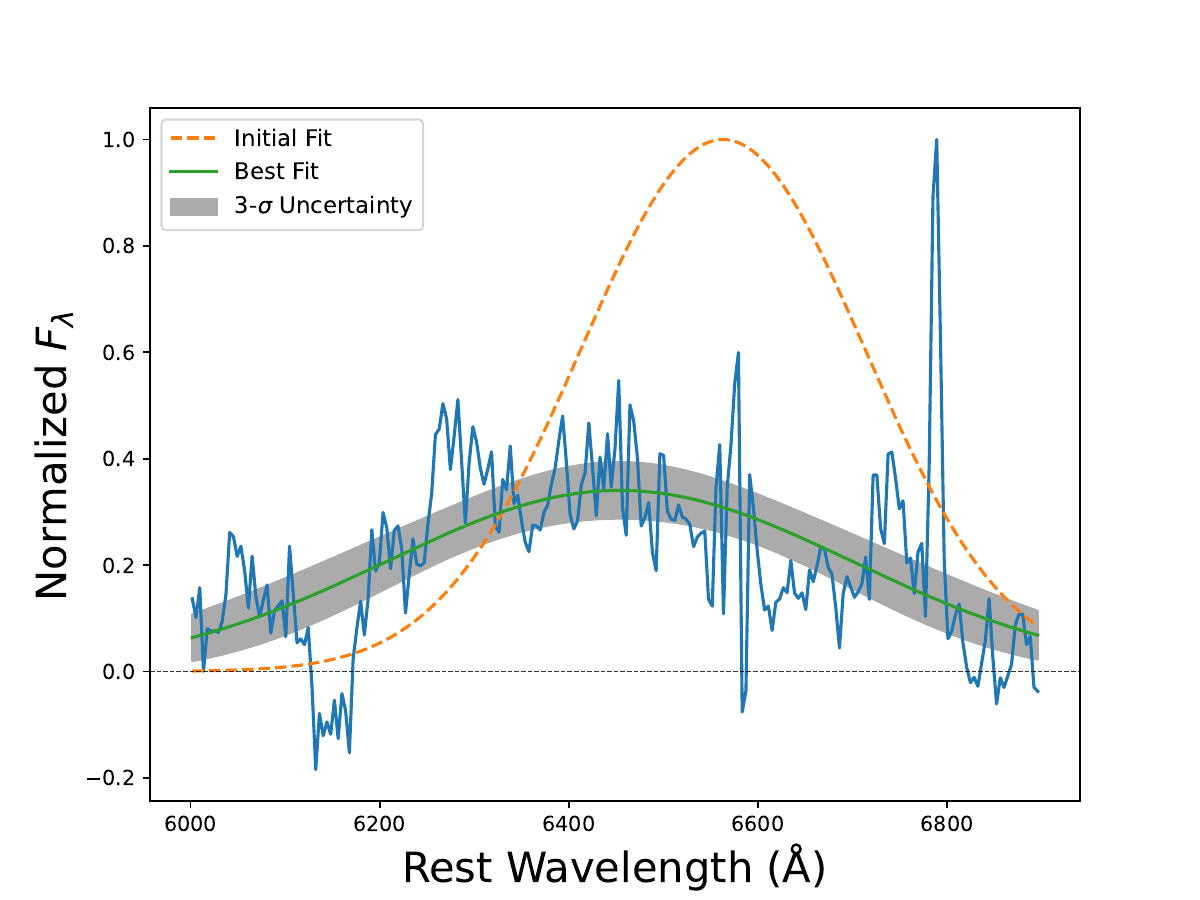}{0.35\textwidth}{Phase: -25d}
}
\gridline{\fig{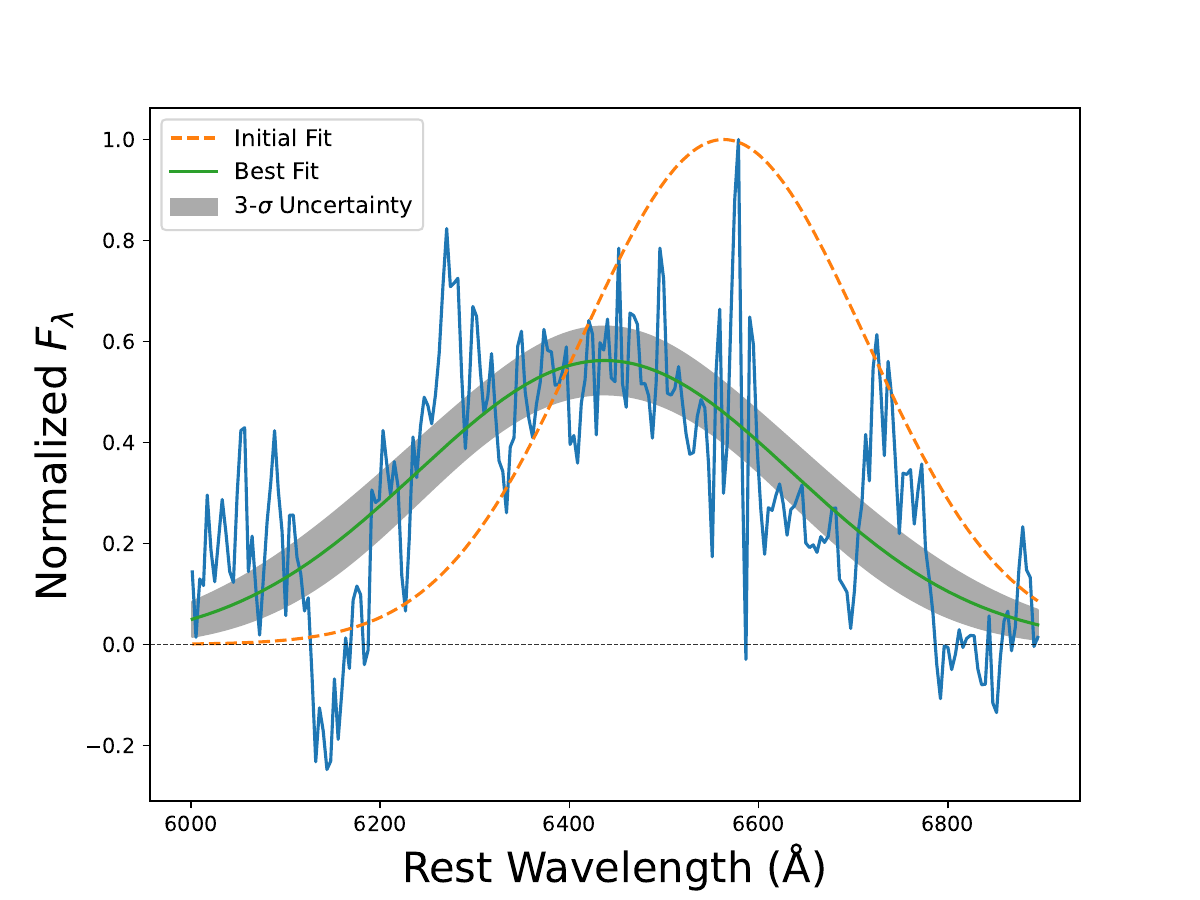}             {0.35\textwidth}{Phase: -24d}
          \fig{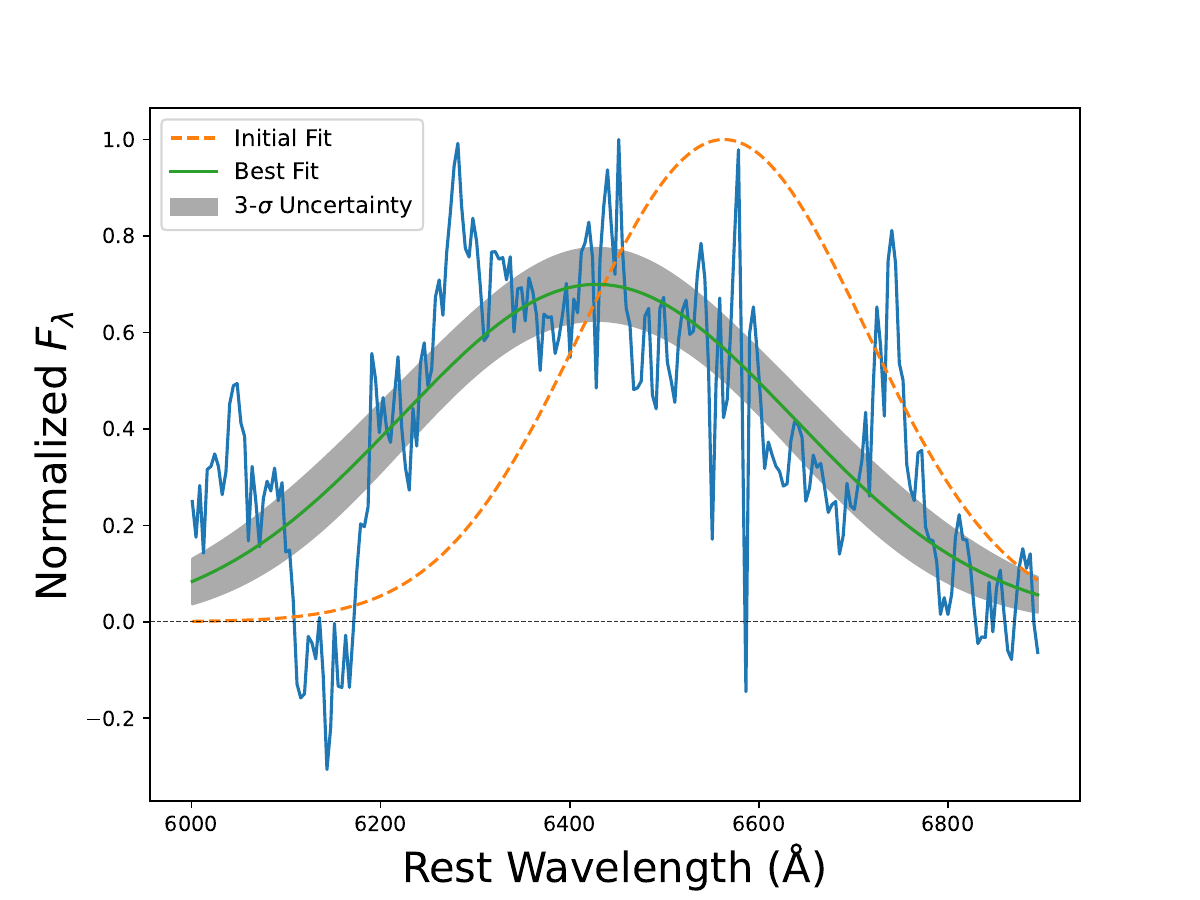}{0.35\textwidth}{Phase: -23d}
          \fig{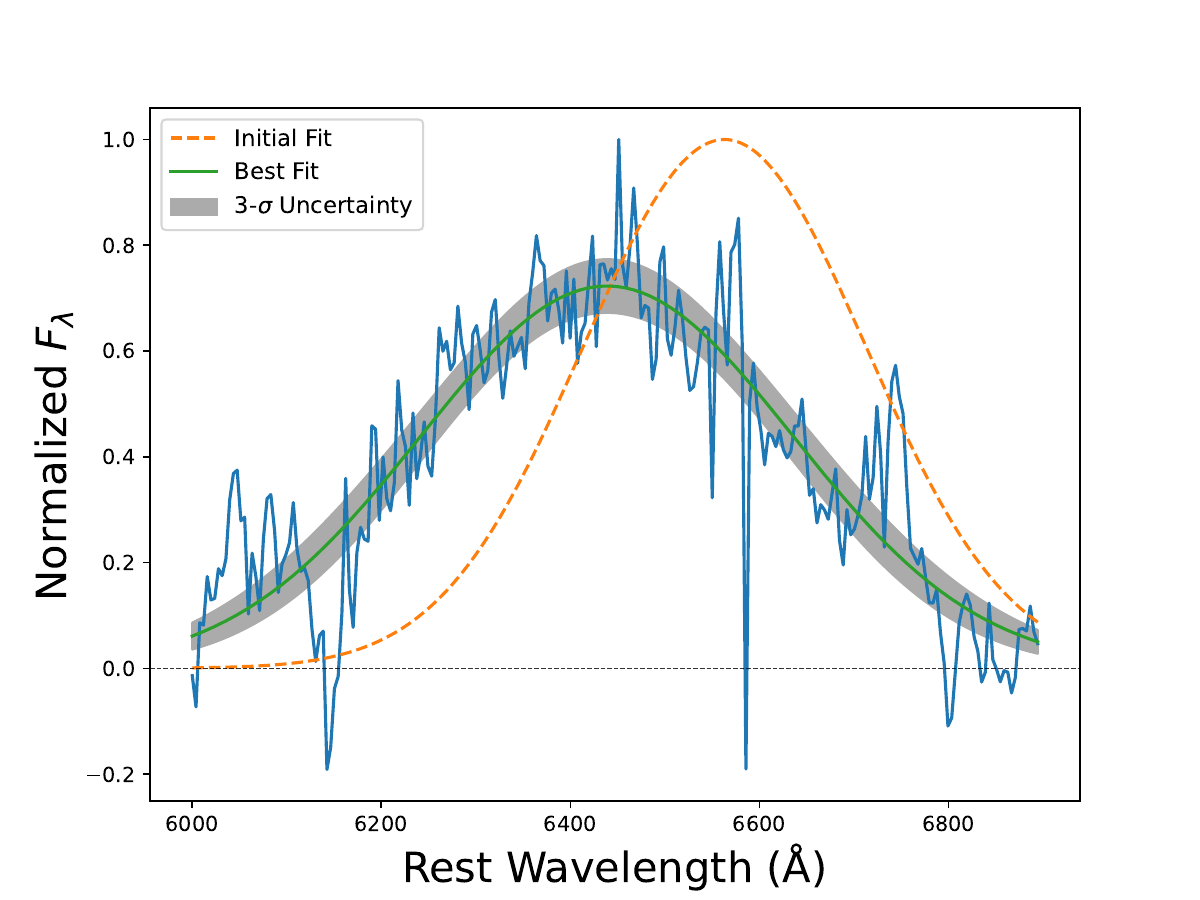}{0.35\textwidth}{Phase: -20d}
}
\gridline{\fig{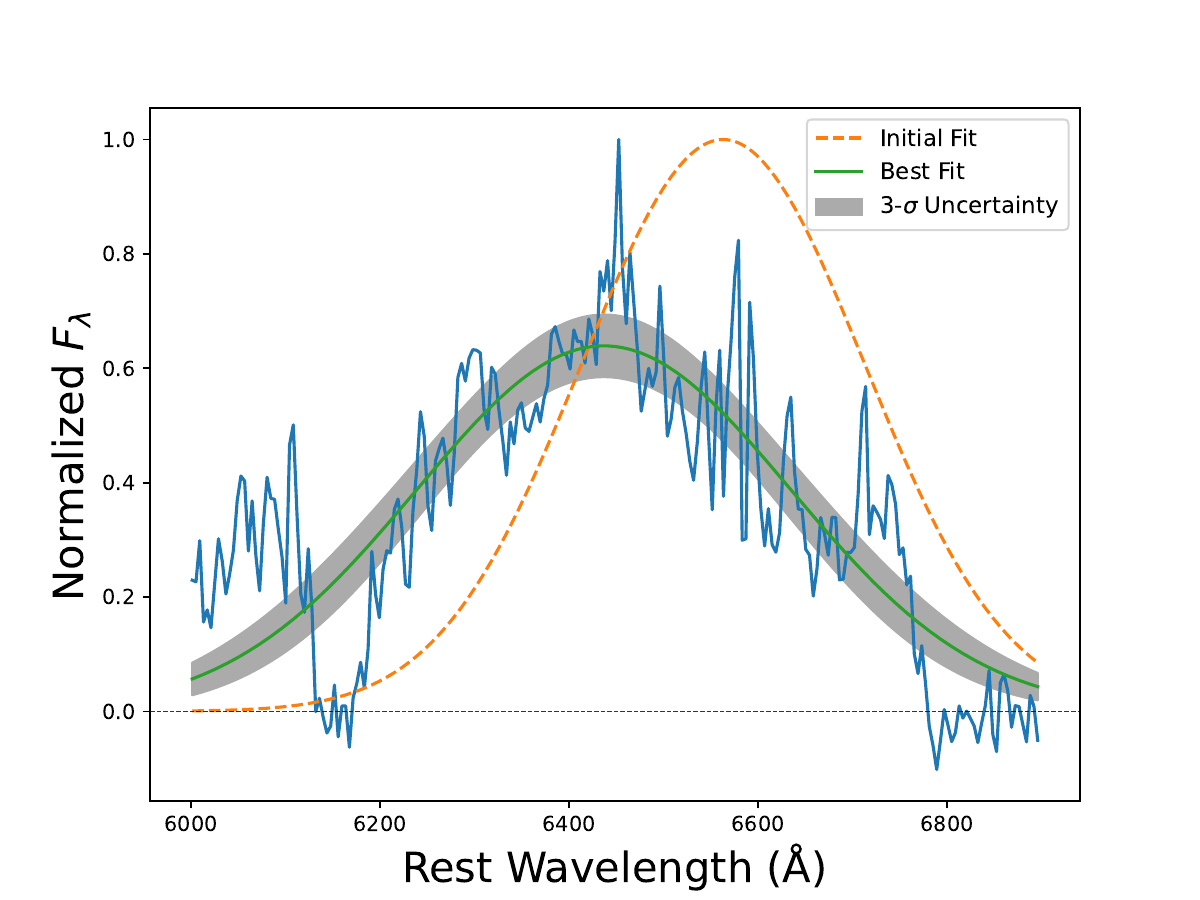}             {0.35\textwidth}{Phase: -17d}
          \fig{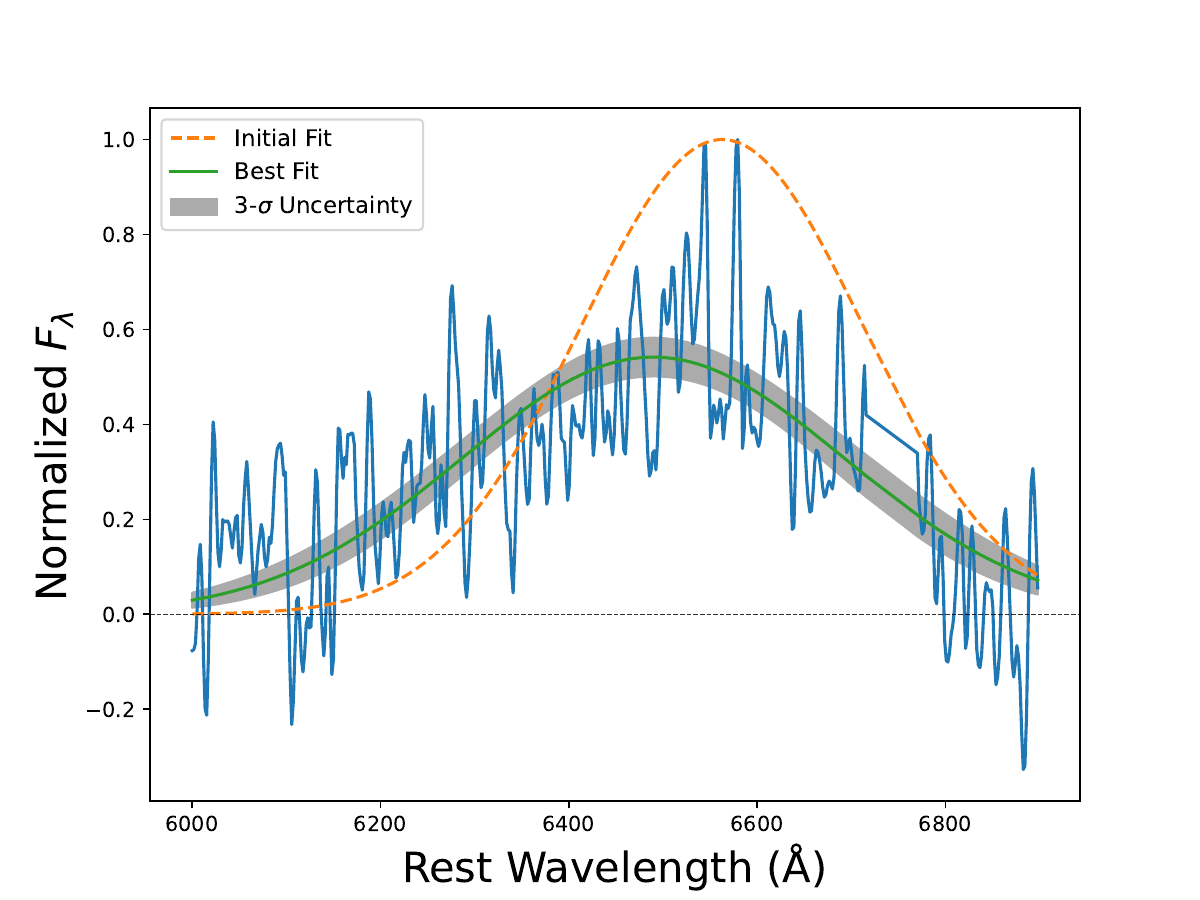}{0.35\textwidth}{Phase: -12d}
          \fig{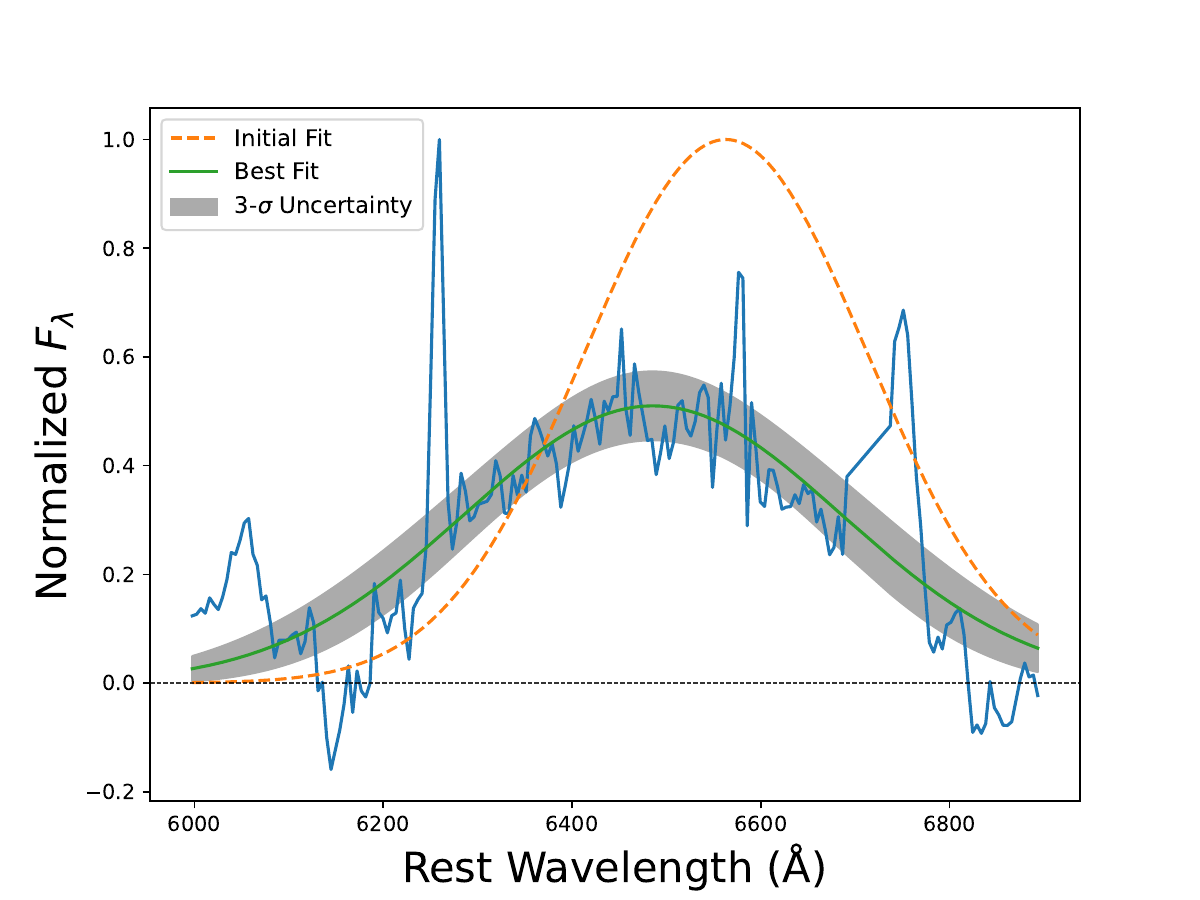}{0.35\textwidth}{Phase: -11d}
          }          
\gridline{\fig{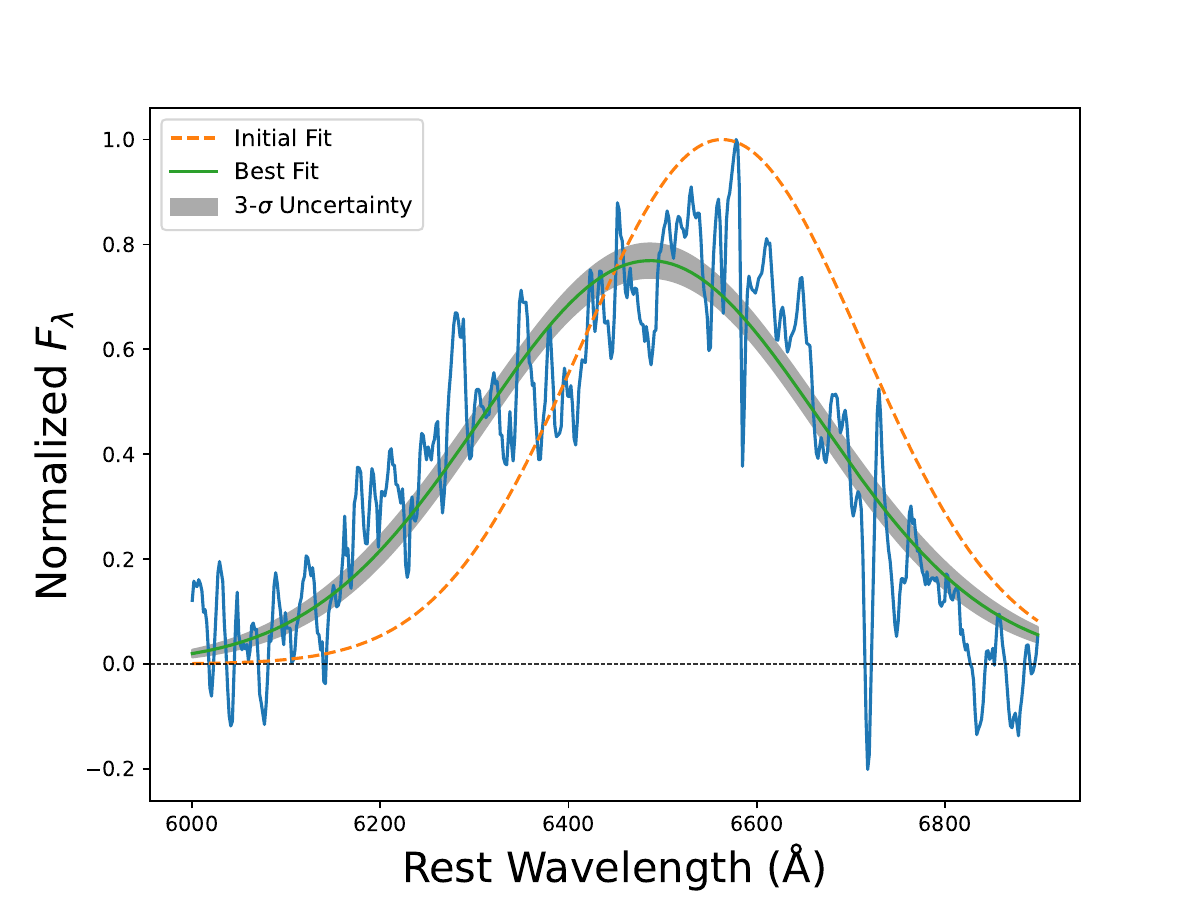}           {0.35\textwidth}{Phase: -11d}
          \fig{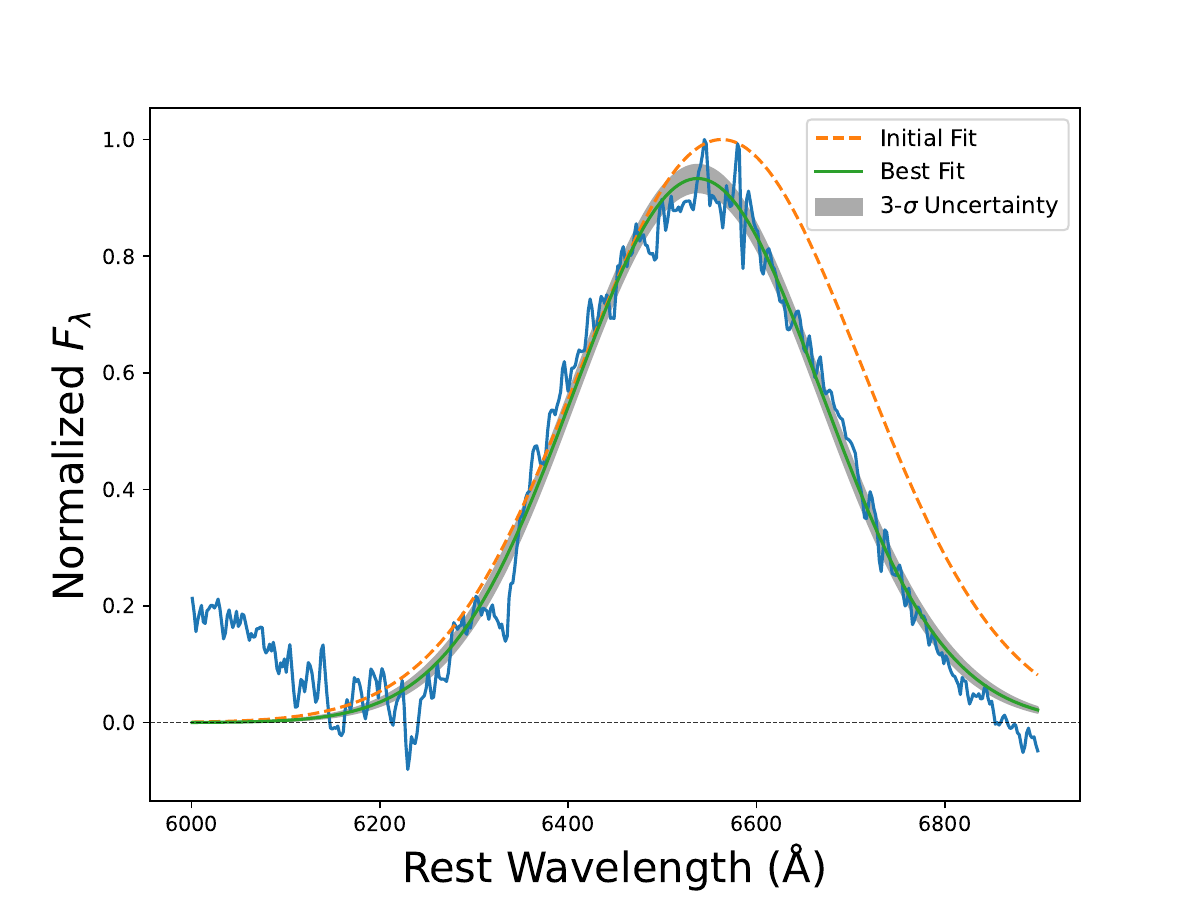}{0.35\textwidth}{Phase: -8d}
          \fig{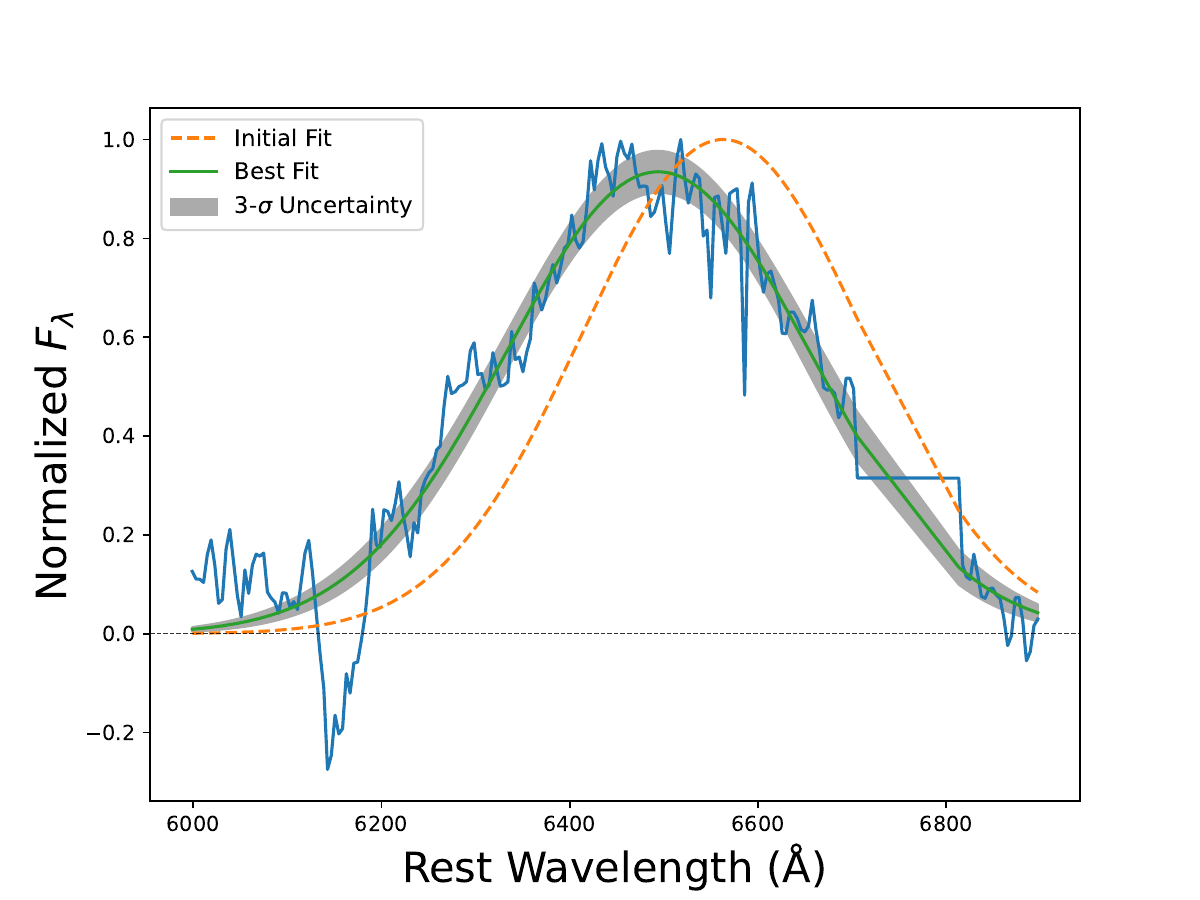}{0.35\textwidth}{Phase: -6d}
}
\end{figure}

\begin{figure}
\gridline{\fig{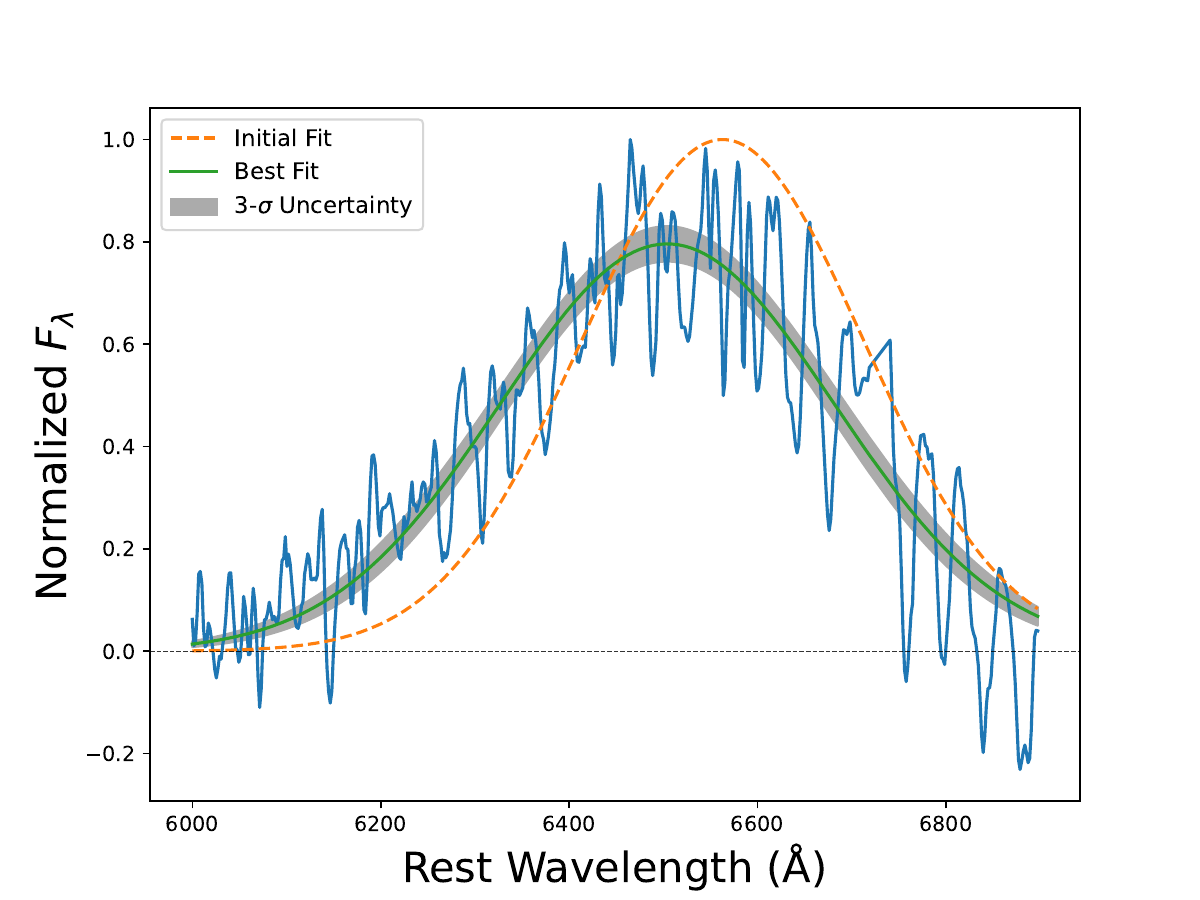}            {0.35\textwidth}{Phase: -6d}
          \fig{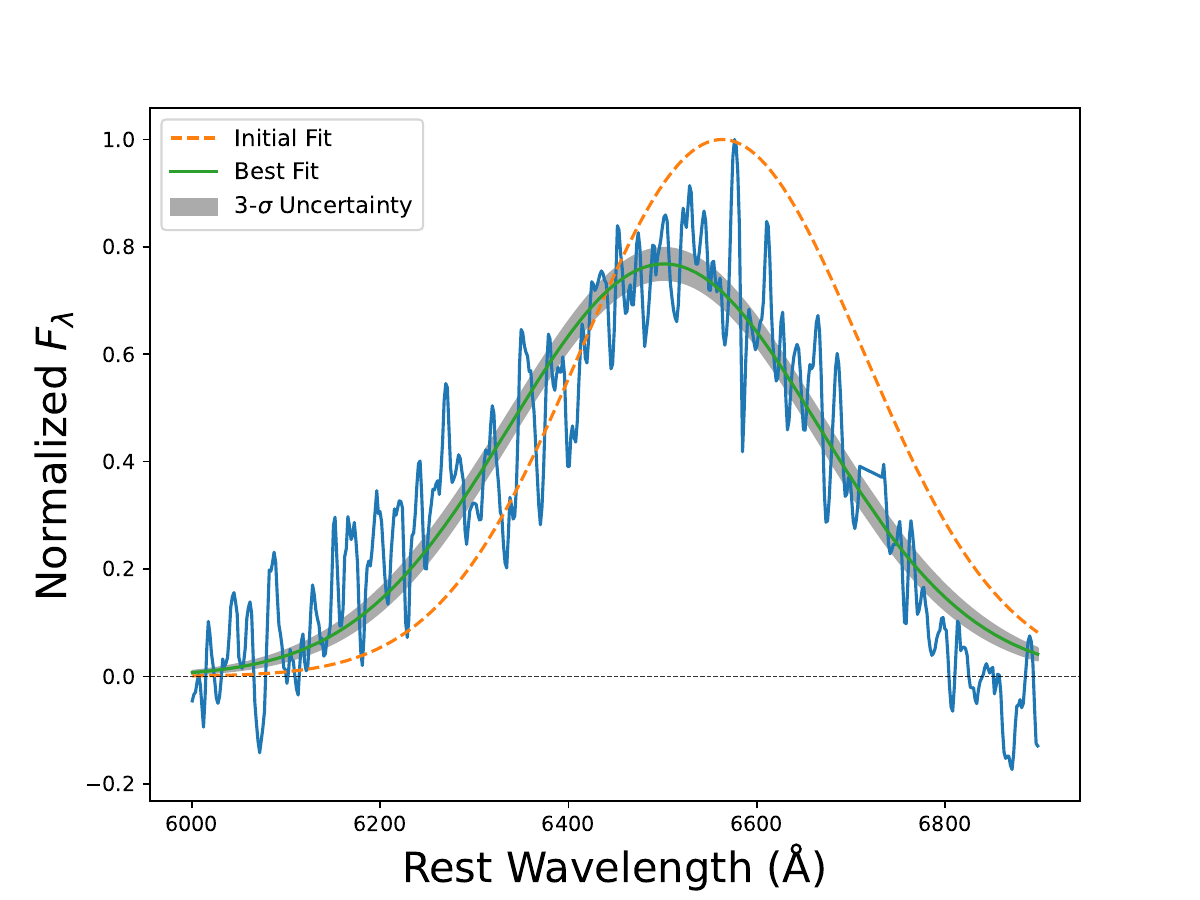}{0.35\textwidth}{Phase: -1d}
          \fig{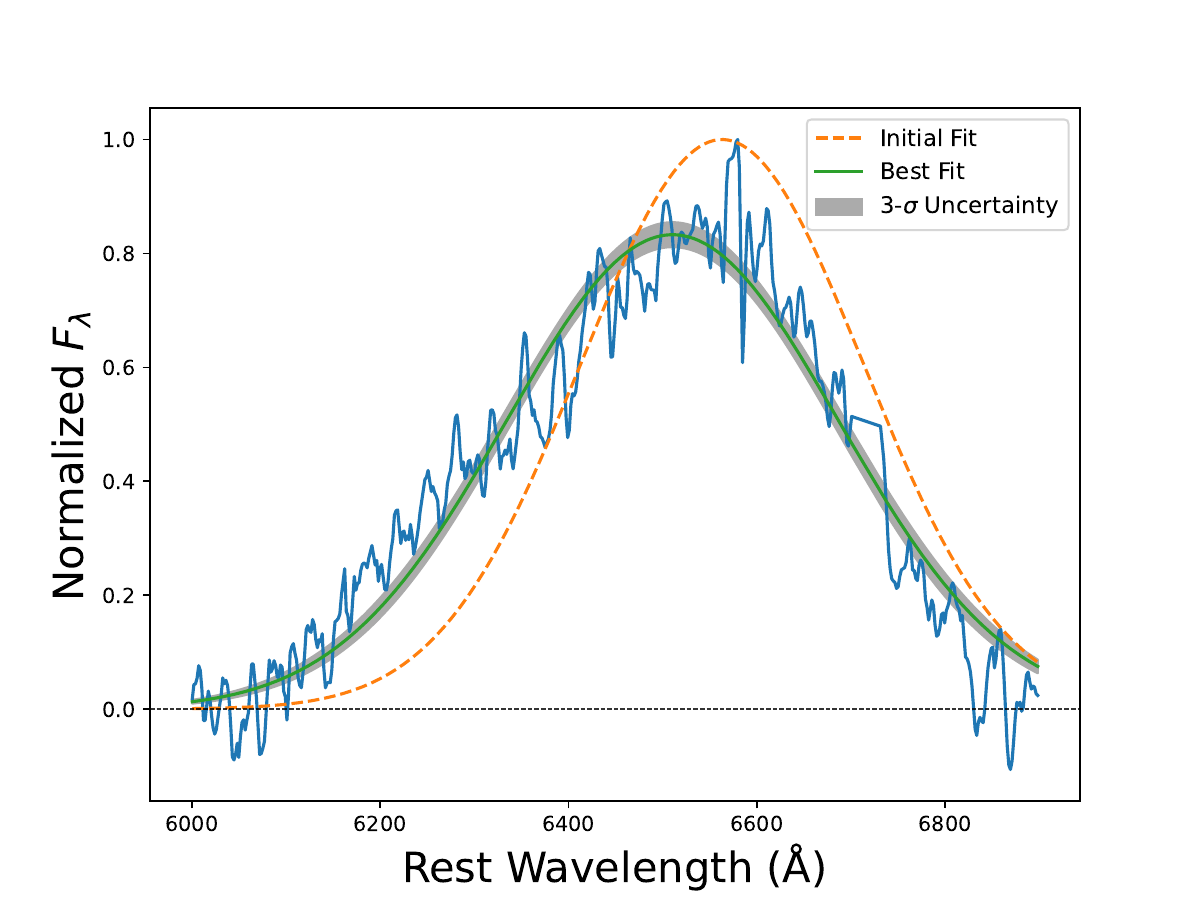}{0.35\textwidth}{Phase: +5d}
}
\gridline{\fig{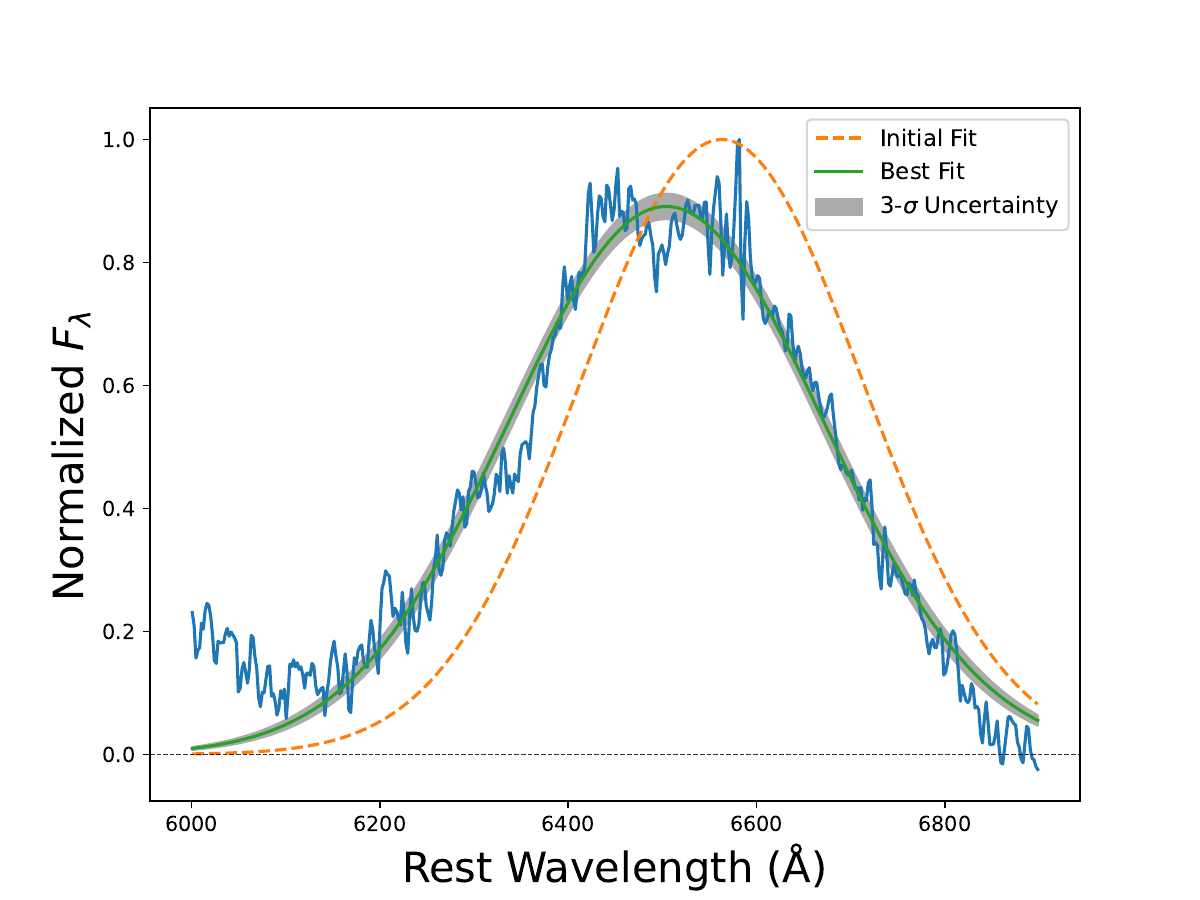}               {0.35\textwidth}{Phase: +6d}
          \fig{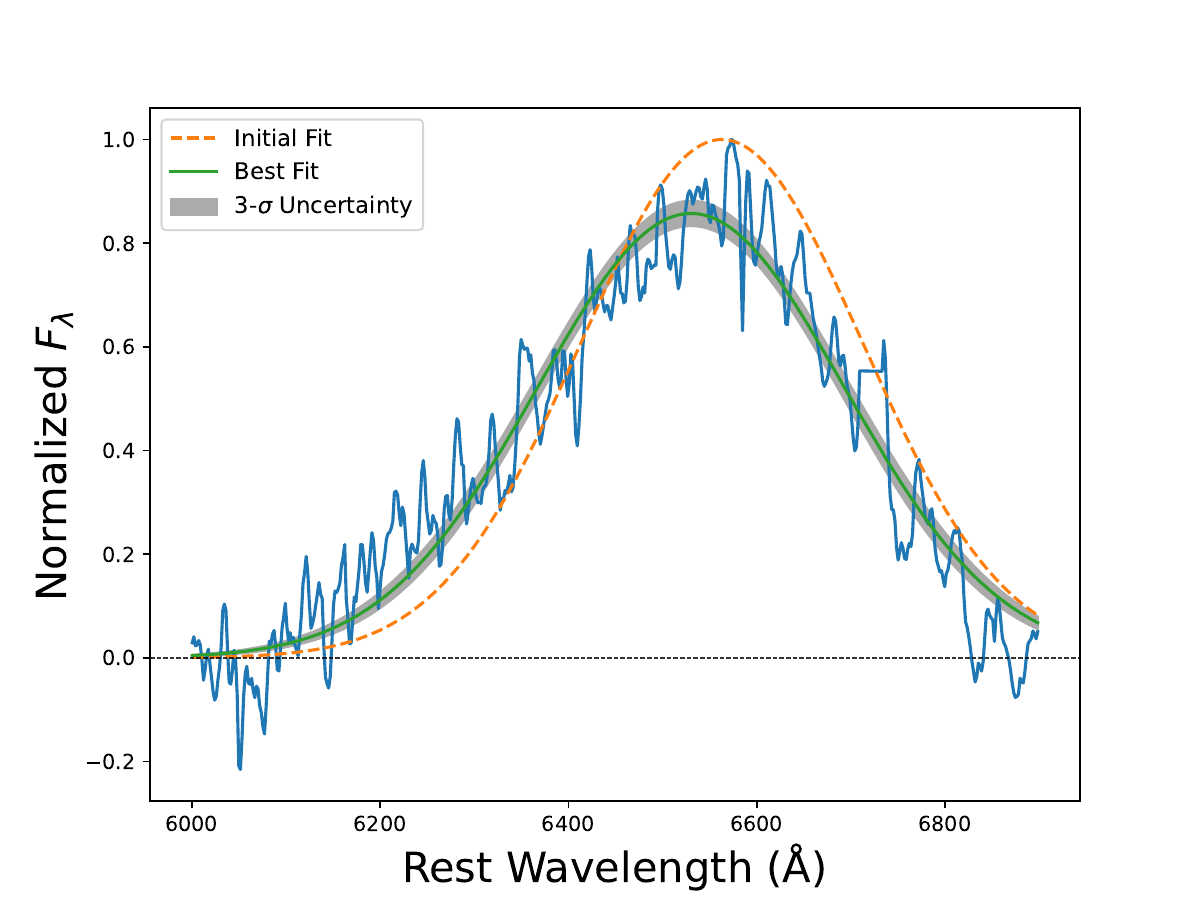}{0.35\textwidth}{Phase: +11d}
          \fig{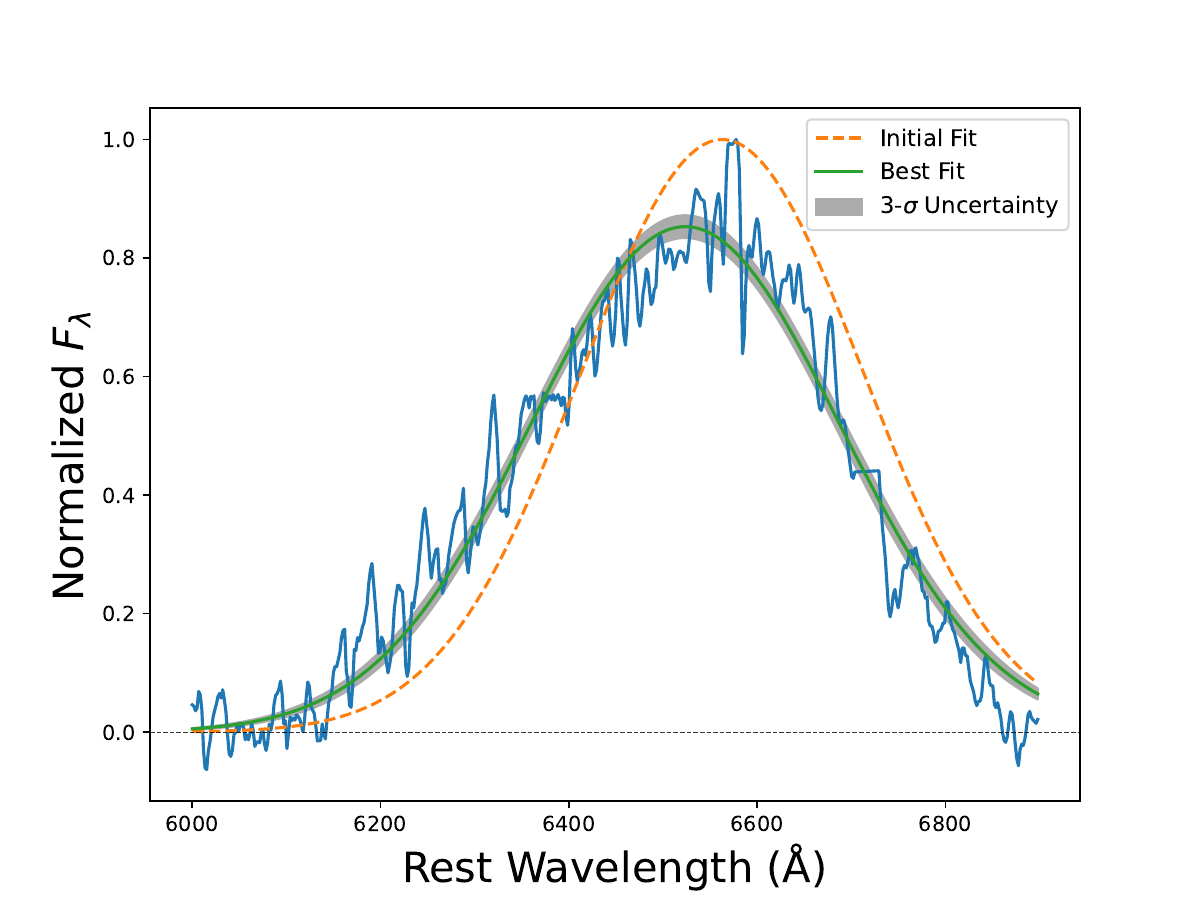}{0.35\textwidth}{Phase: +13d}
}
\gridline{\fig{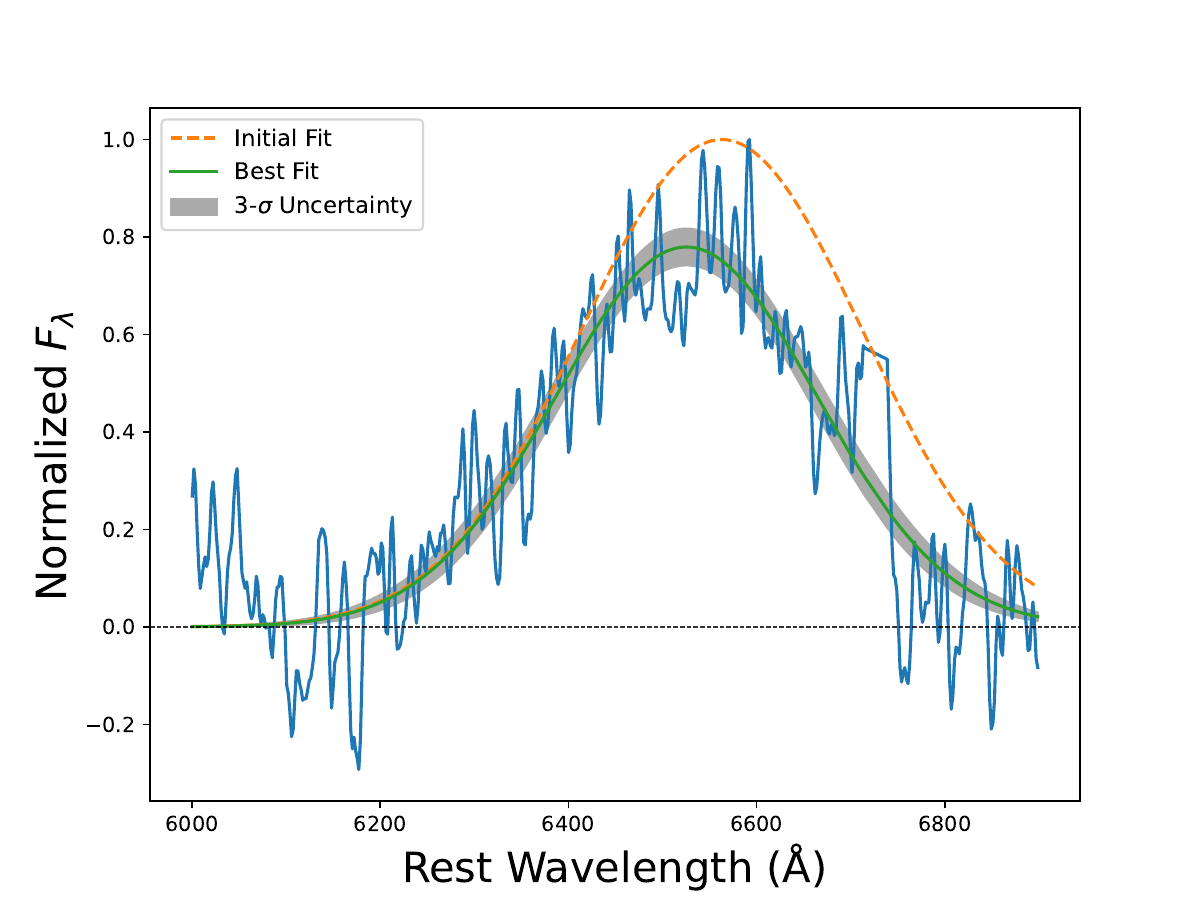}            {0.35\textwidth}{Phase: +15d}
          \fig{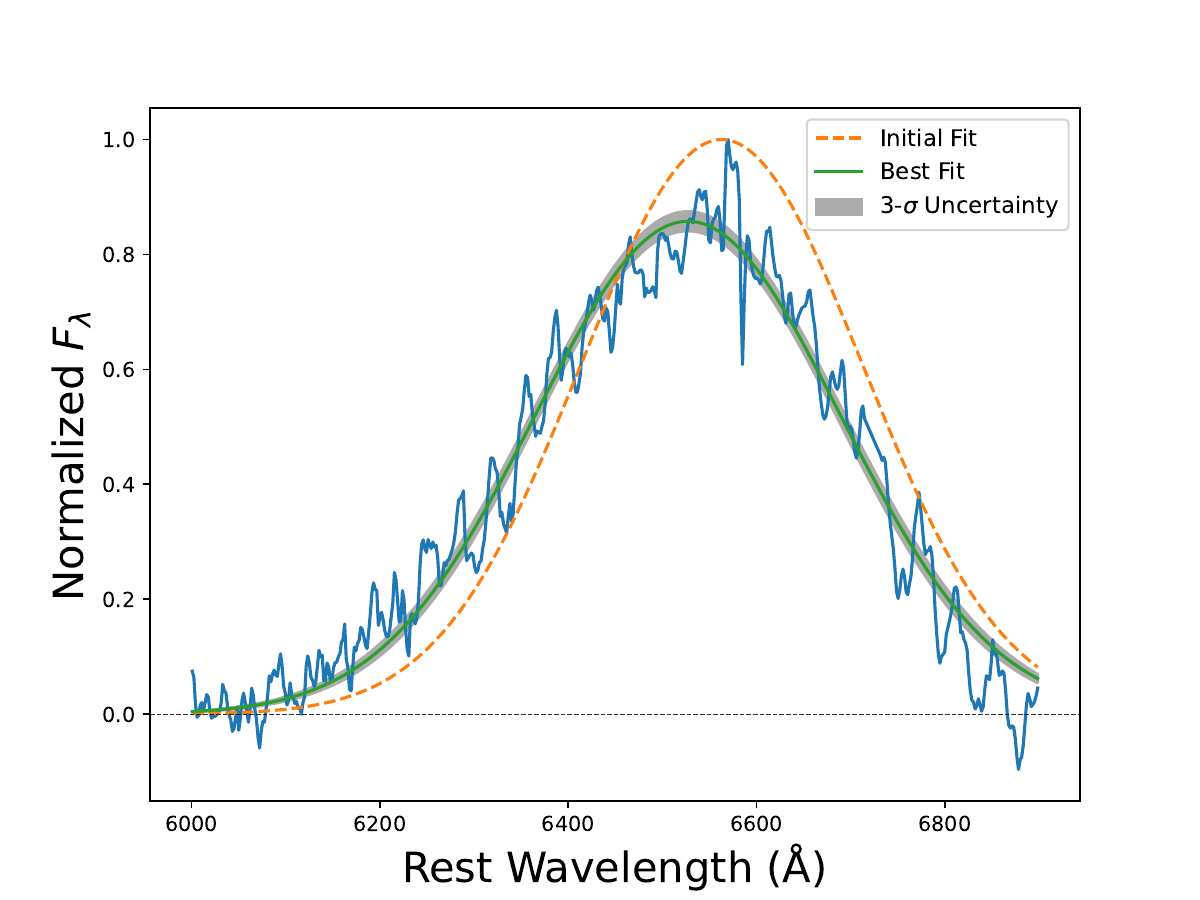}{0.35\textwidth}{Phase: +22d}
          \fig{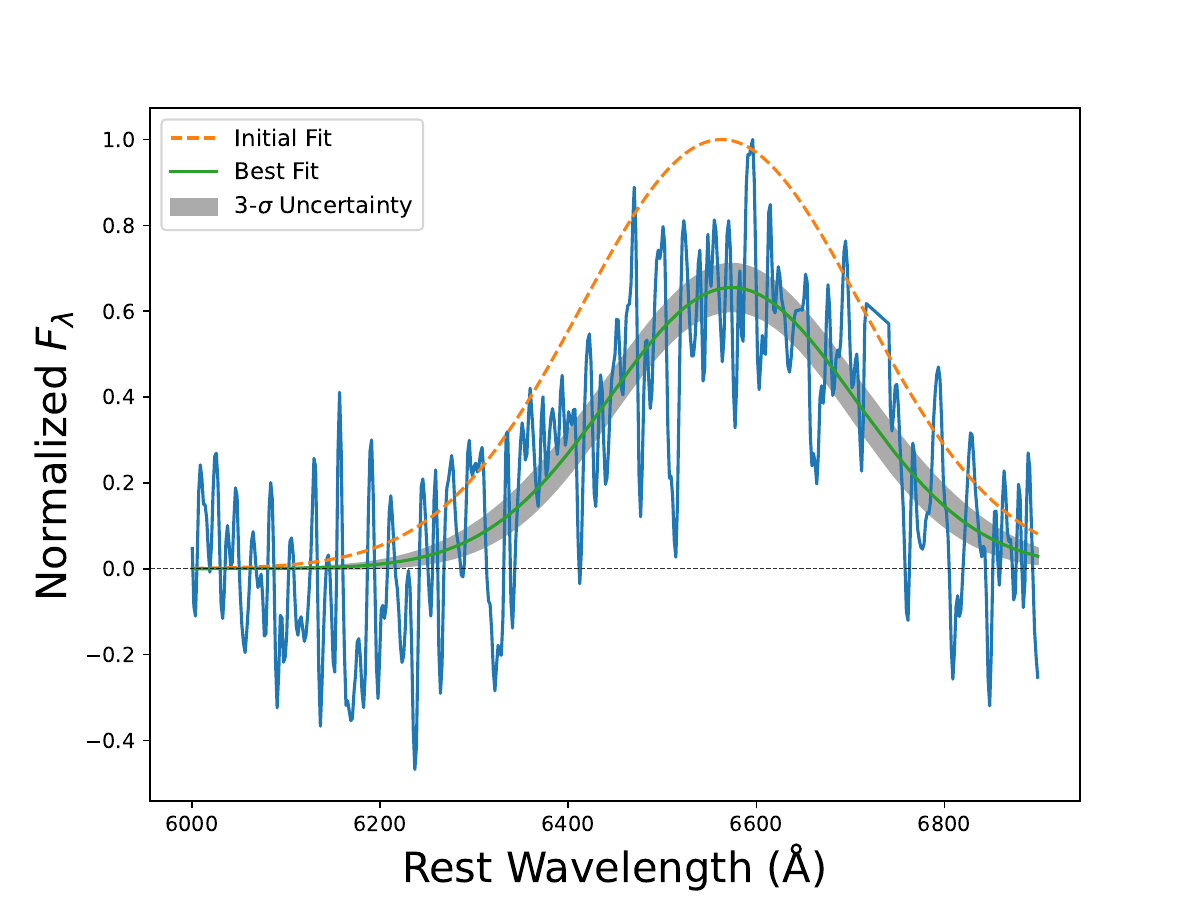}{0.35\textwidth}{Phase: +29d}
}

\end{figure}

\begin{figure}
\gridline{\fig{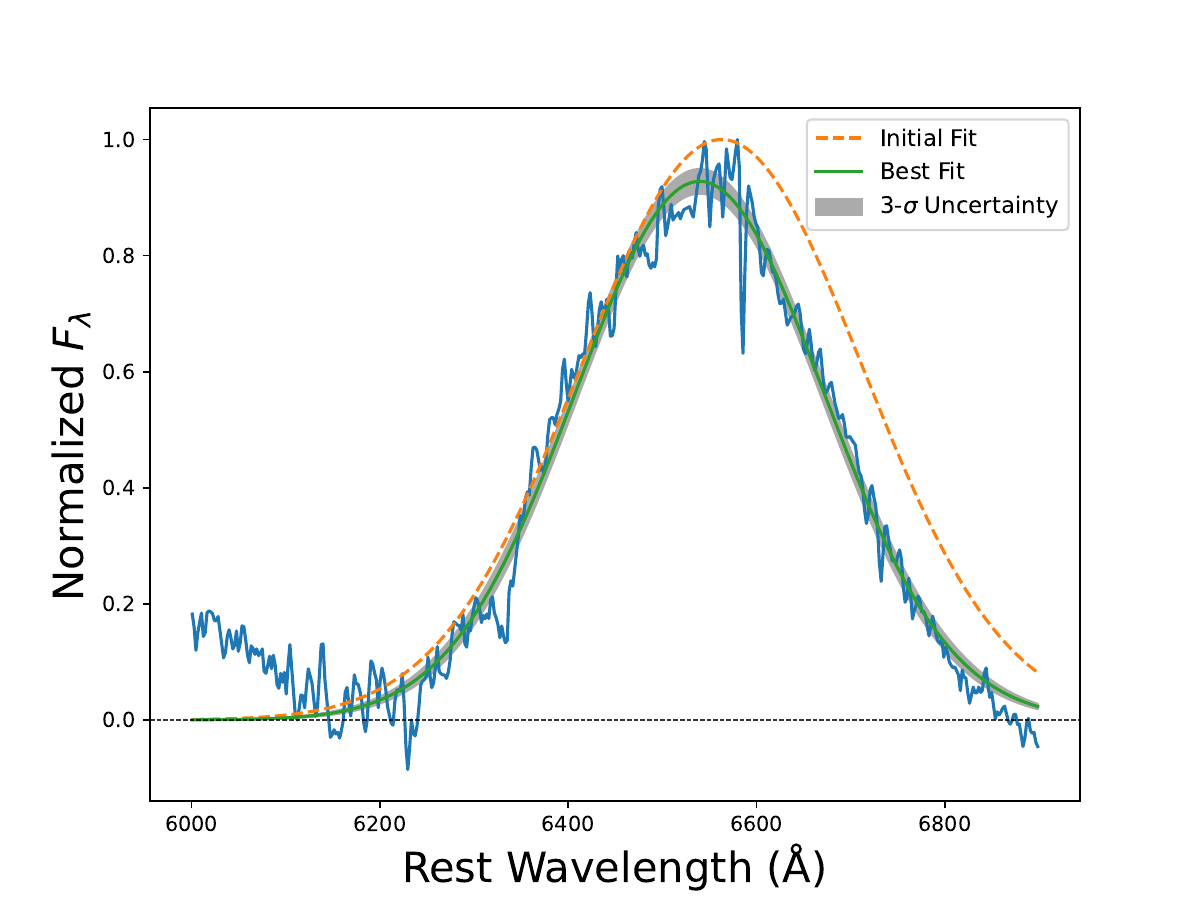}              {0.35\textwidth}{Phase: +31d}
          \fig{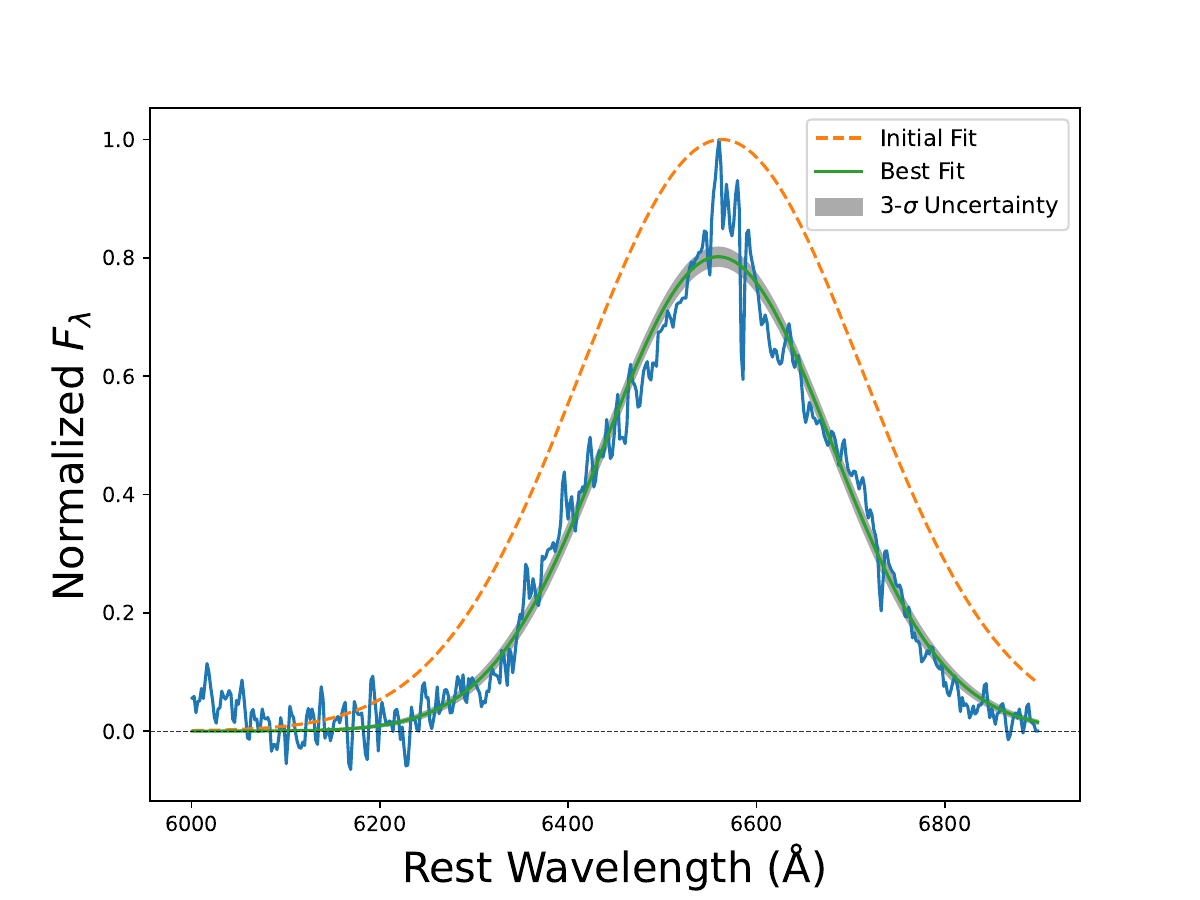}{0.35\textwidth}{Phase: +48d}
          \fig{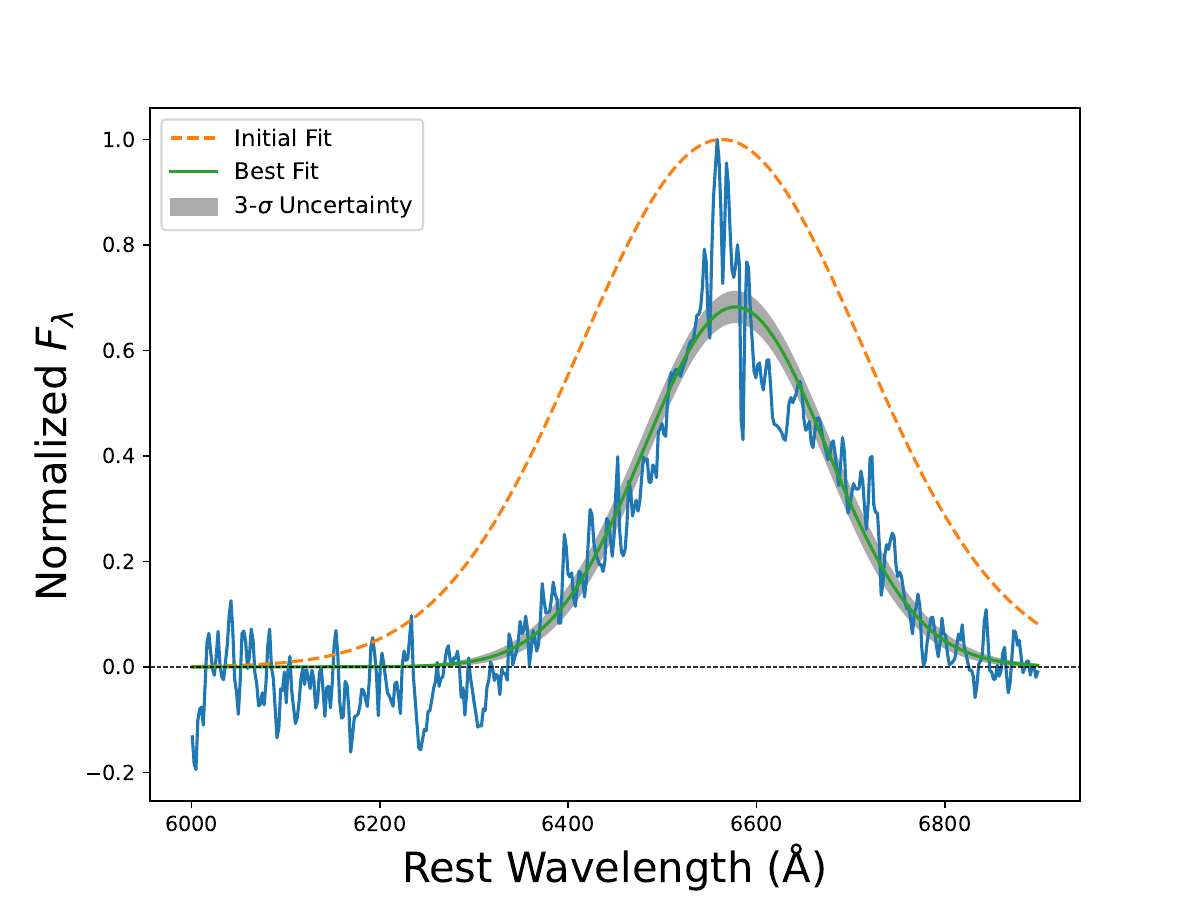}{0.35\textwidth}{Phase: +72d}}
\caption{Best-fit results from fitting the \ha\ line with a Gaussian are shown with solid green lines and the 3$\sigma$ errors are represented in gray. In each case, the dashed line indicates the initial guess for the fit.  Phases are noted in rest-frame days relative to the $g$-band light-curve peak.}
\label{fig:ha-fits}
\end{figure}

\begin{figure}[t]
    \centering
    \includegraphics[width=0.5\textwidth]{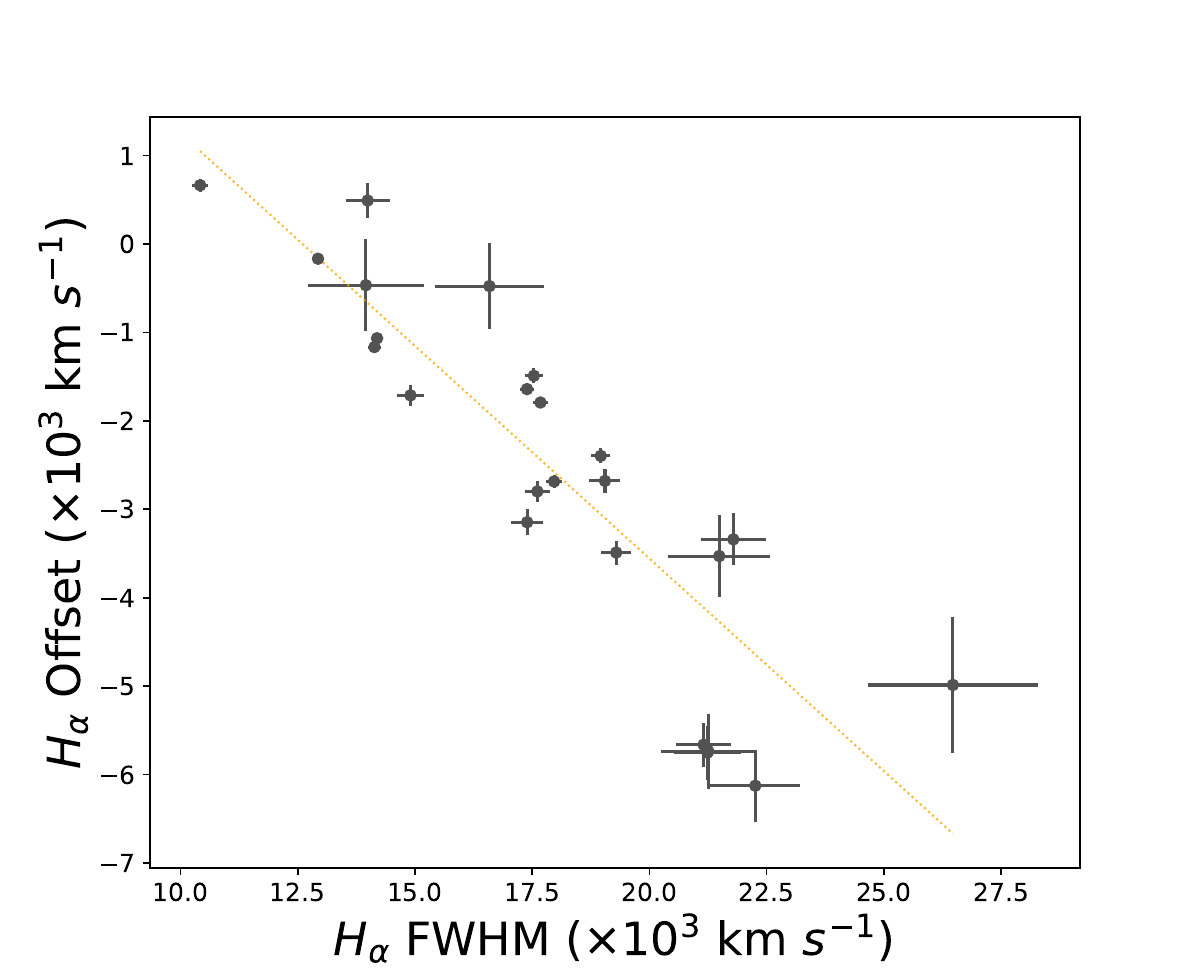}
    \caption{The anticorrelation between the \ha\ offset and FWHM in AT\,2019azh, with a Pearson coefficient of $-0.876$. A linear fit is shown in the dashed orange line, having a slope of $-0.481$.}
    \label{fig:offset-fwhm}
\end{figure}

Figure \ref{fig:offset-fwhm} shows the relation between the offset of the \ha\ line and its FWHM for AT\,2019azh. The plot reveals a strong anticorrelation between these two properties.

\end{document}